\documentclass[fleqn,usenatbib]{mnras}
    
\usepackage{newtxtext,newtxmath}
\usepackage[T1]{fontenc}
\usepackage{ae,aecompl}
\usepackage{placeins}

\usepackage{graphicx}	 
\usepackage{amsmath}	
\usepackage{color}
\usepackage{gensymb}

\usepackage{scalerel,tikz}
\usetikzlibrary{svg.path}
\definecolor{orcidlogocol}{HTML}{A6CE39}
\tikzset{orcidlogo/.pic={
 \fill[orcidlogocol] svg{M256,128c0,70.7-57.3,128-128,128C57.3,256,0,198.7,0,128C0,57.3,57.3,0,128,0C198.7,0,256,57.3,256,128z};
 \fill[white] svg{M86.3,186.2H70.9V79.1h15.4v48.4V186.2z}
 svg{M108.9,79.1h41.6c39.6,0,57,28.3,57,53.6c0,27.5-21.5,53.6-56.8,53.6h-41.8V79.1z M124.3,172.4h24.5c34.9,0,42.9-26.5,42.9-39.7c0-21.5-13.7-39.7-43.7-39.7h-23.7V172.4z}
 svg{M88.7,56.8c0,5.5-4.5,10.1-10.1,10.1c-5.6,0-10.1-4.6-10.1-10.1c0-5.6,4.5-10.1,10.1-10.1C84.2,46.7,88.7,51.3,88.7,56.8z};
}}
\newcommand\orcidicon[1]{\href{https://orcid.org/#1}{\mbox{\scalerel*{
\begin{tikzpicture}[yscale=-1,transform shape]
\pic{orcidlogo};
\end{tikzpicture}
}{|}}}}

\newcommand{\aref}[1]{\hyperref[#1]{Appendix~\ref{#1}}}

\usepackage[normalem]{ulem}

\definecolor{darkgreen}{rgb}{0.13, 0.55, 0.13}
\definecolor{brown}{rgb}{0.65, 0.16, 0.16}

\title[Metallicity gradients in galaxies]{The physics of gas phase metallicity gradients in galaxies}

\author[P. Sharda et al.]{Piyush Sharda$^{\orcidicon{0000-0003-3347-7094}\,1,2}$\thanks{piyush.sharda@anu.edu.au (PS)},
Mark R. Krumholz$^{\orcidicon{0000-0003-3893-854X}\,1,2}$\thanks{mark.krumholz@anu.edu.au (MRK)},
Emily Wisnioski$^{\orcidicon{0000-0003-1657-7878}\,1,2}$\thanks{emily.wisnioski@anu.edu.au (EW)},
John C. Forbes$^{\orcidicon{0000-0002-1975-4449}\,3}$, 
\newauthor
Christoph Federrath$^{\orcidicon{0000-0002-0706-2306}\,1,2}$, and
Ayan Acharyya$^{\orcidicon{0000-0003-4804-7142}\,1,2,4}$\\
$^{1}$Research School of Astronomy and Astrophysics, Australian National University, Canberra, ACT 2611, Australia\\
$^{2}$Australian Research Council Centre of Excellence for All Sky Astrophysics in 3 Dimensions (ASTRO 3D), Australia\\
$^{3}$Center for Computational Astrophysics, Flatiron Institute, New York, NY 10010, USA\\
$^{4}$Department of Physics and Astronomy, Johns Hopkins University, Baltimore, MD 21218, USA
}

\date{Accepted 2021 January 26. Received 2021 January 26; in original form 2020 November 18}
    
\pubyear{2021}

\begin{document}
\label{firstpage}
\pagerange{\pageref{firstpage}--\pageref{lastpage}}
\maketitle

\begin{abstract}
We present a new model for the evolution of gas phase metallicity gradients in galaxies from first principles. We show that metallicity gradients depend on four ratios that collectively describe the metal equilibration timescale, production, transport, consumption, and loss. Our model finds that most galaxy metallicity gradients are in equilibrium at all redshifts. When normalized by metal diffusion, metallicity gradients are governed by the competition between radial advection, metal production, and accretion of metal-poor gas from the cosmic web. The model naturally explains the varying gradients measured in local spirals, local dwarfs, and high-redshift star-forming galaxies. We use the model to study the cosmic evolution of gradients across redshift, showing that the gradient in Milky Way-like galaxies has steepened over time, in good agreement with both observations and simulations. We also predict the evolution of metallicity gradients with redshift in galaxy samples constructed using both matched stellar masses and matched abundances. Our model shows that massive galaxies transition from the advection-dominated to the accretion-dominated regime from high to low redshifts, which mirrors the transition from gravity-driven to star formation feedback-driven turbulence. Lastly, we show that gradients in local ultraluminous infrared galaxies (major mergers) and inverted gradients seen both in the local and high-redshift galaxies may not be in equilibrium. In subsequent papers in this series, we show that the model also explains the observed relationship between galaxy mass and metallicity gradients, and between metallicity gradients and galaxy kinematics.
\end{abstract}

\begin{keywords}
galaxies: evolution – galaxies: ISM – galaxies: abundances –  ISM: abundances – (\textit{ISM:}) \ion{H}{ii} regions – galaxies: fundamental parameters
\end{keywords}

\section{Introduction}
\label{s:intro}
Metals act as tracers of the formation and assembly history of galaxies. Tracking their evolution is crucial to understanding the various pathways a galaxy takes while it forms \citep{2002ARA&A..40..487F}. Metals are produced in galaxies through supernovae \citep{2000ARA&A..38..191H,2002RvMP...74.1015W}, asymptotic giant branch (AGB) stars \citep{2003ARA&A..41..391V,2005ARA&A..43..435H}, neutron star mergers \citep{2017ARNPS..67..253T}, etc., and are consumed by low mass stars and retained in stellar remnants \citep{2006ApJ...653.1145K,2016ApJ...821...38S}. Apart from this in-situ metal production and consumption \citep{1975MNRAS.172...13P}, metals can also be lost through outflows in the form of galactic winds \citep{1990ApJS...74..833H,2005ARA&A..43..769V,2018Galax...6..138R,2018MNRAS.481.1690C}, or transported into the galaxy from the circumgalactic (CGM) and the intergalactic medium (IGM, \citealt{2017ApJ...837..169P,2017ARA&A..55..389T}), or during interactions with other galaxies, like flybys or mergers (e.g., \citealt{2012ApJ...746..108T,2020MNRAS.tmp.2024G}). All of these processes can be classified into four main categories: metal production (through star formation and supernovae), metal consumption (through stellar remnants and low mass stars), metal transport (through advection, diffusion, and accretion), and metal loss (galactic winds and outflows).

The distribution of metals within galaxies places important constraints on galaxy formation \citep{2014A&ARv..22...71S}. One of the strongest pieces of evidence for the inside-out galaxy formation scenario is the existence of negative metallicity gradients (in the radial direction) in both the gas and stars in most galaxies. The presence of such negative radial gradients is easy to understand: in the inside-out scenario, the centre, \textit{i.e.,} the nucleus of the galaxy forms first, and the disc subsequently forms and evolves in time. The nucleus undergoes greater astration, leading to the presence of more metals in the centre as compared to the disc, thus establishing a negative gradient. Such gradients were first observed and quantified through nebular emission lines in H~\textsc{ii} regions in the interstellar medium (ISM) by \cite{1942ApJ....95...52A}, \cite{1971ApJ...168..327S} and \cite{1983MNRAS.204...53S}. The decrease in metallicity is approximately exponential with galactocentric radius \citep{1989ApJ...339..700W,1992ApJ...390L..73Z}, yielding a linear gradient in logarithmic space, with units of $\rm{dex\,kpc^{-1}}$. Since these early works, metallicity gradients have been measured for thousands of galaxies, both in stars and gas (see recent reviews by \citealt{2019ARA&A..57..511K}, \citealt{2019A&ARv..27....3M} and \citealt{2020ARA&A..5812120S}). The stellar metallicity gradients are typically characterised by the abundance of iron, and are written in the form $d\log_{10}\mathrm{(Fe/H)}/dr$, whereas the gas phase metallicity gradients are characterised by the abundance of oxygen, and written as $d\log_{10}\mathrm{(O/H)}/dr$, where $r$ is the galactocentric radius. Hereafter, we will only discuss the gas phase metallicity gradients in galaxies.

In the local Universe, samples of metallicity gradient measurements have been dramatically expanded by three major surveys: CALIFA (Calar Alto Legacy Integral Field Area, \citealt{2012A&A...538A...8S}), MaNGA (Mapping nearby Galaxies at Apache Point Observatory, \citealt{2015ApJ...798....7B}), and SAMI (Sydney-AAO Multi-object Integral-field spectrograph, \citealt{2015MNRAS.447.2857B}). These surveys show that local galaxies contain predominantly negative metallicity gradients, with typical values ranging between $0$ and $-0.1\,\rm{dex\,kpc^{-1}}$. Measurements in high-$z$ ($ z \lesssim 3$) galaxies are more challenging, and the samples are correspondingly smaller, but rapidly expanding \citep{2012A&A...539A..93Q,2012MNRAS.426..935S,2013ApJ...765...48J,2014MNRAS.443.2695S,2014A&A...563A..58T,2016ApJ...820...84L,2016ApJ...827...74W,2018ApJS..238...21F,2019ApJ...882...94W,2020ApJ...900..183W,2020MNRAS.492..821C,2020arXiv201103553S,2021MNRAS.500.4229G}.  

Theoretical efforts to understand these observations are still in their infancy, with most effort thus far dedicated to understanding galaxies' global metallicities \citep[e.g.,][]{2008ApJ...674..151E,2011MNRAS.417.2962P,2012MNRAS.421...98D, 2013ApJ...772..119L,2013MNRAS.430.2891D,2016MNRAS.463.2020H,2020arXiv200901935W,2021MNRAS.500.3394F} rather than their metallicity gradients. Early work on metallicity gradients was tuned to reproduce the present-day Milky Way metallicity gradient \citep[e.g.][]{1997ApJ...477..765C,2001ApJ...554.1044C, 2000MNRAS.313..338P}, and thus offers relatively little insight into how metallicity gradients have evolved over cosmic time and in galaxies with differing histories. More recent work has attempted to address the broader sample of galaxies, using physical models that include a range of processes: cosmological accretion, mass-loaded galactic winds, \textit{in situ} metal production by stars, and radial gas flows \citep[e.g.,][]{2013MNRAS.435.2918M, 2013ApJ...765...48J, 2015MNRAS.448.2030H,2015MNRAS.451..210C, 2015MNRAS.450..342K,2016MNRAS.455.2308P,2019MNRAS.487..456B,2021arXiv210106833K}. However, these works generally suffer from a problem of being under-constrained: the models generally involve multiple free functions (e.g., the radial inflow velocity or mass loading factor as a function of radius and time) that are not constrained by any type of independent physical model, and fits of these free functions to the data are often non-unique, leaving in doubt which physical processes are important in which types of galaxies. 

Moreover, not all models include all possible physical processes, making them difficult to compare. For example, some of the models above assume that galactic wind metallicities are equal to ISM metallicities, contrary to observational evidence \citep[e.g.,][]{2002ApJ...574..663M,Strickland09a, 2018MNRAS.481.1690C}. Many models do not include metal transport processes like advection (flow of metals that are carried by a bulk flow of gas) or turbulent diffusion (metal flow that occurs due to turbulent mixing of gas with non-uniform metal concentration; e.g., \citealt{de-Avillez02a,2009MNRAS.392.1381G, 2012ApJ...758...48Y,2013MNRAS.434.3142A,Kubryk13a, Petit15a,2018MNRAS.481.5000A,2019MNRAS.483.3810R}), despite modelling showing that these effects play an important role in setting metallicity gradients \citep{2012ApJ...754...48F}.

These problems have been partly alleviated by recent radially-resolved semi-analytic models \citep{2013MNRAS.430.1447K,2013MNRAS.434.1531F,2014MNRAS.438.1552F,2019MNRAS.487.3581F,2020MNRAS.491.5795H,2020arXiv201104670Y} and cosmological simulations with enough resolution to capture disc radial structure \citep{2012A&A...540A..56P, 2013A&A...554A..47G, 2017MNRAS.466.4780M, 2019MNRAS.482.2208T, 2020arXiv200710993H,2021arXiv210206220B}, which do at least attempt to model the dynamics of gas in galaxies self-consistently. The general result of these models (as summarised in Figure 8 of \citealt{2020MNRAS.492..821C}), is that there is only a mild evolution in metallicity gradients between $0 \leq z \leq 4$, with a slight steepening toward the present day. However, the physical origin of these results, and insights from the numerical results in general, have yet to be distilled into analytic models that we can use to understand the overall trends in the data. Thus, to this date, we lack a model that can explain the occurrence of metallicity gradients in a diverse range of galaxies from first principles. This leaves many interesting questions around gas phase metallicity gradients unanswered.  

Motivated by this, we present a new theory of gas phase metallicity gradient evolution in galaxies from first principles. As with all theories of metallicity gradients, ours requires a galaxy evolution model that describes the gas in galactic discs as an input. For the purposes of developing the theory, we use the unified galactic disc model of \cite{2018MNRAS.477.2716K}, which has been shown to reproduce a large number of observations of gas kinematics relevant to metallicity gradients, including the radial velocities of gas in local galaxies \citep{2016MNRAS.457.2642S}, the correlation between galaxies' gas velocity dispersions and star formation rates \citep[SFRs; e.g.,][]{Johnson18a,2019MNRAS.486.4463Y,2020MNRAS.495.2265V}, and the evolution of velocity dispersion with redshift \citep[e.g.,][]{2019ApJ...880...48U}. However, our metallicity model is a standalone model into which we can incorporate any galaxy evolution model. In this paper, we present the basic formalism and results of the model, and use them to explain the evolution of metallicity gradients with redshift; in two follow-up papers we first use the model developed here to explain the dependence of metallicity gradients on galaxy mass that is observed in the local Universe \citep{2020aMNRAS.xxx..xxxS}, and then use the model to predict the existence of a correlation between galaxy kinematics and metallicity gradients, which we validate against observations \citep{2020bMNRAS.xxx..xxxS}.

We arrange the rest of the paper as follows: \autoref{s:metalevolve} describes the theory of metal evolution, \autoref{s:gradients} describes the equilibrium metallicity gradients generated by the theory for different types of galaxies both in the local and the high-$z$ Universe, \autoref{s:cosmic} combines the local and high-$z$ predictions of the model to describe the cosmic evolution of metallicity gradients, and \autoref{s:modellimitations} discusses the limitations of the model, including special cases where the metallicity gradients may not be in equilibrium and thus the model may or may not apply. Finally, we present our conclusions in \autoref{s:conclusions}. For the purpose of this paper, we use $Z_{\odot} = 0.0134$, corresponding to $12 + \log_{10}\rm{O/H} = 8.69$ \citep{2009ARA&A..47..481A}, Hubble time $t_{\rm{H(0)}} = 13.8\,\rm{Gyr}$ \citep{2018arXiv180706209P}, and follow the flat $\Lambda$CDM cosmology: $\Omega_{\rm{m}} = 0.27$, $\Omega_{\rm{\Lambda}} = 0.73$, $h=0.71$, and $\sigma_8 = 0.81$ \citep{2003MNRAS.339..289S}.

\section{Evolution of metallicity}
\label{s:metalevolve}

For convenience, we collect all of the symbols we define in this section in \autoref{tab:tab1} and \autoref{tab:tab2}.

\begin{table*}
\centering
\caption{List of fiducial parameters in the model that are common to all galaxies. All of these parameters are adopted from Table~1 in \citet{2018MNRAS.477.2716K}, except for $y$ and $f_{\mathrm{R,inst}}$, which we adopt from \protect\cite{2019MNRAS.487.3581F}.}
\begin{tabular}{|l|l|l|r}
\hline
Parameter & Description & Reference equation & Fiducial value \\
\hline
$y$ & Metal yield & \autoref{eq:source} & 0.028\\
$f_{\mathrm{R,inst}}$ & Fraction of metals produced that are locked in stars & \autoref{eq:source} & 0.77\\
$\phi_y$ & Yield reduction factor & \autoref{eq:phiy} & $0.1$--$1.0$ \\
$r_0$ & Reference radius per kpc & \autoref{eq:main_nondimx} & 1\\
$Q_{\rm{min}}$ & Minimum Toomre $Q$ parameter & \autoref{eq:toomreQ} & $1-2$\\
$\phi_Q$ & 1 $+$ ratio of gas to stellar Toomre $Q$ parameter & \autoref{eq:gas_scale_height} & 2\\
$\epsilon_{\mathrm{ff}}$ & Star formation efficiency per free-fall time & \autoref{eq:sigma_SFR} & 0.015\\
$\phi_{\mathrm{mp}}$ & Ratio of the total to the turbulent pressure at the disc midplane & \autoref{eq:sigma_SFR} & 1.4\\
$f_{\mathrm{B}}$ & Universal baryonic fraction & \autoref{eq:cosmicaccr} & 0.17\\
$\eta$ & Scaling factor for the rate of turbulent dissipation & \autoref{eq:radialinflow} & 1.5 \\
$\phi_{\mathrm{nt}}$ & Fraction of velocity dispersion due to non-thermal motions &  \autoref{eq:radialinflow} & 1\\
\hline
\end{tabular}
\label{tab:tab1}
\end{table*}

\begin{table*}
\centering
\caption{List of fiducial parameters that are specific to a galaxy type as listed in the last 4 columns.}
\begin{tabular}{|l|l|l|c|c|c|c|r}
\hline
Parameter & Description & Reference & Units & Local & Local & Local & High-$z$\\
 & & equation & & spiral & dwarf & ULIRG & ($z=2$)\\
\hline
$x_{\rm{max}}$ & Outer edge of the star-forming disc & ... & $r_0$ & 15 & 6 & 3 & 10\\
$v_{\phi}$ & Rotational velocity of the galaxy & \autoref{eq:orbital} & $\mathrm{km\,s^{-1}}$ & 200 & 60 & 250 & 200 \\
$f_{g,Q}$ & Effective gas fraction in the disc & \autoref{eq:toomreQ} & ... & 0.5 & 0.9 & 1 & 0.7\\
$\sigma_g$ & Gas velocity dispersion & \autoref{eq:toomreQ1} & $\mathrm{km\,s^{-1}}$ & 10 & 7 & 60 & 40\\
$\beta$ & Galaxy rotation curve index & \autoref{eq:gas_surface_density} & ... & 0 & 0.5 & 0.5 & 0\\
$f_{\mathrm{sf}}$ & Fraction of star-forming molecular gas & \autoref{eq:sigma_SFR} & ... & 0.5 & 0.2 & 1 & 1\\
$f_{g,P}$ & Fraction of mid-plane pressure due to disc self-gravity & \autoref{eq:sigma_SFR} & ... & 0.5 & 0.9 & 1 & 0.7\\
$M_{\rm{h}}$ & Halo mass & \autoref{eq:haloaccr} & $\mathrm{M_{\odot}}$ & $10^{12}$ & $10^{10}$ & $10^{12}$ & $5\times10^{11}$ \\
$c$ & Halo concentration parameter & \autoref{eq:halomass} & ... & 10 & 15 & 10 &  13\\
$\sigma_{\mathrm{sf}}$ & Gas velocity dispersion due to star formation feedback & \autoref{eq:radialinflow} & $\mathrm{km\,s^{-1}}$ & 7 & 5 & 9 & 8.5\\
\hline
\end{tabular}
\label{tab:tab2}\\
\end{table*}

\subsection{Evolution equations}
\label{s:modelevolve_equations1}
Let us start by defining $\rho_Z$ to be the volume density of metals at some point in space; this is related to the metallicity, $Z$, and gas density, $\rho_g$, by
\begin{equation}
    \rho_Z = Z \rho_g.
\end{equation}
The density of the metals can change due to transport -- via advection with the gas or diffusion through the gas -- and due to sources and sinks (e.g., production of new metals by stars or consumption during star formation). The conservation equation for metal mass is then
\begin{equation}
    \frac{\partial \rho_Z}{\partial t} + \nabla \cdot \left(\mathbf{v}\rho_Z + \mathbf{j}_Z \right) = s_Z.
\end{equation}
Here, $\mathbf{v}$ is the gas velocity, $\mathbf{j}$ is the flux density of metals as a result of diffusion, and $s_Z$ represents the source and sink terms. The central assumption of diffusion is that the diffusive flux is proportional to minus the gradient of the quantity being diffused (e.g., \citealt{2012ApJ...758...48Y,2018MNRAS.475.2236K}). The slight subtlety here is that what should diffuse is not the density of metals, but the concentration of metals, \textit{i.e.,} the flux  only depends on the gradient of $Z$. We can therefore write down the diffusive flux as
\begin{equation}
    \mathbf{j}_Z = -\kappa \rho_g \nabla Z\,,
\end{equation}
where $\kappa$ is the diffusion coefficient (with dimensions of mass/length$^2$). Inserting this into the continuity equation, we now have
\begin{equation}
    \frac{\partial \rho_Z}{\partial t} + \nabla \cdot \left(\mathbf{v}\rho_Z - \kappa \rho_g \nabla Z \right) = s_Z\,.
\end{equation}

We can now specialise to the case of a disc. Firstly, we assume that the disc is thin, so we can write $\rho_g$ in terms of the surface density as $\Sigma_g = \int \rho_g dz$. We choose our coordinate system so that the disc lies in the $xy$ plane. Integrating all quantities in the \textit{z} direction, the equation of mass conservation becomes
\begin{equation}
    \frac{\partial \Sigma_Z}{\partial t} + \nabla \cdot \left(\mathbf{v}\Sigma_Z - \kappa \Sigma_g \nabla Z \right) = S_Z,
\end{equation}
where $\Sigma_Z$ is the metal surface density, $\nabla$ contains only the derivatives in the $xy$ plane, and $S_Z=\int s_Z\, dz$. Assuming cylindrical symmetry, this reduces to,
\begin{equation}
    \frac{\partial \Sigma_Z}{\partial t} + \frac{1}{r} \frac{\partial}{\partial r} \left(r v \Sigma_Z - r \kappa \Sigma_g \frac{\partial Z}{\partial r} \right) = S_Z,
\end{equation}
where $v$ represents the radial component of the velocity. It is helpful to rewrite the velocity in terms of the inward mass flux across the circle at radius $r$, which is
\begin{equation}
    \dot{M} = -2\pi r v \Sigma_g,
\label{eq:mrk8}
\end{equation}
where we have adopted a sign convention whereby $\dot{M} > 0$ corresponds to inward mass flow\footnote{This is the opposite of the sign convention used in \cite{2012ApJ...754...48F,2014MNRAS.443..168F}, but consistent with the one used in \cite{2018MNRAS.477.2716K}.}. This gives
\begin{equation}
        \frac{\partial \Sigma_Z}{\partial t} - \frac{1}{2\pi r}\frac{\partial}{\partial r}\left(\dot{M} Z\right) - \frac{1}{r}\frac{\partial}{\partial r}\left(r\kappa\Sigma_g \frac{\partial Z}{\partial r}\right) = S_Z.
        \label{eq:cont_metals}
\end{equation}

Similarly, since star formation is the process that is responsible for the source term, it is convenient to parameterize $S_Z$ in terms of the star formation rate. We adopt the instantaneous recycling approximation \citep{1980FCPh....5..287T}, whereby some fraction, $f_{\rm R,inst}$, of the mass incorporated into stars is assumed to be left in long-lived remnants (compact objects and low-mass stars), and the remainder of the mass is returned instantaneously to the ISM through Type II supernovae, enriched by newly formed metals with a yield $y$. Under this approximation, we have
\begin{equation}
    S_Z = \left(y - f_{\rm R,inst} Z - \mu Z_w\right) \dot{\Sigma}_\star,
\label{eq:source}
\end{equation}
where $\dot{\Sigma}_\star$ is the star formation rate surface density. The last term in \autoref{eq:source} represents loss of metals into a galactic wind; here $\mu$ is the mass loading factor of the wind (\textit{i.e.,} the wind mass flux is $\mu \dot{\Sigma}_\star$) and $Z_w$ is the metallicity of the wind. Following \citet[equation~41]{2019MNRAS.487.3581F}, we further parameterize the wind metallicity as
\begin{equation}
    Z_w = Z + \xi \frac{y}{\mathrm{max}(\mu,1-f_{\mathrm{R,inst}})},
\label{eq:Zw}
\end{equation}
where the $1 - f_{\rm{R,inst}}$ limit specifies the minimum mass that can be ejected if some metals are ejected directly after production. The parameter $\xi$, which is bounded in the range $0\leq \xi \leq 1$, specifies the fraction of metals produced that are directly ejected from the galaxy before they are mixed into the ISM. So, $\xi=0$ corresponds to a situation when the metallicity of the wind equals the metallicity of the ISM, whereas $\xi=1$ corresponds to the regime when all the metals produced in the galaxy get ejected in winds. \cite{2014MNRAS.438.1552F,2014MNRAS.443..168F} introduced $\xi$ to relax the assumption that metals fully mix with the ISM before winds are launched, so that $Z_w=Z$. A number of authors have shown that this assumption leads to severe difficulties in explaining observations, particularly in low-mass systems \citep{1993A&A...277...42P,1994MNRAS.270...35M,1999ApJ...513..142M,2001MNRAS.322..800R,2008A&A...489..555R,2002ApJ...574..663M,2017ApJ...835..136R}. 

We can further simplify by writing down the continuity equation for the total gas surface density $\Sigma_g$, which is \autoref{eq:cont_metals} with $Z$ fixed to unity and $y=0$, with an additional term for cosmic accretion,\footnote{Note that \autoref{eq:sigmagas_evol} is identical to equation 1 of \cite{2019MNRAS.487.3581F} except that \citeauthor{2019MNRAS.487.3581F} adopt instantaneous recycling only for Type II supernovae, and not for metals returned on longer timescales (e.g., Type Ia or AGB winds). While this approach is feasible in simulations and semi-analytic models, it renders analytic models of the type we present here intractable. However, this does not make a significant difference for our work because the most common gas phase metallicity tracer, O, comes almost solely from Type II supernovae. One area where our approach might cause concern is at high redshift, where the gradients are often measured through the [\ion{N}{ii}]/\ion{H}{$\alpha$} emission line ratio, because most of the N comes from AGB stars and is released over Gyr or longer timescales \citep{2005ARA&A..43..435H}.}
\begin{equation}
    \frac{\partial \Sigma_g}{\partial t} - \frac{1}{2\pi r} \frac{\partial\dot{M}}{\partial r} = \dot \Sigma_{\mathrm{cos}} -\left(f_{\rm R,inst} + \mu\right)\dot{\Sigma}_\star\,,
    \label{eq:sigmagas_evol}
\end{equation}
where $\dot \Sigma_{\rm{cos}}$ is the cosmic accretion rate surface density onto the disc \citep{2010MNRAS.406.2325O,2010PhR...495...33B}. If we now use this to evaluate $\partial \Sigma_Z/\partial t = \Sigma_g (\partial Z/\partial t) + Z (\partial\Sigma_g/\partial t)$ in \autoref{eq:cont_metals}, the result is
\begin{equation}
    \Sigma_g \frac{\partial Z}{\partial t} - \frac{\dot{M}}{2\pi r}\frac{\partial Z}{\partial r} - \frac{1}{r}\frac{\partial}{\partial r}\left(r \kappa \Sigma_g \frac{\partial Z}{\partial r}\right)
    = \phi_y y \dot{\Sigma}_\star - Z\dot \Sigma_{\mathrm{cos}}\,,
    \label{eq:metal_continuity_2}
\end{equation}
where
\begin{equation}
    \phi_y = 1 - \frac{\mu\xi}{\max(\mu,1-f_{\rm R,inst})}.
\label{eq:phiy}
\end{equation}
We refer to $\phi_y$, which is bounded in the range $0\leq \phi_y\leq 1$, as the yield reduction factor. Note that $f_{\rm{R,inst}}$ only appears in $\phi_y$, implying that metals locked in stars are unimportant for the radial profile of metallicity as long as $\mu > 1-f_{\mathrm{R,inst}}$.

From left to right, we can interpret the terms in \autoref{eq:metal_continuity_2} as follows: the first is the rate of change in the metallicity at fixed gas surface density; the second represents the change due to advection of metals through the disc; the third represents the change due to diffusion of metals; finally, the terms on the right hand side are: (1.) the change in metallicity due to metal production in stars, with an effective yield $\phi_y y$ that is reduced relative to the true yield $y$ by the factor $\phi_y$, and (2.) the change in metallicity in the disc due to cosmic accretion of metal-poor gas. 

The term $\phi_y$ represents the factor by which the effective metal yield is reduced because some fraction of metals directly escape the galaxy before they mix with the ISM. Higher values of $\phi_y$ imply that metals are well-mixed into the ISM, whereas lower values imply that the yield is significantly reduced by preferential ejection of unmixed metals. However, $\phi_y$ does not equate to the mass loading factor $\mu$: galaxies with heavily mass-loaded winds (high $\mu$) may still have $\phi_y$ close to unity if metals mix efficiently before the winds are launched; conversely, galaxies with weakly mass-loaded winds (low $\mu$) may still have small $\phi_y$ if those winds preferentially carry away metals. We discuss the possible range of values for $\phi_y$ in more detail in \autoref{s:metalevolve_krumholz2018} \citep[see also,][]{2020aMNRAS.xxx..xxxS}.

At this point it is helpful to non-dimensionalise the system. We choose a fiducial radius $r_0$, which we will later take to be the inner edge of the disc where the bulge begins to dominate; for now, however, we simply take $r_0$ as a specified constant. We measure position in the disc with the dimensionless variable $x = r/r_0$ and time with $\tau = t\Omega_0$, where $\Omega_0$ is the angular velocity at $r_0$. We further write out the profiles of gas surface density, diffusion coefficient, star formation surface density and cosmic accretion rate surface density as $\Sigma_g = \Sigma_{g0} s_g(x)$, $\kappa = \kappa_0 k(x)$, $\dot{\Sigma}_\star = \dot{\Sigma}_{\star0} \dot{s}_{\star}(x)$, and $\dot{\Sigma}_{\rm{cos}} = \dot{\Sigma}_{\rm{cos}0} \dot c_{\star}(x)$ respectively. Here, the terms subscripted by 0 are the values evaluated at $r = r_0$, and $s_g(x)$, $k(x), \dot{s}_\star(x)$, and $\dot c_{\star} (x)$ are dimensionless functions that are constrained to have a value of unity at $x=1$. Note that, in principle, we could introduce a similar scaling function for $\dot{M}$; we do not do so because both observations \citep{2016MNRAS.457.2642S} and theoretical models \citep{2018MNRAS.477.2716K} suggest that, in steady state, $\dot{M}$ is close to constant with radius within a galactic disc. We express the metallicity as $\mathcal{Z} = Z/Z_{\odot}$.

Using these definitions, we can rewrite \autoref{eq:metal_continuity_2} as a form of the Euler-Cauchy equation \citep{garfken67:math,kreyszig11},
\begin{equation}
    \underbrace{\mathcal{T} s_g \frac{\partial \mathcal{Z}}{\partial \tau}}_{\substack{\text{equilibrium} \\ \text{time}}} - \underbrace{\frac{\mathcal{P}}{x} \frac{\partial \mathcal{Z}}{\partial x}}_\text{advection} - \underbrace{\frac{1}{x}\frac{\partial}{\partial x}\left(x k s_g \frac{\partial \mathcal{Z}}{\partial x}\right)}_\text{diffusion}\,\,
    = \underbrace{\mathcal{S} \dot{s}_\star}_{\substack{\text{production} \\ \text{+} \\ \text{outflows}}} - \underbrace{\mathcal{Z}\mathcal{A}\dot c_{\star}}_\text{accretion}.
    \label{eq:main_nondimx}
\end{equation}
In the above equation, we have suppressed the $x$-dependence of $s_g$, $k, \dot{s}_\star$ and $\dot{c}_\star$ for compactness, and we have defined, 
\begin{eqnarray}
\mathcal{T} & = &
\frac{\Omega_0 r^2_0}{\kappa_0} \\
\mathcal{P} & = &
\frac{\dot{M}}{2\pi \kappa_0 \Sigma_{g0}} \\
\mathcal{S} & = &
r^2_0\frac{\dot{\Sigma}_{\star0} }{\kappa_0\Sigma_{g0}}\left(\frac{\phi_y y}{Z_{\odot}}\right)\\
\mathcal{A} & = &
r^2_0\frac{\dot{\Sigma}_{\mathrm{cos}0}}{\kappa_0\Sigma_{g0}}.
\end{eqnarray}
The four quantities $\mathcal{T},\,\mathcal{P}, \mathcal{S}$ and $\mathcal{A}$ have straightforward physical interpretations: $\mathcal{T}$ is the ratio of the orbital and diffusion timescales, $\mathcal{P}$ is the P\'eclet number of the system, which describes the relative importance of advection and diffusion in fluid dynamics (e.g., \citealt{patankar1980numerical}), $\mathcal{S}$ measures the relative importance of metal production (the numerator) and diffusion (the denominator), and $\mathcal{A}$ measures the relative importance of cosmic accretion and diffusion. $\mathcal{T}$ dictates the time it takes for a given metallicity distribution to reach equilibrium in a galaxy, whereas the other three quantities govern the type and strength of the gradients that form in equilibrium.

We will only look for the steady-state or \textit{equilibrium} solutions to \autoref{eq:main_nondimx}, so we drop the $\partial\mathcal{Z}/\partial \tau$ term. This approach is reasonable because, as we will show below, the equilibration timescale for metals is less than the Hubble time, $t_{\rm{H(z)}}$, for most galaxies. In our model, the time it takes for the metallicity gradient to approach an equilibrium state, $t_{\rm{eqbm}}$, is based on the time it takes for the metal surface density to adjust to changes in metallicity triggered by each of the terms in \autoref{eq:main_nondimx},
\begin{equation}
\frac{1}{t_{\mathrm{eqbm}}} = \Omega_0 \frac{\big\lvert \frac{\mathcal{P}}{x}\frac{\partial \mathcal{Z}}{\partial x}\big\rvert + \big\lvert\frac{1}{x}\frac{\partial}{\partial x}\left(x k s_g\frac{\partial \mathcal{Z}}{\partial x}\right)\big\rvert + \big\lvert\mathcal{S}\dot s_{\star}\big\rvert + \big\lvert\mathcal{Z}\mathcal{A} \dot c_{\star}\big\rvert}{\mathcal{Z}s_g\mathcal{T}}\,.
\label{eq:teqbm}    
\end{equation}
If $t_{\rm{eqbm}} > t_{\rm{H(z)}}$, the metallicity gradient in the galaxy cannot attain equilibrium within a reasonable time, and the model we present below does not apply. While this is a necessary condition for metal equilibrium, it may not be sufficient. This is because if input parameters to the metallicity model (e.g., accretion rate, surface density, etc.) change on timescales much shorter than $t_{\rm{H(z)}}$, the equilibrium of metals will depend on that timescale. For a steady-state model like ours, it is safe to assume this is not the case, since the input galactic disc model in the next Section we use is an equilibrium model. We discuss this condition in more detail in \autoref{s:gradients} and \autoref{s:noneqbm} where we also compare $t_{\rm{eqbm}}$ with the molecular gas depletion time that dictates the star formation timescale.

The accretion of material from the CGM can also impact metallicity in the galactic disc \citep{2012ApJ...745...50W,2012ApJ...753...16K,2012ApJ...754...48F,2015MNRAS.448.1835T,2017ARA&A..55..389T,2019ApJ...884..156S}. While this is an important consideration, in the absence of which `closed-box' galaxy models overestimate metallicity gradients (e.g., \citealt{2007ApJ...658..941D,2013ApJ...771L..19Z,2015MNRAS.450..342K}), it typically adds a floor metallicity at the outer edge of the galactic disc, and is of concern for simulations where the entire (star-forming as well as passive) disc up to tens of $\rm{kpc}$ is considered. CGM metallicity can also be important for long term $\left(0.1-1\,t_{\rm{H(z)}}\right)$ wind recycling \citep{2013MNRAS.431.3373H,2020ApJ...905....4P}, which we do not take into account in this model. As we show later in \autoref{s:modelevolve_solution}, we make use of this effect only as a boundary condition on the metallicity at the outer edge of the disc, and do not include it directly in the evolution equation. 

This completes the basic formulation of the theory of metallicity gradients in galaxies. To further solve for the equilibrium metallicity, we now need a model of the galactic disc. We use the unified galactic disc model of \cite{2018MNRAS.477.2716K} for this purpose. However, we remind the reader that the metallicity evolution described by \autoref{eq:main_nondimx} can be used with other galactic disc models as well.

\subsection{Galactic disc model}
\label{s:metalevolve_krumholz2018}
We use the unified galactic disc model of \cite{2018MNRAS.477.2716K} to further solve for metallicity. This model self-consistently incorporates all of the ingredients that we require as inputs: profiles of $\Sigma_g$, $\dot M$, $\kappa$ and $\dot \Sigma_{\star}$, and the relationship between them. We refer the reader to \cite{2018MNRAS.477.2716K} for full details of the model, and here, we simply extract the portions that are relevant for this work.

Firstly, note that the angular velocity at $r_0$ is simply,
\begin{equation}
    \Omega_0 = \frac{v_{\phi}}{r_0}
\label{eq:orbital}
\end{equation}
where $v_{\phi}$ is the rotational velocity of gas in the galactic disc. We can solve for the gas surface density $\Sigma_g$ by requiring that the Toomre $Q$ parameter for stars and gas is close to 1; formally, following \cite{2014MNRAS.438.1552F}, we take $Q=Q_{\rm min}$, where $Q_{\rm min}\approx 1-2$ is the minimum $Q$ parameter below which gravitational instability prevents discs from falling (e.g., \citealt{2001ApJ...555..301M,2002ApJ...574..663M,2010MNRAS.407.2091G,2013MNRAS.429.2537M,2013MNRAS.433.1389R,2016MNRAS.456.2052I,2016MNRAS.457.1888S,2017MNRAS.469..286R}). This can be re-written as \citep[equation~8]{2018MNRAS.477.2716K},
\begin{equation}
    Q_{\rm min} = f_{g,Q} \times Q_g
\label{eq:toomreQ}
\end{equation}
where $Q_g$ is the Toomre $Q$ parameter for the gas alone, and $f_{g,Q}$ is the effective gas fraction in the disc \citep[equation~9]{2018MNRAS.477.2716K}, which, based on the estimates of $\Sigma_g$ \citep{2015ApJ...814...13M} and gas velocity dispersion $\sigma_g$ \citep{2009ARA&A..47...27K} is $\approx 0.5$ in the Solar neighbourhood. Writing down the Toomre equation \citep{1964ApJ...139.1217T}, this becomes,
\begin{equation}
    f_{g,Q} \frac{\omega_c \sigma_g}{\pi G \Sigma_g} = Q_{\rm min}\,.
\label{eq:toomreQ1}
\end{equation}
Here, $\omega_c$ is the epicyclic frequency given by $\omega_c=\sqrt{2(\beta+1)}\Omega = \sqrt{2(\beta+1)}v_{\phi}/r$, where $\beta$ is the index of the rotation curve given by $\beta = d\ln v_{\phi}/d\ln r$. Following \cite{2018MNRAS.477.2716K} and results from time-dependent numerical solutions for energy equilibrium in galactic discs \citep{2014MNRAS.438.1552F}, we can assume that in the steady-state, $\beta$ and $\sigma_g$ are independent of radius. Thus, we obtain
\begin{equation}
    \Sigma_g = \frac{\sqrt{2(\beta+1)}f_{g,Q}\sigma_g v_{\phi}}{\pi Gr Q_{\rm min}}\,.
\label{eq:gas_surface_density}
\end{equation}
This solution provides a $1/r$ dependence for $\Sigma_g$ that is somewhat at odds with observations that find an exponential dependence of $\Sigma_g$ \citep{2012ApJ...756..183B}. However, these observations trace the entire disc (using CO as well as \ion{H}{i}) and the $\Sigma_g$ profiles show a large scatter in the inner disc, which is the focus of our work. Given these findings, we cannot conclude that a $1/r$ profile of $\Sigma_g$ is unrealistic, and therefore continue to use it for our work. The quantities $\Sigma_{g0}$ and $s_g(x)$ that we defined in \autoref{s:modelevolve_equations1} are thus given by
\begin{eqnarray}
\Sigma_{g0} & = &
\frac{\sqrt{2\left(\beta+1\right)}f_{g,Q}\sigma_g v_{\phi}}{\pi Gr_0 Q_{\rm min}} \\
s_g(x) & = &
\frac{1}{x}.
\end{eqnarray}

We can express the diffusion coefficient due to turbulent diffusion as $\kappa \approx h_g \sigma_g/3$, where $h_g$ represents the gas scale height \citep{2013RvMP...85..809K,2018MNRAS.475.2236K} given by \citep[equations 24 and 27]{2018MNRAS.477.2716K},
\begin{equation}
h_g = \frac{\sigma^2_g}{\pi G\left(\Sigma_g + \left(\frac{\sigma_g}{\sigma_{\star}}\right)\Sigma_{\star}\right)} = \frac{\sigma^2_g}{\pi G \Sigma_g \phi_Q}\,,
\label{eq:gas_scale_height}
\end{equation}
where $\Sigma_{\star}$ and $\sigma_{\star}$ is the stellar surface density and velocity dispersion, respectively, and $\phi_Q-1$ is the ratio of gas to stellar Toomre $Q$ parameters. This gives
\begin{equation}
\kappa = \frac{\sigma^3_g}{3\pi G \Sigma_g \phi_Q}
\end{equation}
Hence, the factors $\kappa_0$ and $k(x)$ that we defined in \autoref{s:modelevolve_equations1} are given by
\begin{eqnarray}
\kappa_0 & = &
\frac{Q_{\rm min} r_0 \sigma^2_g}{3\phi_Q \sqrt{2\left(\beta+1\right)}f_{g,Q}v_{\phi}} \\
k(x) & = &
x.
\end{eqnarray}
Thus, the product $\kappa_0\Sigma_{g0} \propto \sigma^3_{g}/G$ describes an effective metal flow rate in the disc due to diffusion. 

To derive $\dot \Sigma_{\star}$, we can use equations 31 and 32 of \cite{2018MNRAS.477.2716K},
\begin{equation}
    \dot \Sigma_{\star} = \frac{4v_{\phi}f_{g,Q}\epsilon_{\mathrm{ff}}f_{\mathrm{sf}}\Sigma_g}{\pi r \sqrt{\frac{3f_{g,P}\phi_{\mathrm{mp}}}{2(1+\beta)}}}\,,
\label{eq:sigma_SFR}
\end{equation}
where $\epsilon_{\mathrm{ff}}$ is the star formation efficiency per free-fall time \citep{2005ApJ...630..250K,2012ApJ...745...69K,2013MNRAS.436.3167F,2018MNRAS.477.4380S,2019MNRAS.487.4305S}, $f_{\mathrm{sf}}$ is the fraction of gas in the cold, molecular phase that is not supported by thermal pressure, and thus forms stars \citep{2008ApJ...689..865K,2009ApJ...693..216K,2013MNRAS.436.2747K}, $f_{g,P}$ is the fraction of the mid-plane pressure due to self-gravity of the gas only, and not stars or dark matter \citep{2018MNRAS.477.2716K}, and $\phi_{\mathrm{mp}}$ is the ratio of the total to the turbulent pressure at the mid-plane. Following \autoref{eq:sigma_SFR}, we can derive $\dot \Sigma_{\star0}$ and $\dot s_{\star}(x)$ as,
\begin{eqnarray}
\dot \Sigma_{\star0} & = &
\frac{8\left(\beta+1\right)f^2_{g,Q}\epsilon_{\rm{ff}}f_{\rm{sf}}\sigma_g v^2_{\phi}}{\pi^2 r^2_0 G Q_{\rm min} \sqrt{3f_{g,P}\phi_{\rm{mp}}}} \\
\dot s_{\star}(x) & = &
\frac{1}{x^2}.
\end{eqnarray}

Next, we consider the cosmic accretion of gas onto the disc. The functional form of $\dot c_{\star}(x)$ is not provided in the \cite{2018MNRAS.477.2716K} model. Within the framework of inside-out galaxy formation, $\dot\Sigma_{\rm{cos}}$ decreases with radius, as has been noted in several works \citep{1997ApJ...477..765C,2001ApJ...554.1044C,2009ApJ...696..668F,2010AIPC.1240..131C,2014MNRAS.443..168F,2016MNRAS.455.2308P,2016MNRAS.462.1329M}. In particular, we find from \citet[see their Figure 2]{2008A&A...483..401C} that $\dot c_{\star} \approx 1/x^2$ is necessary to reproduce the present day total surface mass density along the disc in the Milky Way. Additionally, a $1/x^2$ accretion profile is also identical to $\dot s_{\star}$, implying a direct correlation between star formation and accretion, as has been noticed in simulations \citep{2011MNRAS.416.1354D}. Such a profile also means that more accretion is expected in more massive parts of the disc due to higher gravitational potential \citep{2000MNRAS.313..338P}. Keeping these results in mind, we set $\dot c_{\star} (x) = 1/x^2$. However, we show in \aref{s:app_cosmicaccr} that changing the functional form of $\dot c_{\star} (x)$ has only modest effects on the qualitative results. Following \cite{2014MNRAS.438.1552F}, we define
\begin{equation}
    \dot \Sigma_{\mathrm{cos}0} = \frac{\dot M_{\mathrm{h}}f_{\mathrm{B}}\epsilon_{\mathrm{in}}}{2\pi r^2_0\int^{x_{\mathrm{max}}}_{x_{\mathrm{min}}} x \dot c_{\star}dx}
\label{eq:cosmicaccr}
\end{equation}
where $f_{\rm{B}} \approx 0.17$ is the universal baryonic fraction \citep{1995MNRAS.273...72W,2010ApJ...725.2324B,2016A&A...594A..13P}, and $\epsilon_{\rm{in}}$ is the baryonic accretion efficiency given by \citet[equation~22]{2014MNRAS.438.1552F}, which is based on cosmological simulations performed by \cite{2011MNRAS.417.2982F}. $\dot M_{\rm{h}}$ is the dark matter accretion rate \citep{2008MNRAS.383..615N,2010ApJ...718.1001B,2013MNRAS.435..999D} given by \citet[equation~65]{2018MNRAS.477.2716K},
\begin{equation}
    \dot M_{\mathrm{h}} \approx 39 \left(\frac{M_{\mathrm{h}}}{10^{12}\mathrm{M_{\odot}}}\right)^{1.1}\,\left(1+z\right)^{2.2}\,\mathrm{M_{\odot}\,yr^{-1}},
\label{eq:haloaccr}
\end{equation}
where the halo mass, $M_{\mathrm{h}}$, can be written in terms of $v_{\phi}$ by assuming a \cite{1997ApJ...490..493N} density profile for the halo as \citep[equations 69 to 71]{2018MNRAS.477.2716K},
\begin{equation}
  \frac{M_{\mathrm{h}}}{10^{12}\mathrm{M_{\odot}}} = \left(\frac{v_{\phi}/\mathrm{km\,s^{-1}}}{76.17\sqrt{\frac{c}{\ln(1+c) - c/(1+c)}}}\right)^3\,\left(1+z\right)^{-3/2}
\label{eq:halomass}
\end{equation}
where $c$ is the halo concentration parameter \citep[section~7.5]{2010gfe..book.....M}. It is now known that $c$ scales inversely with halo mass \citep{2007MNRAS.378...55M,2009ApJ...707..354Z,2014MNRAS.441.3359D}. For the purposes of this work, we simply adopt $c=10,\,15$ and $13$ for local spirals, local dwarfs and high-$z$ galaxies, respectively, rather than adopting more complex empirical relations (e.g., \citealt{2019MNRAS.487.3581F}). Finally, note that the numerator in \autoref{eq:cosmicaccr} is simply the baryonic accretion rate, $\dot M_{\rm{ext}}$.

The inflow rate required to maintain a steady state is given by the balance between radial transport, turbulent dissipation and star formation feedback \citep[equation~49]{2018MNRAS.477.2716K}
\begin{equation}
    \dot M = \frac{4(1+\beta)\eta\phi_Q \phi^{3/2}_{\mathrm{nt}}}{(1-\beta)G Q^2_{\mathrm{min}}}\,f^2_{g,Q}\sigma^3_g\,\bigg(1 - \frac{\sigma_{\mathrm{sf}}}{\sigma_g}\bigg).
\label{eq:radialinflow}
\end{equation}
Here $\sigma_{\mathrm{sf}}$ is the gas velocity dispersion that can be maintained by star formation feedback alone, $\eta$ is the scaling factor for the rate of turbulent dissipation \citep{2010ApJ...724..895K}, and $\phi_{\mathrm{nt}}$ is the fraction of gas velocity dispersion that is turbulent as opposed to thermal. While a cosmological equilibrium dictates that $\dot M \lesssim \dot M_{\rm{ext}}$ (and also $\dot{M}_\star \lesssim \dot{M}_{\rm ext}$, with the former being the star formation rate), it is unclear if these conditions in fact hold for observed galaxies at high redshift. We discuss this in detail in \aref{s:app_obsuncertainties}, showing that these uncertainties do not affect our qualitative results on metallicity gradients.

Finally, we revisit the yield reduction factor $\phi_y$ that we introduced in \autoref{eq:phiy}. Both the mass loading factor $\mu$ and the direct metal ejection fraction $\xi$ that are incorporated into $\phi_y$ are largely unknown \citep{2013MNRAS.429.1922C,2015MNRAS.446.2125C,2018ApJ...867..142C}. A number of authors have proposed models for $\mu$ (e.g., \citealt{2013MNRAS.429.1922C,2014MNRAS.443..168F,2019MNRAS.484.5587T,2020MNRAS.497..698T}), and it is believed to scale inversely with halo mass. However, there are no robust observational constraints, with current estimates ranging from 0 to 30 \citep{2012MNRAS.426..801B,2012ApJ...761...43N,2014ApJ...792L..12K,2015ApJ...804...83S,2019MNRAS.490.4368S,2017MNRAS.469.4831C,2019ApJ...873..122D,2019ApJ...875...21F,2019ApJ...886...74M}. $\xi$ is even less constrained by observations and theory, although observations and simulations suggest non-zero values in dwarf galaxies \citep[e.g.,][]{2018MNRAS.481.1690C,2018ApJ...869...94E,2019MNRAS.482.1304E}. For this reason we leave $\phi_y$ as a free parameter in the model and present solutions for metallicity evolution for a range of values. As we show in a companion paper \citep{2020aMNRAS.xxx..xxxS}, galaxies tend to prefer a particular value of $\phi_y$ based on their stellar mass, $M_{\star}$.

We list fiducial values of all the parameters used in the \cite{2018MNRAS.477.2716K} model in \autoref{tab:tab1} and \autoref{tab:tab2}. Plugging in these parameters in equations 15$-$18, we get,
\begin{equation}
    \mathcal{T} = \frac{3\phi_Q\sqrt{2\left(\beta+1\right)}f_{g,Q}}{Q_{\rm min}}\left(\frac{v_{\phi}}{\sigma_g}\right)^2
    \label{eq:physicalChi}
\end{equation}
\begin{equation}
    \mathcal{P} = \frac{6\eta\phi^2_Q\phi^{3/2}_{\rm{nt}} f^2_{g,Q}}{Q^2_{\rm{min}}}\left(\frac{1+\beta}{1-\beta}\right)\left(1 - \frac{\sigma_{\mathrm{sf}}}{\sigma_g}\right)
    \label{eq:physicalP}
\end{equation}
\begin{equation}
    \mathcal{S} = \frac{24 \phi_Q  f^2_{g,Q} \epsilon_{\rm{ff}} f_{\rm{sf}}}{\pi Q_{\rm min} \sqrt{3f_{g,P} \phi_{\rm{mp}}}}\left(\frac{\phi_y y}{Z_{\odot}}\right)\left(1+\beta\right)\left(\frac{v_{\phi}}{\sigma_g}\right)^2
    \label{eq:physicalS}
\end{equation}
\begin{equation}
    \mathcal{A} = \frac{3G\dot M_{h}f_{\mathrm{B}}\epsilon_{\mathrm{in}}\phi_Q}{2\sigma^3_g\int^{x_{\mathrm{max}}}_{x_{\mathrm{min}}} x \dot c_{\star}dx}
    \label{eq:physicalA}
\end{equation}
where we have explicitly retained the dependence of the radial profile of cosmic accretion rate surface density in $\mathcal{A}$. Note that none of these ratios depend on $r_0$. Some of these parameters are dependent on other parameters: e.g., $\dot M_{\rm{h}}$ can be expressed as a function of $v_{\phi}$ as is clear from \autoref{eq:haloaccr} and \autoref{eq:halomass}.

\subsection{Solution for the equilibrium metallicity}
\label{s:modelevolve_solution}
Now, we can combine the metallicity evolution model from \autoref{s:modelevolve_equations1} and the galactic disc model from \autoref{s:metalevolve_krumholz2018} to obtain an analytic solution to \autoref{eq:main_nondimx} in steady-state ($\partial \mathcal{Z}/\partial \tau = 0$). The solution is
\begin{eqnarray}
    \lefteqn{\mathcal{Z}(x) = \frac{\mathcal{S}}{\mathcal{A}} + c_1 x^{\frac{1}{2}\left[\sqrt{\mathcal{P}^2+\,4\mathcal{A}}-\mathcal{P}\right]}
    }
    \nonumber \\
    & & {}  + \left(\mathcal{Z}_{r_0} - \frac{\mathcal{S}}{\mathcal{A}} - c_1\right) x^{\frac{1}{2}\left[-\sqrt{\mathcal{P}^2+\,4\mathcal{A}}-\mathcal{P}\right]},
    \label{eq:main_nondimx_solution}
\end{eqnarray}
where $c_1$ is a constant of integration and $\mathcal{Z}_{r_0} \equiv \mathcal{Z}(r=r_0)$. We remind the reader that $\mathcal{Z} = Z/Z_{\odot}$ and $x = r/r_0$ as we define in \autoref{s:modelevolve_equations1}. In writing the above analytic solution, we have assumed that the metallicity at the inner edge of the disc (to which we shall hereafter refer as the central metallicity), $\mathcal{Z}_{r_0}$, is known. We show below (\autoref{s:gradients}) that this approach is reasonable, because the solutions naturally tend toward a particular value of $\mathcal{Z}_{r_0}$. Thus, in practice, $c_1$ is the only unknown parameter in the solution. We also show later in \autoref{s:gradients} that $c_1$ can be expressed as a function of the metallicity gradient at $r_0$.

We now turn to constraining $c_1$. Firstly, note that $\mathcal{Z} >0$ for all $x$. In practice, we ask that $\mathcal{Z} > \mathcal{Z}_{\rm{min}}$ for some fiducial $\mathcal{Z}_{\rm{min}} \approx 0.01$. For $x \gg 1$, this gives
\begin{equation}
    c_1 > \left(\mathcal{Z}_{\mathrm{min}}-\frac{\mathcal{S}}{\mathcal{A}}\right)\,x^{-\frac{1}{2}\left[\sqrt{\mathcal{P}^2+\,4\mathcal{A}} - \mathcal{P}\right]}_{\mathrm{max}}\,,
    \label{eq:bc_c1_1}
\end{equation}
where $x_{\rm max}$ is the outer radius of the disc at which we apply this condition\footnote{The inequality is such that applying this condition at $x_{\rm max}$ ensures that it is also satisfied everywhere else in the disc.}. Secondly, the total metal flux into the disc across the outer boundary cannot exceed that supplied by advection of gas with metallicity $\mathcal{Z}_{\rm CGM}$ into the disc, since otherwise this would imply the presence of a metal reservoir external to the disc that is supplying metals to it, which is only true in special circumstances, e.g., during or after a merger \citep{2012ApJ...746..108T,2018MNRAS.475.1160H}, or due to long term wind recycling through strong galactic fountains \citep{2019MNRAS.490.4786G}. Mathematically, this condition can be written as
\begin{equation}
    -\underbrace{\frac{\dot M \mathcal{Z}}{2\pi x}}_\text{adv. flux} - \underbrace{\kappa\Sigma_g\frac{\partial \mathcal{Z}}{\partial x}}_\text{diff. flux} \geq -\underbrace{\frac{\dot M \mathcal{Z}_{\rm{CGM}}}{2\pi x}}_\text{CGM flux}\,.
\label{eq:outerbc}
\end{equation}
For $x \gg 1$, this translates to,
\begin{equation}
    c_1 \leq \frac{2\mathcal{P}\left(\mathcal{Z}_{\mathrm{CGM}} - \mathcal{S}/\mathcal{A}\right)}{\mathcal{P}+\sqrt{\mathcal{P}^2+\,4\mathcal{A}}}\,x^{-\frac{1}{2}\left[\sqrt{\mathcal{P}^2+\,4\mathcal{A}} - \mathcal{P}\right]}_{\mathrm{max}},
    \label{eq:bc_c1_2}
\end{equation}
Thus, we find that $c_1$ is bounded within a range dictated by the two conditions above. Given a value of $c_1$, we can also calculate the $\Sigma_g$-weighted and $\dot\Sigma_{\star}$-weighted mean metallicity in the model,
\begin{eqnarray}
\overline{\mathcal{Z}}_{\Sigma_g} & = &
\frac{\int^{x_\mathrm{max}}_{x_{\mathrm{min}}} 2\pi x \Sigma_{g0} s_g \mathcal{Z}  dx}{\int^{x_\mathrm{max}}_{x_{\mathrm{min}}} 2\pi x \Sigma_{g0} s_g dx},\\
\label{eq:meanweightedZ1}
\overline{\mathcal{Z}}_{\dot\Sigma_{\star}} & = &
\frac{\int^{x_\mathrm{max}}_{x_{\mathrm{min}}} 2\pi x \dot \Sigma_{\star0} \dot s_{\star} \mathcal{Z} dx}{\int^{x_\mathrm{max}}_{x_{\mathrm{min}}} 2\pi x \dot \Sigma_{\star0} \dot s_{\star} dx}.
\label{eq:meanweightedZ2}
\end{eqnarray}
Finding $\overline{\mathcal{Z}}$ is helpful because we can use it to produce a mass-metallicity relation (MZR) that can serve as a sanity check for the model. We show in a companion paper that our model can indeed reproduce the MZR \citep{2020aMNRAS.xxx..xxxS}.

\section{Equilibrium metallicity gradients}
\label{s:gradients}
We apply our model to four different classes of galaxies: local spirals, local ultra-luminous infrared galaxies (ULIRGs), local dwarfs, and high-$z$ galaxies. The fiducial dimensional parameters we adopt for each of these galaxy types are listed in \autoref{tab:tab1} and \autoref{tab:tab2}. We remind the reader that the metallicity evolution model can only be applied to those galaxies where the metallicity gradient can reach equilibrium. This condition is approximately satisfied if  $t_{\rm eqbm} < t_{\rm{H(z)}}$, where $t_{\rm{H(z)}}$ is the Hubble time at redshift $z$. We also compare $t_{\rm{eqbm}}$ with the molecular gas depletion timescale $t_{\rm{dep,H_2}}$, since we expect that $t_{\rm{dep,H_2}}$ controls the metal production timescale (hence, $\mathcal{S}$) and can potentially impact metallicity gradients. Thus, the metallicity gradients may also not be in equilibrium if $t_{\mathrm{eqbm}} \gg t_{\rm{dep,H_2}}$. An exception to this is for local ULIRGs, where we compare $t_{\rm{eqbm}}$ with $t_{\rm{merge}}$, the merger timescale. This is because the dynamics of the galaxy (as dictated by its rotation curve and orbital time) are dictated by $t_{\rm{merge}}$ for local ULIRGs.

Before checking whether equilibrium is satisfied for each individual galaxy class, it is helpful to put our work in context. Considering galaxies' total metallicity (as opposed to metallicity gradient), \citet[see their Figure 15]{2014MNRAS.443..168F} predict that galaxies with halo masses $ M_{\mathrm{h}} \geq 10^{10.5}\,\mathrm{M}_{\odot}$ (corresponding to $M_{\star} \geq 10^9\,\mathrm{M}_{\odot}$ -- \citealt{2010ApJ...710..903M}, their Figure 4) reach equilibrium by $z\approx 2.5$. \citet{2015MNRAS.449.3274F} use a linear stability analysis to show that the metal equilibration time is at most of order the gas depletion time $t_{\rm{dep}}$, which is small compared to $t_{\rm H(z)}$ for all massive main sequence galaxies. Similar arguments have been made by \cite{2011MNRAS.416.1354D,2012MNRAS.421...98D} and \cite{2013ApJ...772..119L} where the authors find that the metallicity attains equilibrium on very short timescales as compared to $t_{\rm{dep}}$, and is thus in equilibrium both in the local and the high-$z$ Universe. In contrast, \citet{2018MNRAS.475.2236K} study metallicity fluctuations, and find that these attain equilibrium on an even shorter timescale, $\sim 300$ Myr. Our naive expectation is that equilibration times for metallicity gradients should be intermediate between those for total metallicity and those for local metallicity fluctuations, and thus should generally be in equilibrium. We show later in \autoref{s:modellimitations} that, while these expectations are in general satisfied, some galaxy classes, namely, local dwarfs with no radial inflow, local ULIRGs, and galaxies with inverted gradients, \textit{can} be out of equilibrium. Thus, our model cannot be applied to these galaxies.

For the rest of the galaxies where the equilibrium model can be applied, we use the fiducial parameters that we list in \autoref{tab:tab2}, and solve the resulting differential equation to obtain $\mathcal{Z}(x)$, for different yield reduction factors. We list the resulting values of $\mathcal{T},\,\mathcal{P},\,\mathcal{S}$ and $\mathcal{A}$ for different galaxies in \autoref{tab:tab3}. To mimic the process followed in observations and simulations (e.g., \citealt{2018MNRAS.478.4293C,2020MNRAS.495.2827C}) as well as existing models (e.g., \citealt{2009ApJ...696..668F}), we linearly fit the resulting metallicity profiles using least squares with equal weighting in logarithmic space
\begin{equation}
    \log_{10}\mathcal{Z}\,(x) = \log_{10}\mathcal{Z}_{r_0} + x\nabla \left[\log_{10}\mathcal{Z}\,(x)\right]\,,
\end{equation}
between $x=1$ and $x_{\rm{max}}$, thereby excluding the innermost galactic disc where the rotation curve index is not constant, and where factors such as stellar bars can affect the central metallicity \citep{2012A&A...543A.150F,2020MNRAS.tmp.2303Z}. While it is clear from \autoref{eq:main_nondimx_solution} that the functional form of $\mathcal{Z}$ is such that $\log_{10}\mathcal{Z}$ may not be a linear function of $x$ in certain cases, we will continue to use the linear fit as above in order to compare with observations. We show in \aref{s:app_xmax} how the gradients change if we vary $x_{\rm{min}}$ or $x_{\rm{max}}$. For each class of galaxy that we discuss in the subsections below, we plot a range of gradients that results from the constraints on the constant of integration $c_1$ (see \autoref{s:modelevolve_solution}), as well as the weighted mean metallicities, $\overline{\mathcal{Z}}_{\Sigma_g}$ and $\overline{\mathcal{Z}}_{\dot\Sigma_{\star}}$.

\begin{table*}
\centering
\caption{Resulting dimensionless ratios in different types of galaxies from the fiducial model based on the input parameters from \autoref{tab:tab1} and \autoref{tab:tab2}.}
\begin{tabular}{l|l|l|l|c|c|r}
\hline
Dimensionless & Description & Reference & Local & Local & Local & High-$z$\\
Ratio & & equations & spiral & dwarf & ULIRG & \\
\hline
$\mathcal{T}$ & Ratio of orbital to diffusion timescales & \autoref{eq:physicalChi} & 1697 & 458 & 77 & 99 \\
$\mathcal{P}$ & Péclet number (ratio of advection and diffusion) & \autoref{eq:physicalP} & 2.7 & 11 & 41 & 6.2 \\
$\mathcal{S}/\phi_y$ & Ratio of metal production (incl. loss in outflows) and diffusion & \autoref{eq:physicalS} & 16.5 & 2.9 & 2.6 & 2.3 \\
$\mathcal{A}$ & Ratio of cosmic accretion and diffusion & \autoref{eq:physicalA} & 9.9 & 1.6 & 0.1 & 0.7 \\
\hline
\end{tabular}
\label{tab:tab3}
\end{table*}

\begin{figure}
\includegraphics[width=\columnwidth]{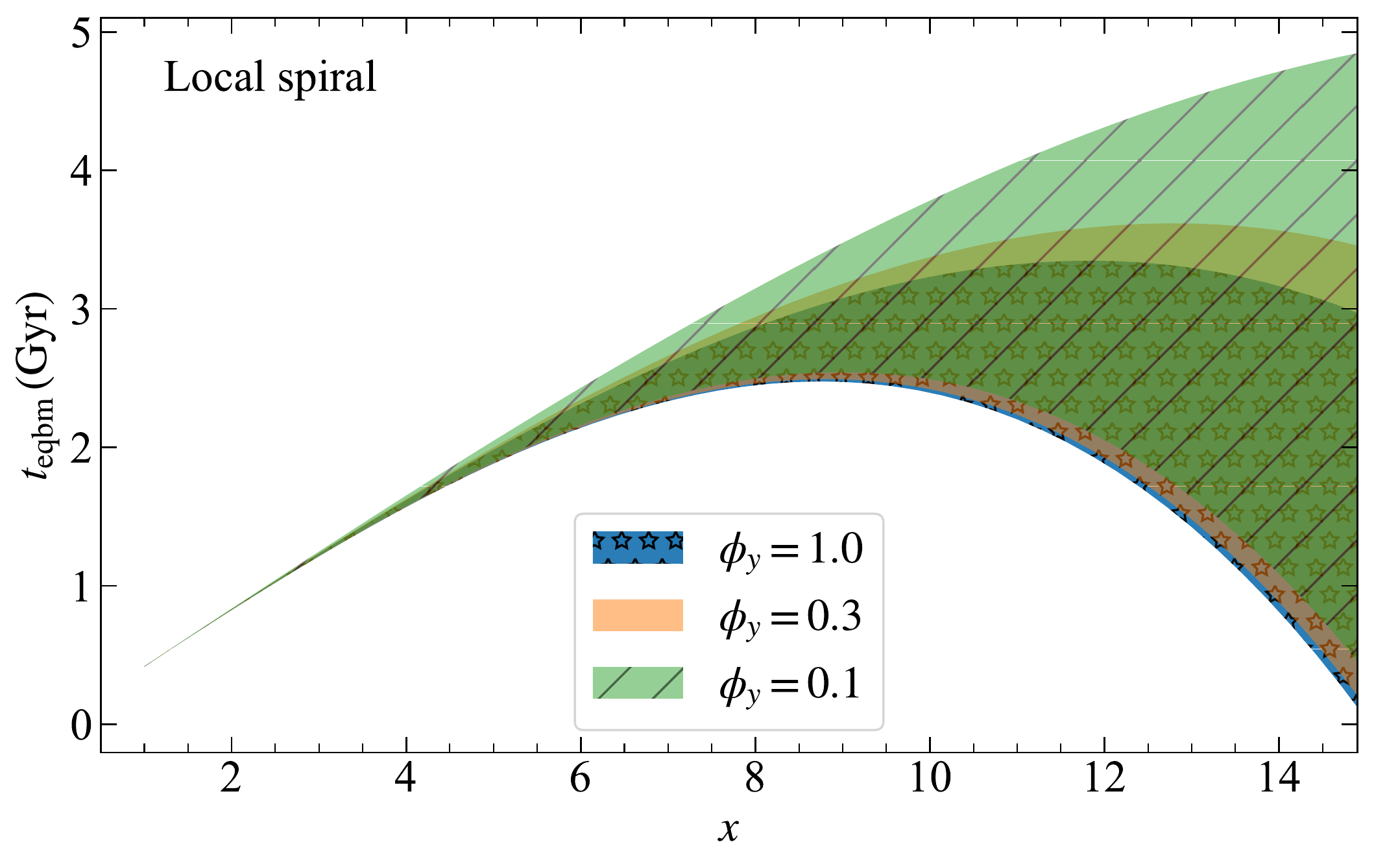}
\caption{Metallicity equilibration time, $t_{\rm{eqbm}}$ plotted as a function of the dimensionless radius $x$ for three different values of the yield reduction factor, $\phi_y$, for a fiducial local spiral galaxy (see \autoref{eq:teqbm}). Here, $x = r/r_0$, where $r_0=1\,\rm{kpc}$. The shaded bands correspond to solutions that cover all allowed values of the constant of integration $c_1$ in the solution to the metallicity equation (see \autoref{s:modelevolve_solution}). Since, $t_{\rm{eqbm}}$ is substantially smaller than the Hubble time $t_{\rm{H(0)}}$ and comparable to the molecular gas depletion time $t_{\rm{dep,H_2}}$, metallicity gradients in local spirals are in equilibrium.}
\label{fig:teqbm_ls}
\end{figure}

\subsection{Local spirals}
\label{s:gradients_localspirals1}
For local spirals, we select the outer boundary of the star-forming disc to be $15\,\rm{kpc}$, thus $x_{\rm{max}} = 15$, reminding the reader that $x=r/r_0$ where $r_0=1\,\rm{kpc}$. We first study the metallicity equilibration time ($t_{\rm{eqbm}}$) to see if metallicity gradients in these galaxies have attained equilibrium, so that the model can be applied to them. \autoref{fig:teqbm_ls} shows the value of $t_{\rm{eqbm}}$ we find from \autoref{eq:teqbm} as a function of $x$ for local spirals for different values of the yield reduction factor, $\phi_y$; the bands shown correspond to solutions covering all allowed values of the integration constant $c_1$. It is clear from \autoref{fig:teqbm_ls} that $t_{\rm{eqbm}} < t_{\rm{H(0)}}$ for all possible $\phi_y$ and $c_1$, so we conclude that the gradients in local spirals are in equilibrium. Additionally, $t_{\mathrm{eqbm}} \sim t_{\mathrm{dep,H_2}}$ for local spirals ($1-3\,\rm{Gyr}$, e.g., \citealt{2002ApJ...569..157W,2008AJ....136.2846B,2012ApJ...758...73S,2013AJ....146...19L,2014MNRAS.443.1329H}), implying that the metallicity distribution reaches equilibrium on timescales comparable to the molecular gas depletion timescale. The model also predicts that central regions of local spirals should achieve equilibrium earlier than the outskirts, however, this is somewhat sensitive to the choice of $c_1$ and $\phi_y$ as we can see from \autoref{fig:teqbm_ls} (see also, Figure 4 of \citealt{2019MNRAS.487..456B}). Our equilibrium timescales are also consistent with our naive expectation as stated above: long compared to the timescale for local fluctuations to damp, but shorter than the time required for the total metallicity to reach equilibrium.

\begin{figure}
\includegraphics[width=\columnwidth]{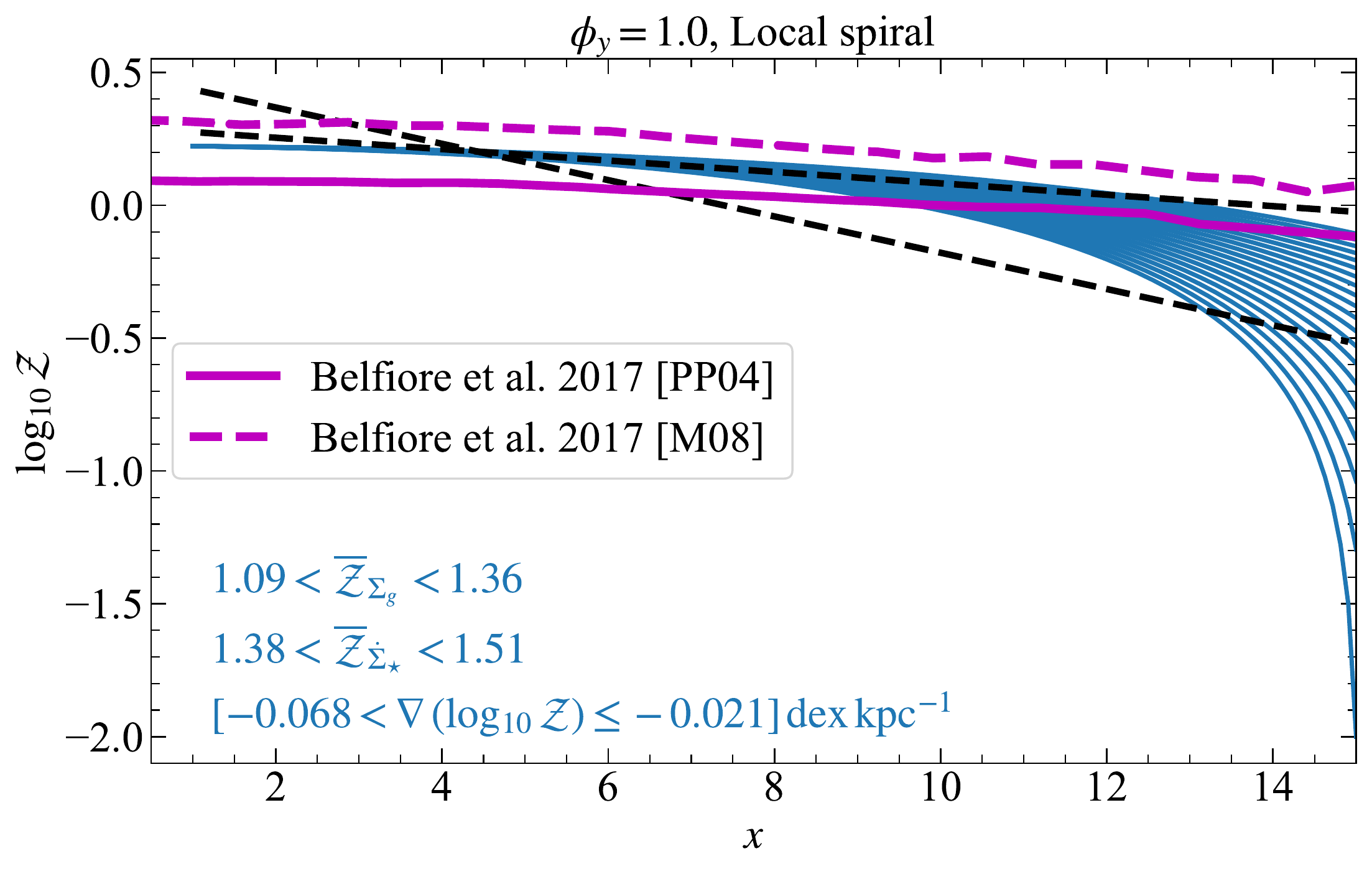}
\includegraphics[width=\columnwidth]{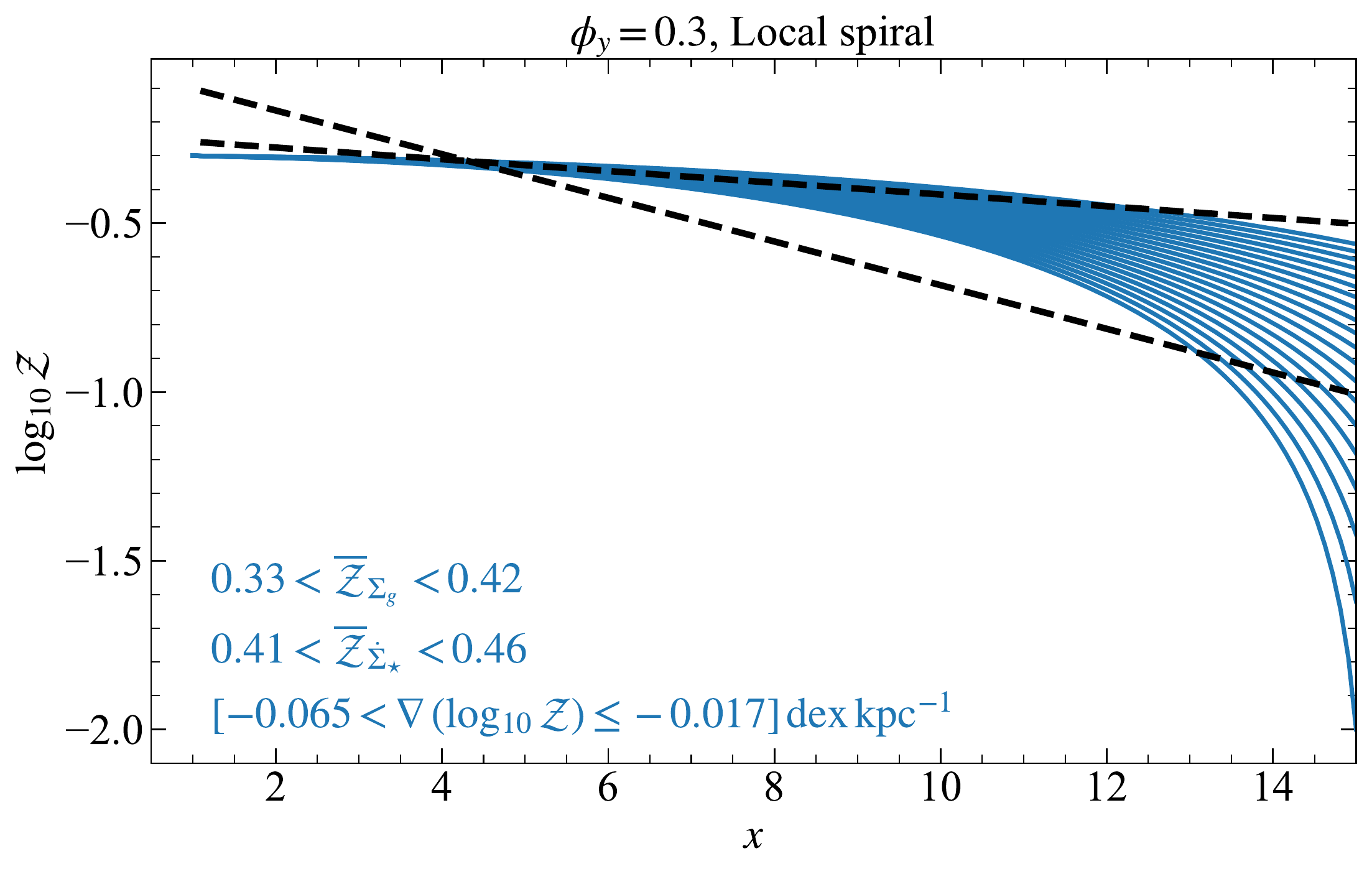}
\includegraphics[width=\columnwidth]{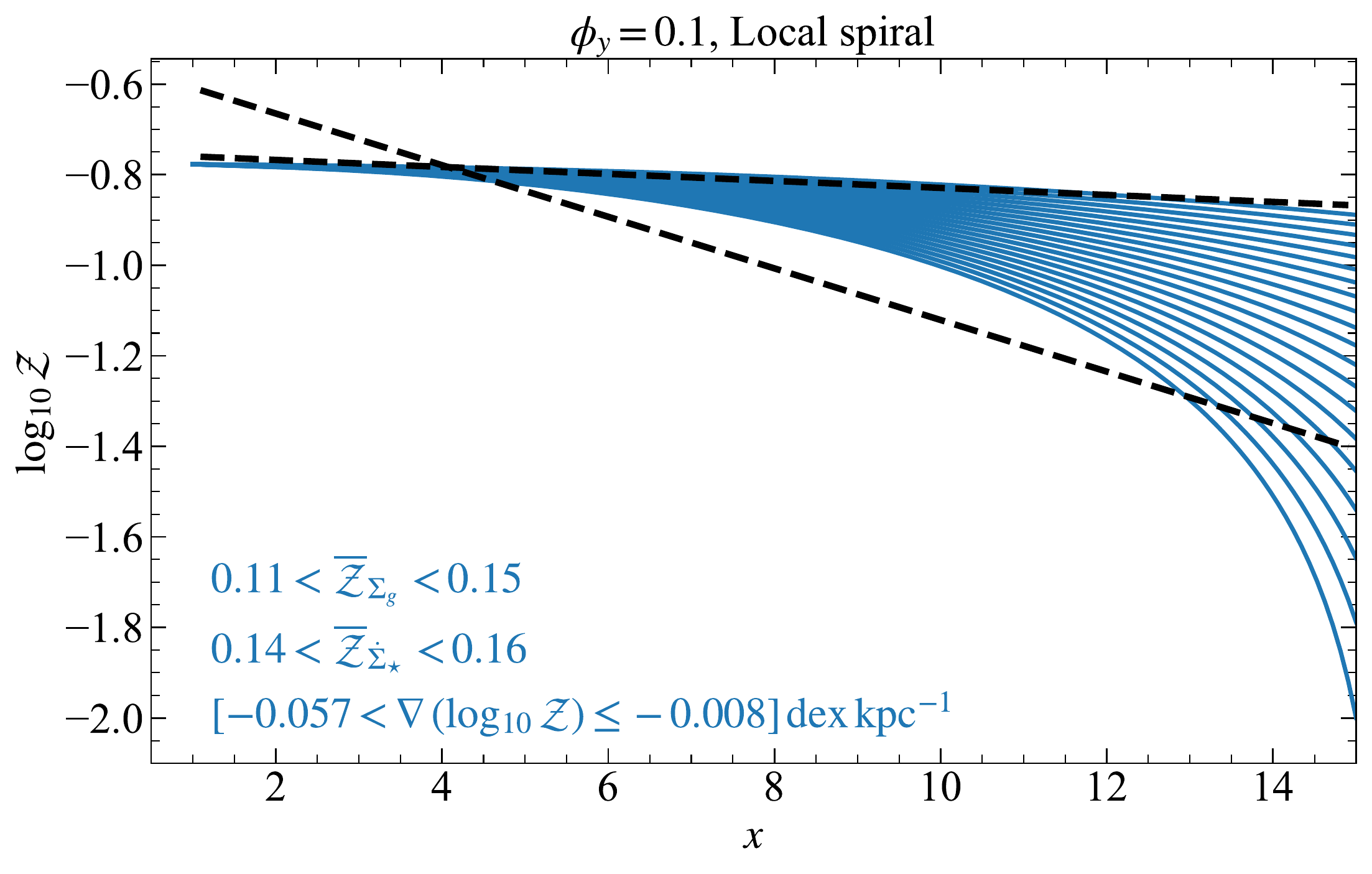}
\caption{Metallicity ($\mathcal{Z} = Z/Z_{\odot}$; blue lines) as a function of dimensionless radius ($x = r/r_0$ with $r_0 = 1\,\rm{kpc}$) produced by the model for a fiducial local spiral galaxy with input parameters listed in \autoref{tab:tab1} and \autoref{tab:tab2}, for different values of the yield reduction factor, $\phi_y$. The analytic solution to the metallicity evolution equation is given by \autoref{eq:main_nondimx_solution}. The slope of the linear fit to the model gradients between $x=1-15$ (black, dashed lines) gives the metallicity gradient that can be compared against simulations and observations. The blue coloured curves show the acceptable parameter space of the gradients based on the constraints on the constant of integration, $c_1$, using the boundary conditions criteria described in \autoref{s:modelevolve_solution}. The metallicity at the inner edge of the disc (referred to as the central metallicity in the text), $\mathcal{Z}_{r_0}$, is set by the balance between source and accretion for local spirals (see \autoref{eq:Zr0_localspirals}). $\overline{\mathcal{Z}}_{\Sigma_g}$ and $\overline{\mathcal{Z}}_{\dot\Sigma_{\star}}$ represent the range of mass-weighted and SFR-weighted mean equilibrium metallicities produced by the solution, respectively (see \autoref{eq:meanweightedZ1}). We expect $\phi_y$ closer to unity for local spirals, implying that metals in these galaxies are well-mixed with the ISM before they are ejected. Finally, in the top panel we overplot the average metallicity profiles observed in local spirals in the MaNGA survey by \protect\cite{2017MNRAS.469..151B} using the PP04 \protect\citep{2004MNRAS.348L..59P} and M08 \protect\citep{2008A&A...488..463M} calibrations, adjusting the normalisation to overlap with the model profiles.}
\label{fig:localspirals}
\end{figure}

\begin{figure}
\includegraphics[width=\columnwidth]{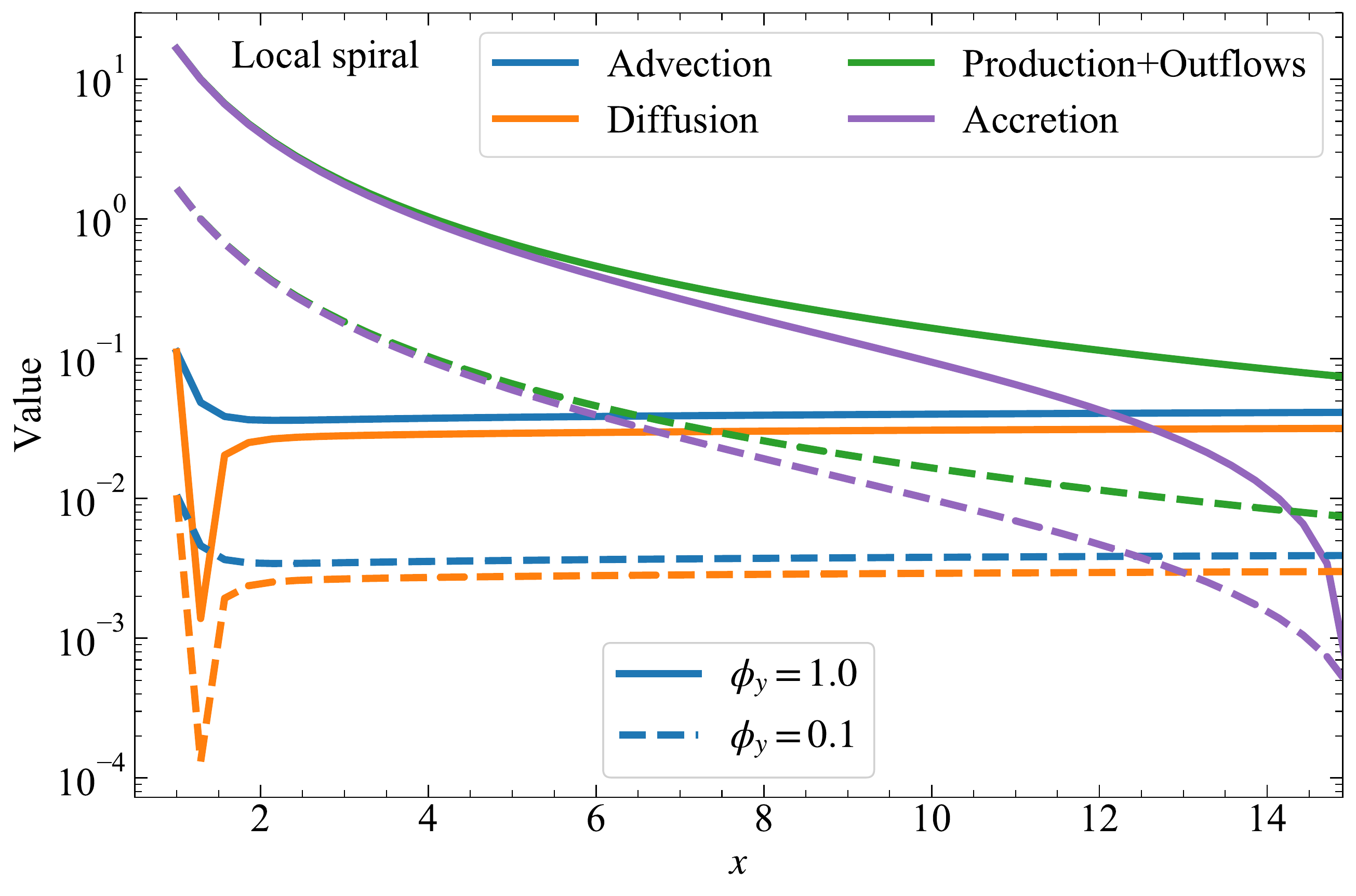}
\caption{Absolute values of different terms in the numerator of \autoref{eq:teqbm} that collectively build the metallicity gradient in local spirals, for a fixed $\mathcal{Z}_{\rm{CGM}} = 0.1$ and fixed $c_1$ for different yield reduction factors, $\phi_y$. These terms are defined in \autoref{eq:main_nondimx}. The leading terms that set the gradients in local spirals are metal production and accretion of gas onto the galaxy, whereas advection and diffusion play a subdominant role in local spirals, due to the small velocity dispersion, $\sigma_g$. Note that the sharp feature in the diffusion term near $x=1.3$ corresponds to the location where this term passes through zero as it changes sign; the term in fact behaves smoothly everywhere, but this behaviour appears as a sharp feature when plotted on a logarithmic axis.}
\label{fig:localspirals_terms}
\end{figure}

\autoref{fig:localspirals} presents the family of radial metallicity distributions we obtain from the model for local spirals; the different lines correspond to varying choices of the outer boundary condition $c_1$, from the minimum to the maximum allowed. We report in the text annotations that accompany these curves the range of gas- and SFR-weighted mean metallicities $\overline{\mathcal{Z}}_{\Sigma_g}$ and $\overline{\mathcal{Z}}_{\dot{\Sigma}_{\star}}$, and metallicity gradients $\nabla(\log_{10}\mathcal{Z})$, spanned by the models shown. To aid in the interpretation of these results, in \autoref{fig:localspirals_terms} we also show the magnitudes of the various terms in the numerator on the right hand side of \autoref{eq:teqbm}, which represent, respectively, the relative importance of advection, diffusion, metal production (reduced by metal ejection in outflows), and cosmological accretion in determining the metallicity gradient. We use this figure to read off which processes are dominant in different parts of the disc. While the source and the accretion terms fall off in the outermost regions due to the $1/x^2$ dependence, the advection and diffusion terms slightly increase with $x$, thereby resulting in a shorter metal equilibration time in the outermost regions as compared to intermediate regions, as we see in \autoref{fig:teqbm_ls}. Thus, transport processes in the outer regions play an important role in establishing metal equilibrium in local spirals. 

There are several noteworthy features in these plots. First, note how the solution asymptotically reaches a particular value of the central metallicity. We choose to set $\mathcal{Z}_{r_0}$ to this value, but we emphasise that the behaviour of the solution does not depend on this choice except very close to $x=1$: if we choose a different value of $\mathcal{Z}_{r_0}$, the solution is (by construction) forced to this value close to $x=1$, but returns to the asymptotic limit for $x \gtrsim 1.1$. Indeed, we shall see that this is a generic feature for all of our cases: the limiting central metallicity is set by a balance between two dominant processes, and can be deduced analytically by equating the two dominant terms in \autoref{eq:main_nondimx_solution}. For the case of local spirals, the two dominant terms throughout the disc are production and accretion, as we can read off from \autoref{fig:localspirals_terms}. The balance between these two processes gives
\begin{equation}
    \mathcal{Z}_{r_0} = \frac{\mathcal{S}}{\mathcal{A}}\,\,\,\,\left[\mathrm{Local}\,\mathrm{spirals}\right]\,.
\label{eq:Zr0_localspirals}
\end{equation}
This matches the conclusions of \cite{2008MNRAS.385.2181F} regarding the total metallicity. However, we show below in \autoref{s:gradients_localdwarfs} that this conclusion holds only for local, massive galaxies, since other processes like metal transport also play a significant role in low mass galaxies as well as at high redshift. Using the above definition of $\mathcal{Z}_{r_0}$, we can now express $c_1$ in a more physically-meaningful way
\begin{equation}
c_1 = \frac{1}{\sqrt{\mathcal{P}^2 + 4\mathcal{A}}}\left.\frac{\partial \mathcal{Z}}{\partial x}\right|_{r=r_0}\,.
\label{eq:c1_localspirals}
\end{equation}
Thus, for local spirals, $c_1$ essentially describes the metallicity gradient at $r_0$.

Second, both the central metallicity $\mathcal{Z}_{r_0}$ and the mean metallicity $\overline{\mathcal{Z}}$ decrease with decreasing $\phi_y$, as expected; we obtain mean metallicities close to Solar, as expected for massive local spirals, for $\phi_y$ fairly close to unity. Thus our models give reasonable total metallicities for local spirals if we assume that there is relatively little preferential ejection of metals, consistent with the results of recent simulations \citep{2017ApJ...837..152D,2020ApJ...899..108T,2020MNRAS.496.4433T}. Note that some semi-analytic models find a high metal ejection fraction for spirals, but self-consistently following the evolution of the CGM subsequently leads to high re-accretion of the ejected metals \citep{2020arXiv201104670Y}. In the language of our model, this essentially implies a high $\phi_y$ when averaging over the metal recycling timescale for local spirals, consistent with our expectations.

Third, and most importantly for our focus in this paper, the value of $\phi_y$ has little effect on the metallicity gradient, as is clear from the similar range of gradients produced by the model for different $\phi_y$. Our models robustly predict a gradient $\nabla (\log_{10}\mathcal{Z})\approx -0.07$ to $0$ dex kpc$^{-1}$, in very good agreement with the range observed in local spirals \citep[e.g.,][]{1994ApJ...420...87Z,2014A&A...563A..49S,2015MNRAS.448.2030H,2016A&A...587A..70S,2017MNRAS.469..151B,2019MNRAS.484.5009E,2020A&A...636A..42M}, and within the range provided by existing simpler models of metallicity gradients \citep{2001ApJ...554.1044C,2009ApJ...696..668F}. 

Apart from the mean gradient, we can also study the detailed shape of the metal distribution with the model. For the given input parameters as in \autoref{tab:tab2}, the model features a nearly-flat metal distribution in the inner galaxy for all allowed values of $c_1$. Such flat gradients in the inner regions are commonly observed in local spirals \citep{2012ApJ...745...66M,2017MNRAS.469..151B,2020A&A...636A..42M}, and have been attributed to metallicity reaching saturation in these regions \citep{2016MNRAS.462.2715Z,2019A&ARv..27....3M}, although the flatness depends on the metallicity calibration used \citep[Figure 4]{2020arXiv201104670Y}. This is also the case for our models of spirals, since the flat region corresponds to the part of the disc where the metallicity is set by the balance between metal injection and dilution by metal-poor infall (c.f.~\autoref{fig:localspirals_terms}). For comparison, we also show in \autoref{fig:localspirals} the measured average metallicity profiles in local spirals observed in the MaNGA survey \citep{2017MNRAS.469..151B} using two different metallicity calibrations \citep{2004MNRAS.348L..59P,2008A&A...488..463M}, where we have adjusted the overall metallicity normalisation by 0.02 dex so that the model profiles overlap with the data. We see that the profiles produced by the model are in reasonable agreement with that seen in the observations (see also, \citealt{2018A&A...609A.119S}).

Several works have also noted that local spirals with higher gas fractions (at fixed mass) show steeper metallicity gradients \citep{2015MNRAS.451..210C,2019A&A...623A...5D,2020arXiv201212887P}. In the language of the \cite{2018MNRAS.477.2716K} model, a higher gas fraction implies a higher value of $f_{g,Q}$ and $f_{g,P}$. Increasing these parameters leads to an increase in the source term $\mathcal{S}$, which gives rise to steeper metallicity gradients in the model, consistent with the above observations. Moreover, a higher gas fraction (\textit{i.e.,} higher $f_{g,Q}$ and $f_{g,P}$) also results in a rather steep metallicity profile in the inner disc, thus giving slightly lower metallicities in the inner disc as compared to the fiducial case above, consistent with the standard picture of galaxy chemical evolution (\citealt{1972A&A....20..383T,1973ApJ...186...35T}; see also, \citealt{2020arXiv201212887P}).

It is difficult to provide robust predictions for the metal distribution in the outer parts of the galaxy without further constraining $c_1$. The outer-galaxy metal distribution in the model is also sensitive to parameters like the galaxy size and the CGM metallicity. The result of these uncertainties is that depending on the choice of $c_1$, the model can produce both nearly-flat and quite steep metal distributions in the outer parts of the galaxy. A steep drop in the metallicity in the outer disc has been observed in several local spirals \citep{2012ApJ...745...66M}, but is dependent on the metallicity calibration used \citep{2015MNRAS.451..210C}. In our models, this region corresponds to where cosmological accretion of metal-poor gas onto the disc becomes less important than inward advection of metal-poor gas through the disc -- a process whose rate we would expect to be correlated with the available mass supply in the far outer disc, as measured by \ion{H}{i}. Note that the gradient can also flatten again in the outermost regions in the disc \citep{Werk11a,2014A&A...563A..49S,2016A&A...587A..70S,2019MNRAS.488.3826B}; however, these regions typically have insufficient spatial resolution \citep{2020MNRAS.495.3819A} as well as significant diffused ionised gas emission, both of which can cause the gradients to appear flatter than their true values \citep[Section~6]{2019ARA&A..57..511K}. Given the uncertainties in the model as well as observations of metallicities in the outer discs in spirals, it is not yet obvious if the metal distribution in the outer disc in the model can be validated against the available observations. Thus, we do not study these regions with our model. This analysis also shows that linear fits to the metallicity profiles is a crude approximation to the true underlying metallicity distribution in local spiral galaxies.

\subsection{Local dwarfs}
\label{s:gradients_localdwarfs}
Our model can also be applied to local dwarf galaxies that can be classified as rotation-dominated, e.g., the Large Magellanic Cloud (LMC), for which $v_{\phi} \sim 60\,\mathrm{km\,s^{-1}}$ and $\sigma_{g} \sim 7\,\mathrm{km\,s^{-1}}$ \citep{2000ApJ...542..789A}. Such galaxies typically lie at the massive end of dwarfs ($M_{\star,\rm{LMC}} = 2\times10^9\,\rm{M_{\odot}}$, as reported in \citealt{2006lgal.symp...47V,2012ApJ...761...42S}), and possess an equilibrium gas disc to which the unified galaxy evolution model of \cite{2018MNRAS.477.2716K} can be applied. We set the outer disc radius to $6\,\mathrm{kpc}$ to find the gradient in the fiducial model, in line with the estimated gas disc size of local dwarfs ($r_{\rm{LMC}} \sim 4.3\,\rm{kpc}$, \citealt{1990A&ARv...2...29W}).

\begin{figure}
\includegraphics[width=\columnwidth]{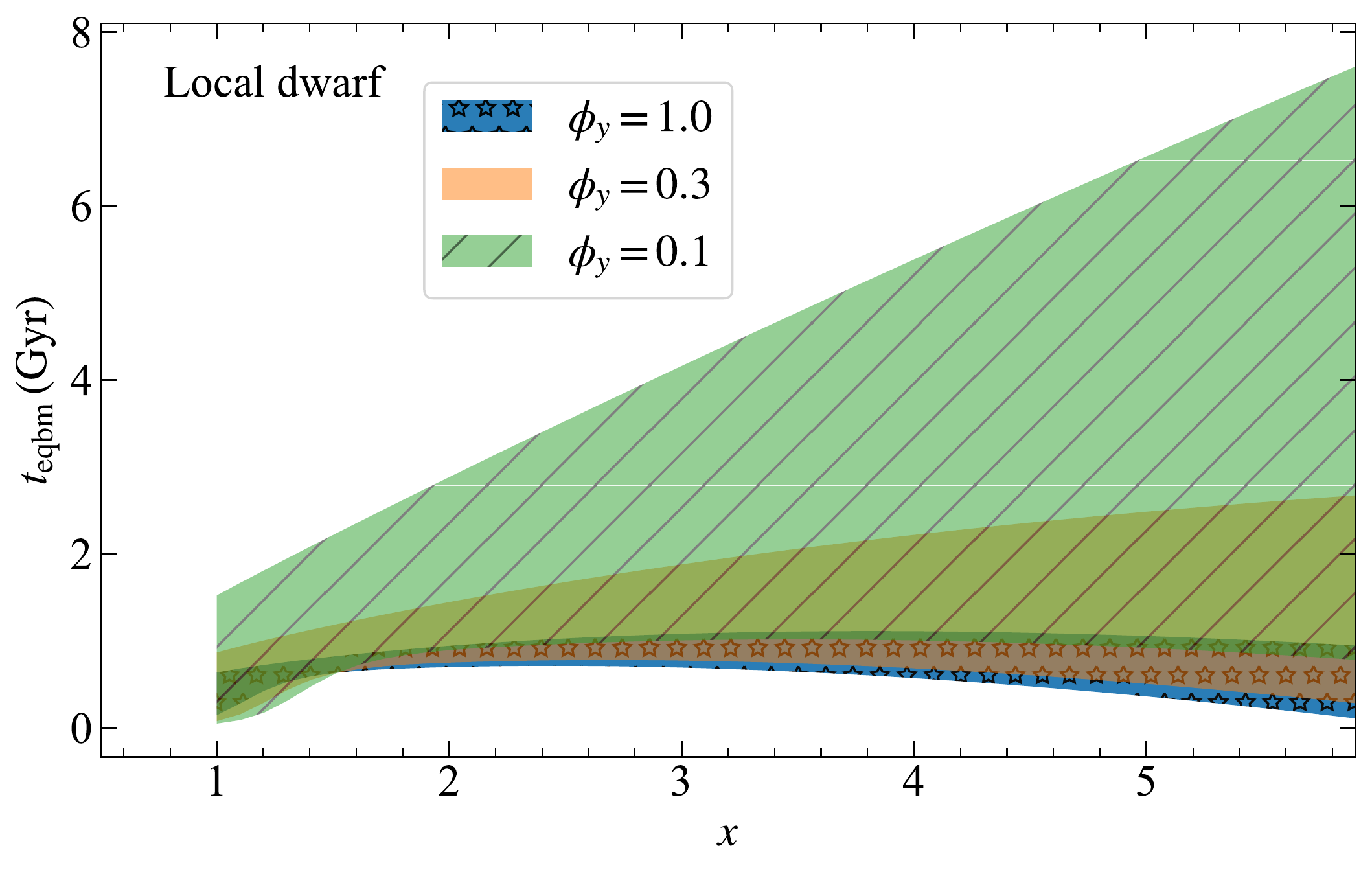}
\caption{Same as \autoref{fig:teqbm_ls}, but for local dwarfs. Here, $t_{\rm{eqbm}} < t_{\rm{H(0)}}$, implying that the metallicity gradients in local dwarfs are also in equilibrium, even in the case of low $\phi_y$ (see the text for a discussion on $t_{\rm{dep,H_2}}$ for local dwarfs). The corresponding metallicity gradients are plotted in \autoref{fig:localdwarfs}.}
\label{fig:teqbm_ld}
\end{figure}

\autoref{fig:teqbm_ld} shows the metal equilibration time, $t_{\rm{eqbm}}$, for local dwarfs based on the parameters we list in \autoref{tab:tab1} and \autoref{tab:tab2}. It is clear that metallicity gradients are in equilibrium in dwarfs, since $t_{\rm{eqbm}} < t_{\rm{H(0)}}$ as in the case of local spirals (see, however, \autoref{s:noneqbm_ld_noadvec} where we show that this may not be the case under certain circumstances). Contrary to local spirals, local dwarfs show a wide range of $t_{\mathrm{dep,H_2}}$, from a few hundred Myr to several Gyr (e.g., \citealt{2011ApJ...741...12B,2014MNRAS.445.2599B,2015A&A...583A.114H,2016ApJ...825...12J,2017ApJ...835..278S}), similar to the scatter we find in $t_{\rm{eqbm}}$ (see also, \autoref{s:noneqbm_ld_noadvec})\footnote{While it is often quoted that $t_{\rm{dep,H_2}}$ is smaller by a factor of $2-5$ in local dwarfs as compared to local spirals, \cite{2017ApJ...835..278S} point out that this may not necessarily be true. This is because it is difficult to trace the entire molecular gas content in dwarfs, and a significant fraction of the molecular gas can be `CO-faint' or `CO-dark' \citep{2011ApJ...741...12B,2018ApJ...853..111J}, or in quiescent molecular clouds that are not targeted in observations \citep{2010ApJ...722.1699S,2014MNRAS.439.3239K}.}. 

\begin{figure}
\includegraphics[width=\columnwidth]{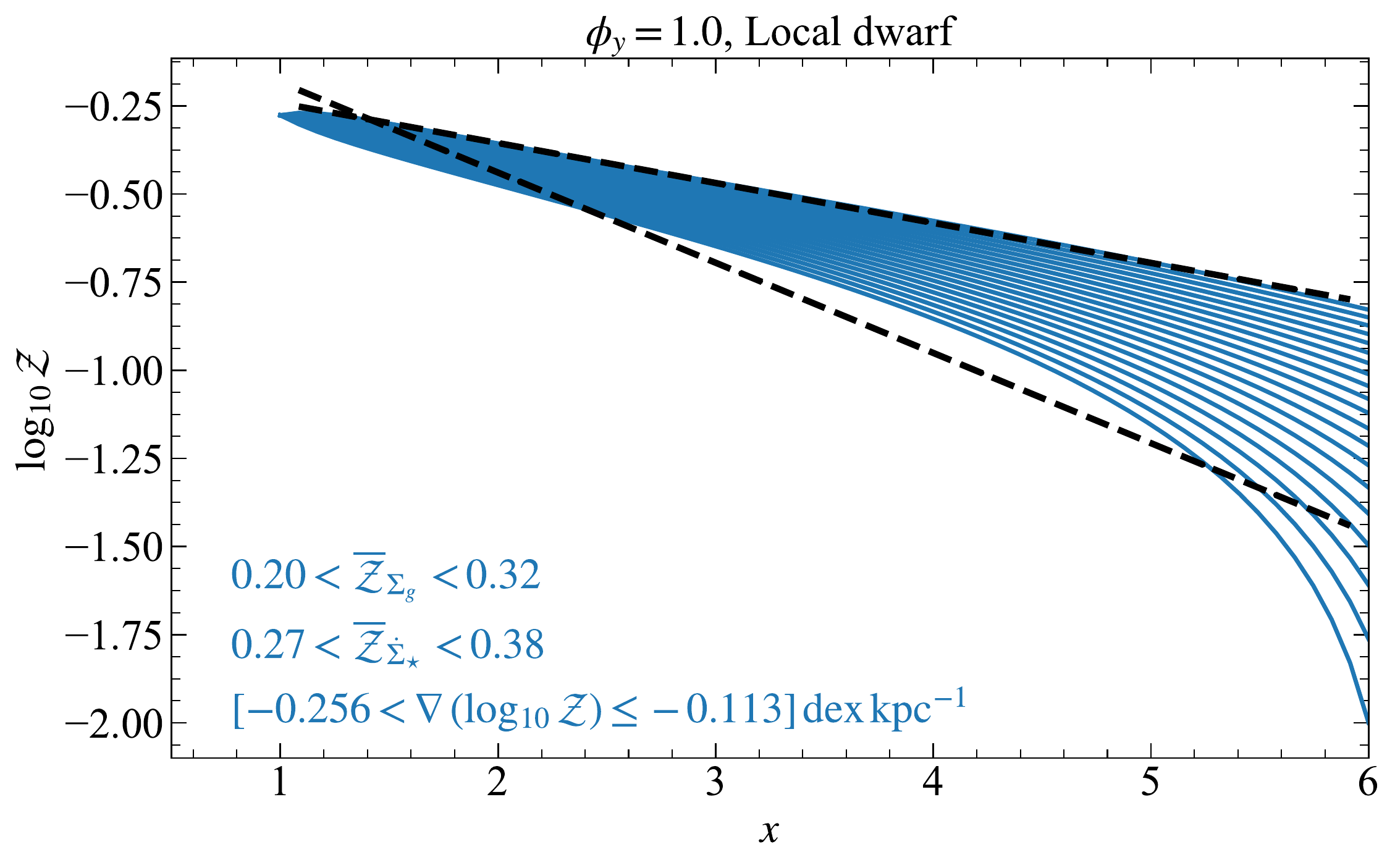}
\includegraphics[width=\columnwidth]{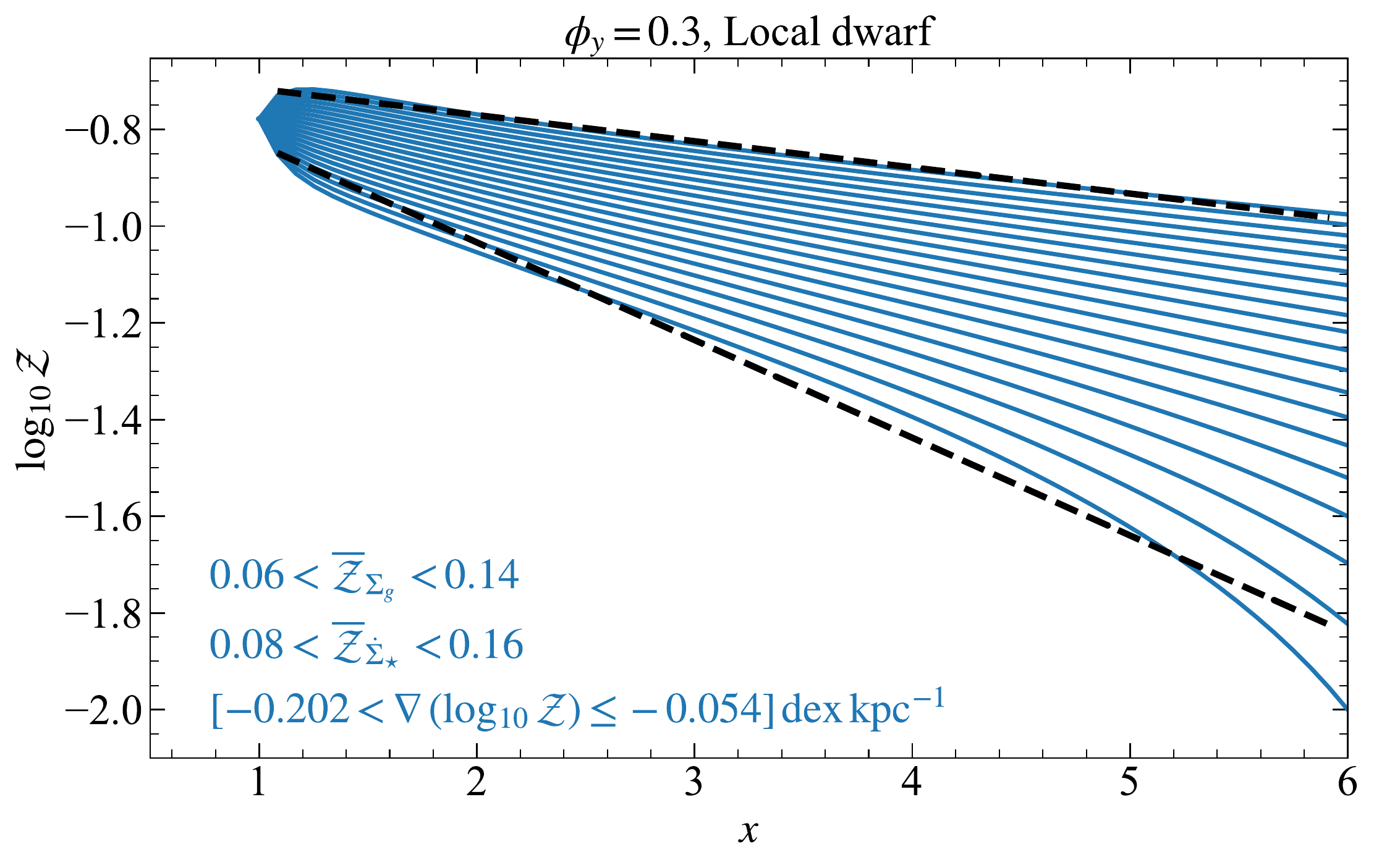}
\includegraphics[width=\columnwidth]{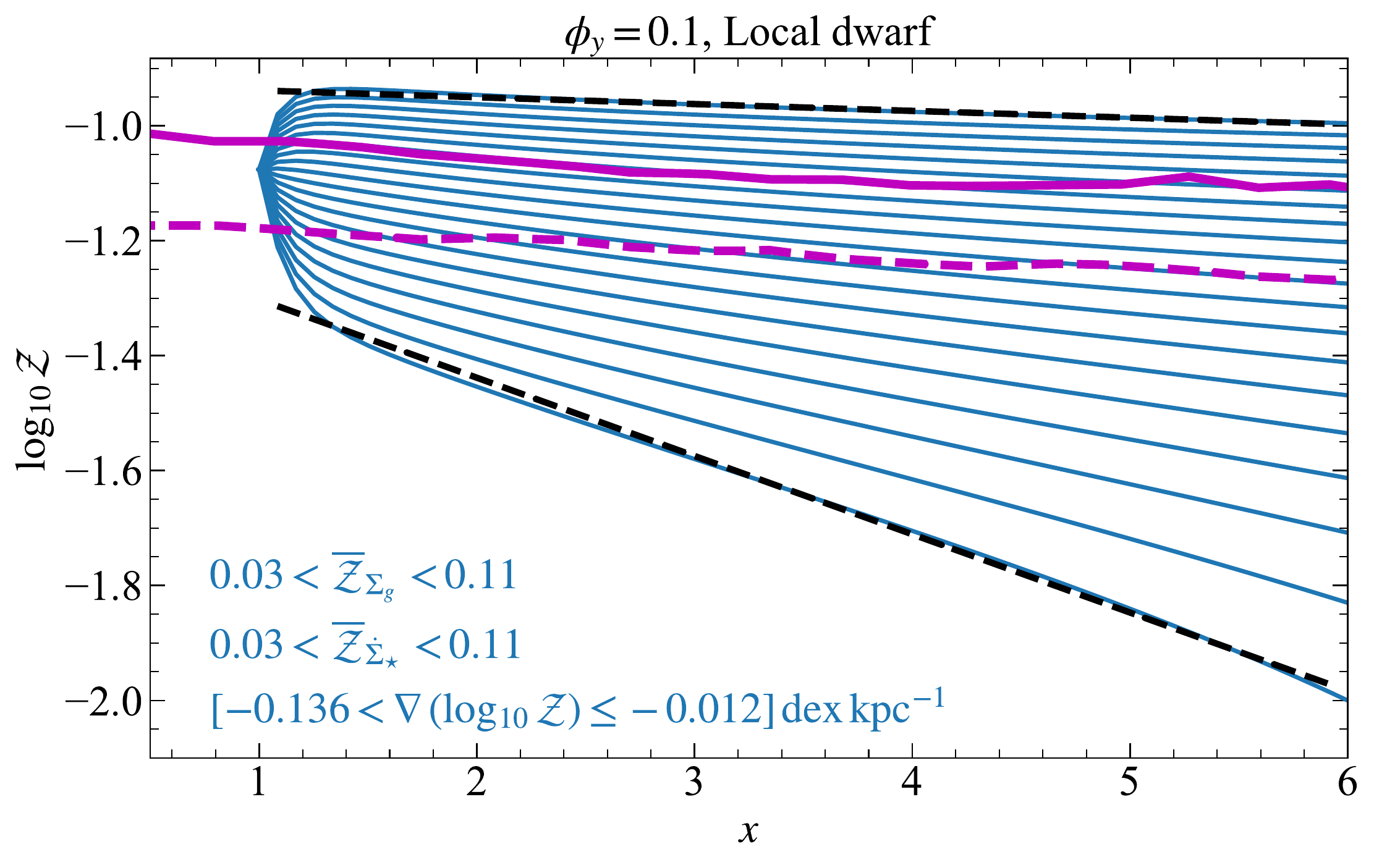}
\caption{Same as \autoref{fig:localspirals}, but for local dwarfs. Here, $\mathcal{Z}_{r_0}$ is set by the balance between advection and diffusion, whereas metallicities in the disc are set by the balance between advection and source. The sharp rise and fall in the profile at $x=1$ is an artefact of the choice of the constant of integration $c_1$ used to calculate $\mathcal{Z}_{r_0}$ (see \autoref{eq:c1_localdwarfs}). The gradients are particularly sensitive to the strength of advection for local dwarfs since turbulence due to star formation feedback is comparable to that due to gravity, $\sigma_{\rm{sf}} \sim \sigma_{g}$. When they are exactly equal, advection vanishes, and the gradients may not be in equilibrium (see \autoref{s:noneqbm_ld_noadvec}). In the last panel we also plot (purple lines) the average metallicity profiles observed in local dwarfs in the MaNGA survey; see \autoref{fig:localspirals}.}
\label{fig:localdwarfs}
\end{figure}

\begin{figure}
\includegraphics[width=\columnwidth]{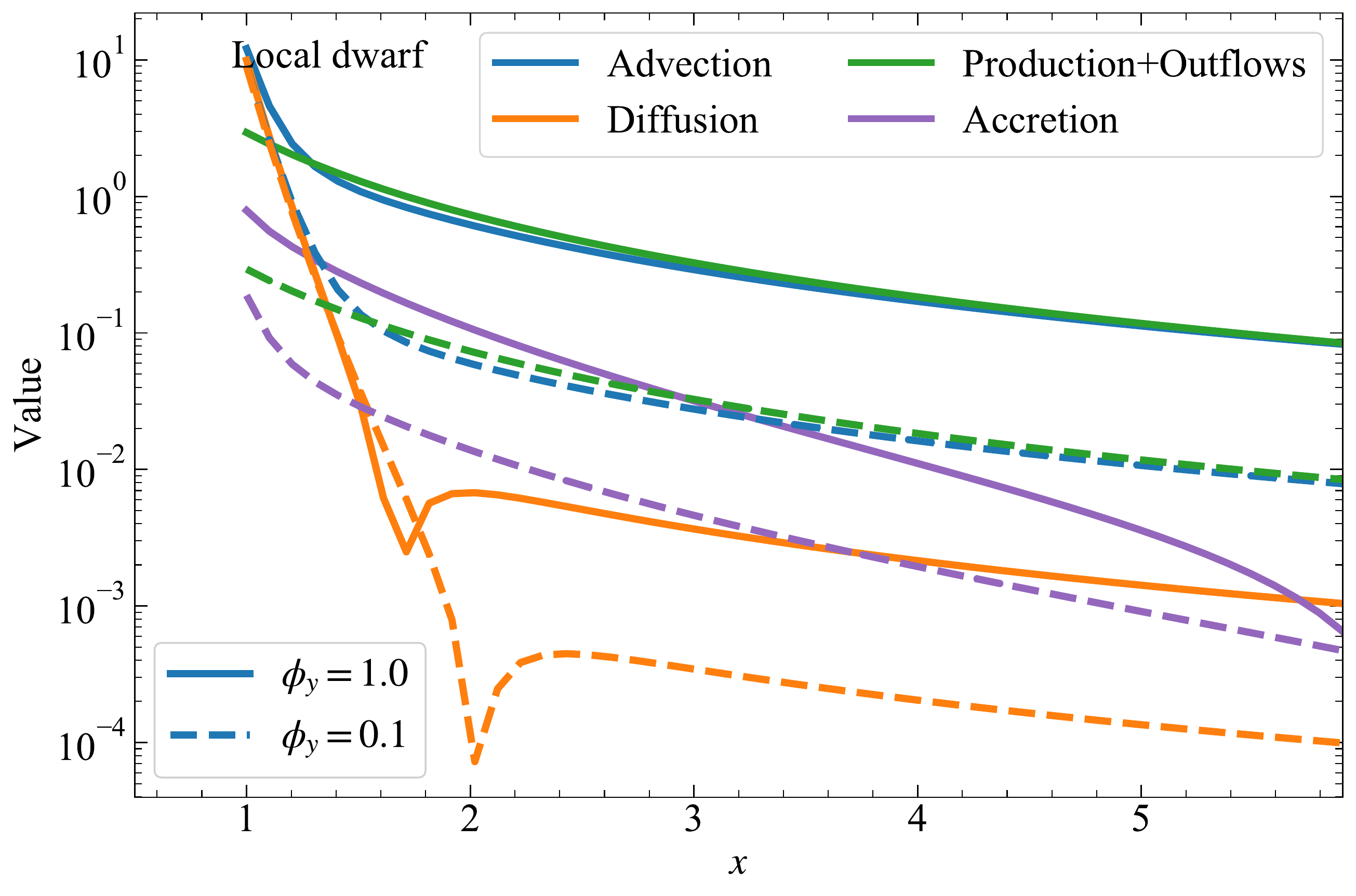}
\caption{Same as \autoref{fig:localspirals_terms}, but for local dwarfs. The dominant terms that set the gradients in local dwarfs are advection and diffusion (in the inner disc) and source and advection (in the outer disc).}
\label{fig:localdwarfs_terms}
\end{figure}

Having established metal equilibrium in local dwarfs, we can now study the gradients produced by the model. \autoref{fig:localdwarfs} shows the resulting metallicity versus radius for different $\phi_y$ (analogous to \autoref{fig:localspirals}), and \autoref{fig:localdwarfs_terms} shows the relative importance of the various processes (analogous to \autoref{fig:localspirals_terms}).

In the case of local dwarfs, we see that $\mathcal{Z}_{r_0}$ is set by the balance between advection and diffusion, giving
\begin{equation}
    \mathcal{Z}_{r_0} = \frac{\mathcal{S}}{\mathcal{A}} + c_1\left(1 + \frac{\sqrt{\mathcal{P}^2+4\mathcal{A}} - \mathcal{P}^2 - \mathcal{A}}{\sqrt{\mathcal{P}^2+4\mathcal{A}} + \mathcal{P}^2 + \mathcal{A}}\right)\,\,\,\,\left[\mathrm{Local}\,\mathrm{dwarfs}\right]\,.
\label{eq:Zr0_localdwarfs}
\end{equation}
Using the above definition of $\mathcal{Z}_{r_0}$, we can express $c_1$ as
\begin{equation}
    c_1 = \frac{\sqrt{\mathcal{P}^2+4\mathcal{A}}+\mathcal{P}^2+\mathcal{A}}{\left[\mathcal{A} + \left(\mathcal{P}-1\right)\mathcal{P}\right]\sqrt{\mathcal{P}^2+4\mathcal{A}}}\left.\frac{\partial \mathcal{Z}}{\partial x}\right|_{r=r_0}\,.
\label{eq:c1_localdwarfs}
\end{equation}

Central metallicities are in the range $\mathcal{Z}_{r_0} \approx 0.2-0.6$ depending on the choice of $\phi_y$, in good agreement with that observed in local dwarfs, e.g., in the SMC and the LMC \citep{1992ApJ...384..508R,1997macl.book.....W}, and M82 \citep{2004ApJ...606..862O}. While $\mathcal{Z}_{r_0}$ depends only on $\mathcal{S}/\mathcal{A}$ in local spirals, it also depends on the choice of $c_1$ for local dwarfs, implying that it is independent of the disc properties in the former case but not in the latter.\footnote{This dependence is also behind the sharp rise and fall near $x=1$ seen in both the diffusion term and the metallicity profile. For the purposes of plotting, we have chosen a single value of $c_1$, which in turn forces all models to converge to a single $\mathcal{Z}_{r_0}$. While we could correct this by choosing different values of $c_1$ for different models so that they remain smooth, since the sharp feature does not affect the metallicity gradient that is our main focus in this paper, we choose for reasons of simplicity to retain the fixed $c_1$.} Similarly, mean metallicities range from $\overline{\mathcal{Z}} \sim 0.1 - 0.5$ as $\phi_y$ varies from $\approx 0.1 - 1$; both observations \citep{2002ApJ...574..663M,Strickland09a, 2018MNRAS.481.1690C} and numerical simulations \citep{2018ApJ...869...94E, 2019MNRAS.482.1304E} suggest that dwarfs suffer considerable direct metal loss, so $\phi_y$ considerably smaller than unity seems likely.

As opposed to spirals, our models predict that gradients are not necessarily flat in the inner regions of dwarfs, which is also consistent with observations \citep{2017MNRAS.469..151B, 2020A&A...636A..42M}. The reason for this difference is due to different physical processes dominating in the two types of galaxies: accretion versus metal production in spirals, and advection versus production in dwarfs. Consequently, we predict linear gradients for local dwarfs that are steeper than the ones for local spirals at fixed $\phi_y$ and $c_1$. For the smaller values of $\phi_y$ expected in local dwarfs, we expect gradients in the range $\sim -0.01$ to $-0.15\,\rm{dex\,kpc^{-1}}$, implying a larger scatter in the gradients measured in local dwarfs as compared to that in local spirals, consistent with observations \citep[Figure 12]{2020A&A...636A..42M}. The metallicity profiles produced by the model for smaller values of $\phi_y$ are also in agreement with that observed in the MaNGA survey \citep{2017MNRAS.469..151B}, as we show in \autoref{fig:localdwarfs}, where we have adjusted the overall metallicity normalization by 0.15 dex to facilitate a comparison of the data and the model profiles. Further, the larger range of gradients in low mass local galaxies as compared to massive galaxies allowed within the framework of our model is also relevant and necessary for reproducing the observed steepening of gradients with decreasing galaxy mass \citep[Figure~10]{2019MNRAS.488.3826B}.

Although this is not illustrated in \autoref{fig:localdwarfs}, we also find that the magnitude of the gradient is quite sensitive to both the ``floor'' velocity dispersion supplied by star formation, $\sigma_{\rm{sf}}$, and the Toomre $Q$ parameter, since these two jointly set the strength of advection and in this case, $\sigma_{\rm{sf}} \sim \sigma_g$. Thus, we expect that gradients for local dwarfs will show more scatter than those for local spirals. It is interesting to note that there is a similarly large scatter in simulations of dwarf galaxies, with some groups  \citep[e.g.,][]{2016MNRAS.456.2982T} finding steeper gradients for dwarfs as compared to spirals whereas others \citep[e.g.][]{2017MNRAS.466.4780M} finding the opposite. This difference between the simulations has been attributed to the strength of feedback, which, in the language of our model, corresponds to variations in $\sigma_{\rm sf}$ and $\phi_y$; thus the sensitivity of our model is at least qualitatively consistent with the strong dependence of feedback strength observed in simulations.

\subsection{High-redshift galaxies}
\label{s:gradients_highz2}
\begin{figure}
\includegraphics[width=\columnwidth]{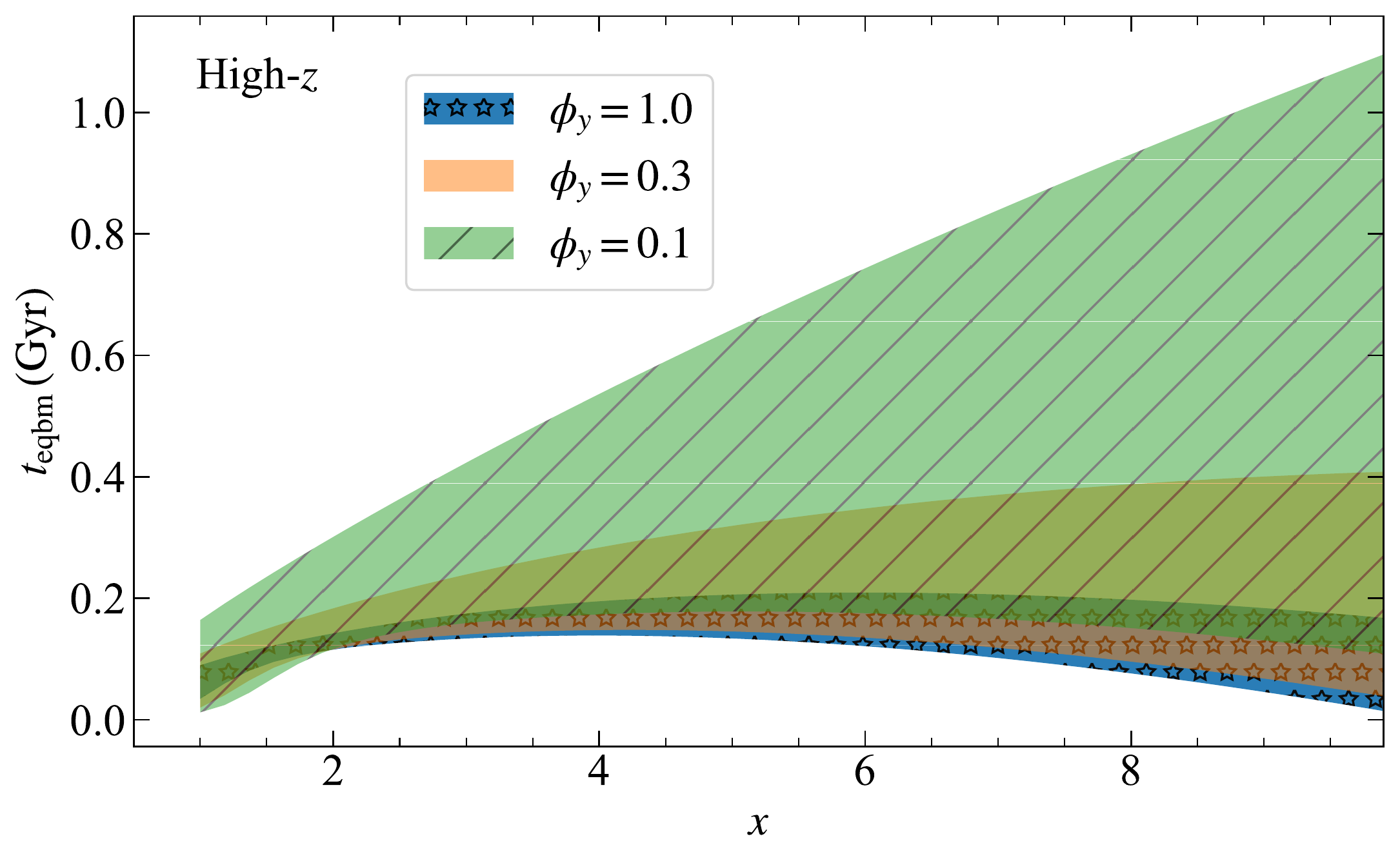}
\caption{Same as \autoref{fig:teqbm_ls}, but for high-$z$ galaxies. The corresponding equilibrium metallicity gradients are plotted in \autoref{fig:highz}.}
\label{fig:teqbm_hz}
\end{figure}

Massive galaxies at high-$z$ are primarily rotation-dominated with underlying disc-like structures \citep{2006ApJ...653.1027W,2009ApJ...706.1364F,2018ApJS..238...21F,2011MNRAS.417.2601W,2015ApJ...799..209W,2019ApJ...886..124W,2011ApJ...742...96W,2016A&A...594A..77D,2017ApJ...843...46S,2019ApJ...880...48U}. Thus, we can apply the model to these galaxies. For high-$z$ galaxies, we set the outer disc radius to $10\,\mathrm{kpc}$ to find the gradient in the fiducial model, acknowledging that galaxies at higher redshifts are smaller than that in the local Universe \citep[e.g.,][]{2012A&A...539A..93Q,2014ApJ...788...28V}. Hereafter, we work with $z =2$ as a fiducial redshift. \autoref{fig:teqbm_hz} shows the metal equilibration time for high-$z$ galaxies. It is clear that $t_{\rm{eqbm}} < t_{\rm{H(z)}}$, so that the equilibrium solution can be applied to these galaxies. Following \cite{2018ApJ...853..179T,2020arXiv200306245T}, if we assume that a main sequence high-$z$ galaxy follows $t_{\rm{dep,H_2}} \propto (1+z)^{-0.6}$, it implies that $t_{\rm{dep,H_2}} \sim 0.5-1.5\,\rm{Gyr}$ for high-$z$ galaxies, which is comparable with $t_{\rm{eqbm}}$ as above.

\begin{figure}
\includegraphics[width=\columnwidth]{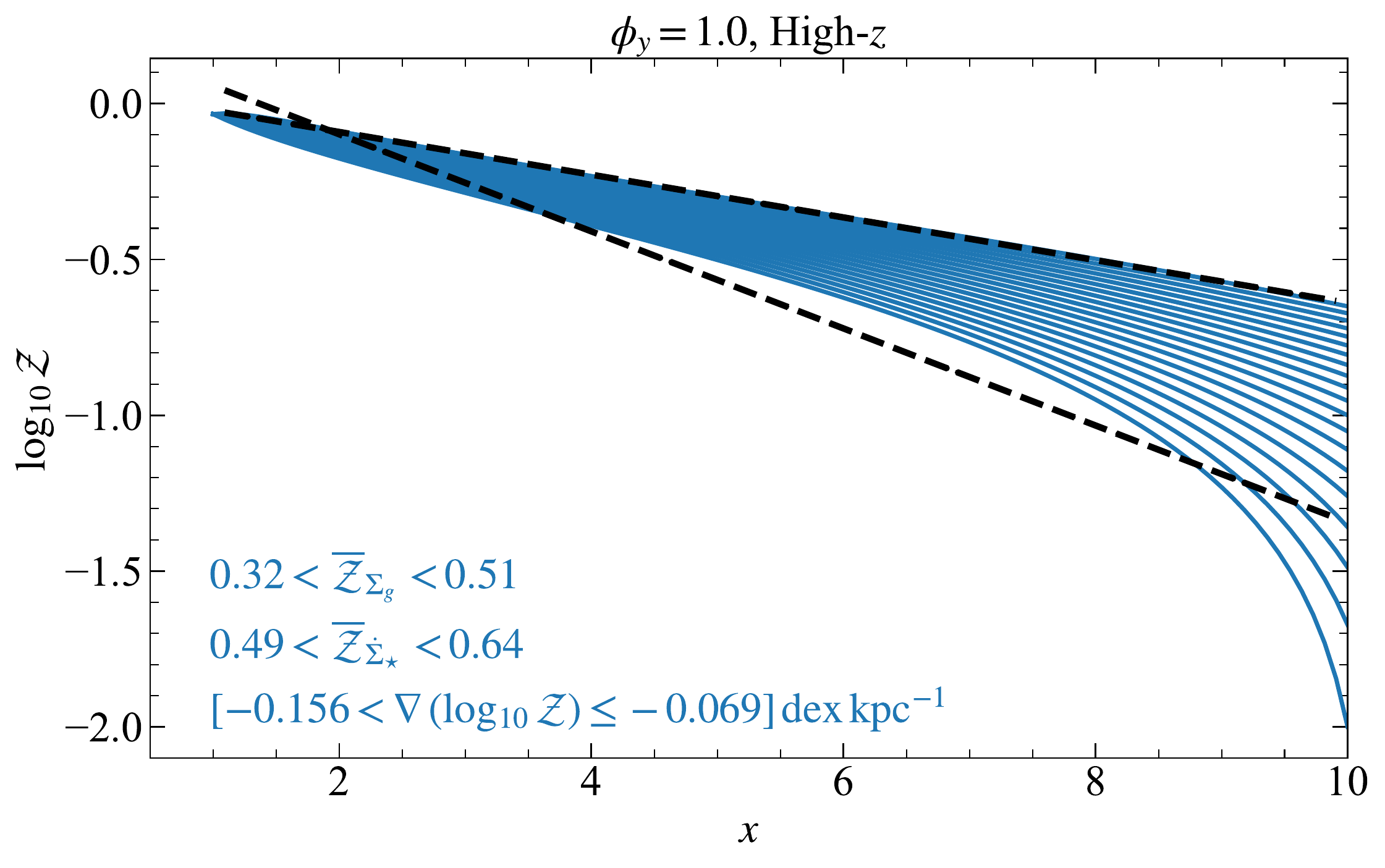}
\includegraphics[width=\columnwidth]{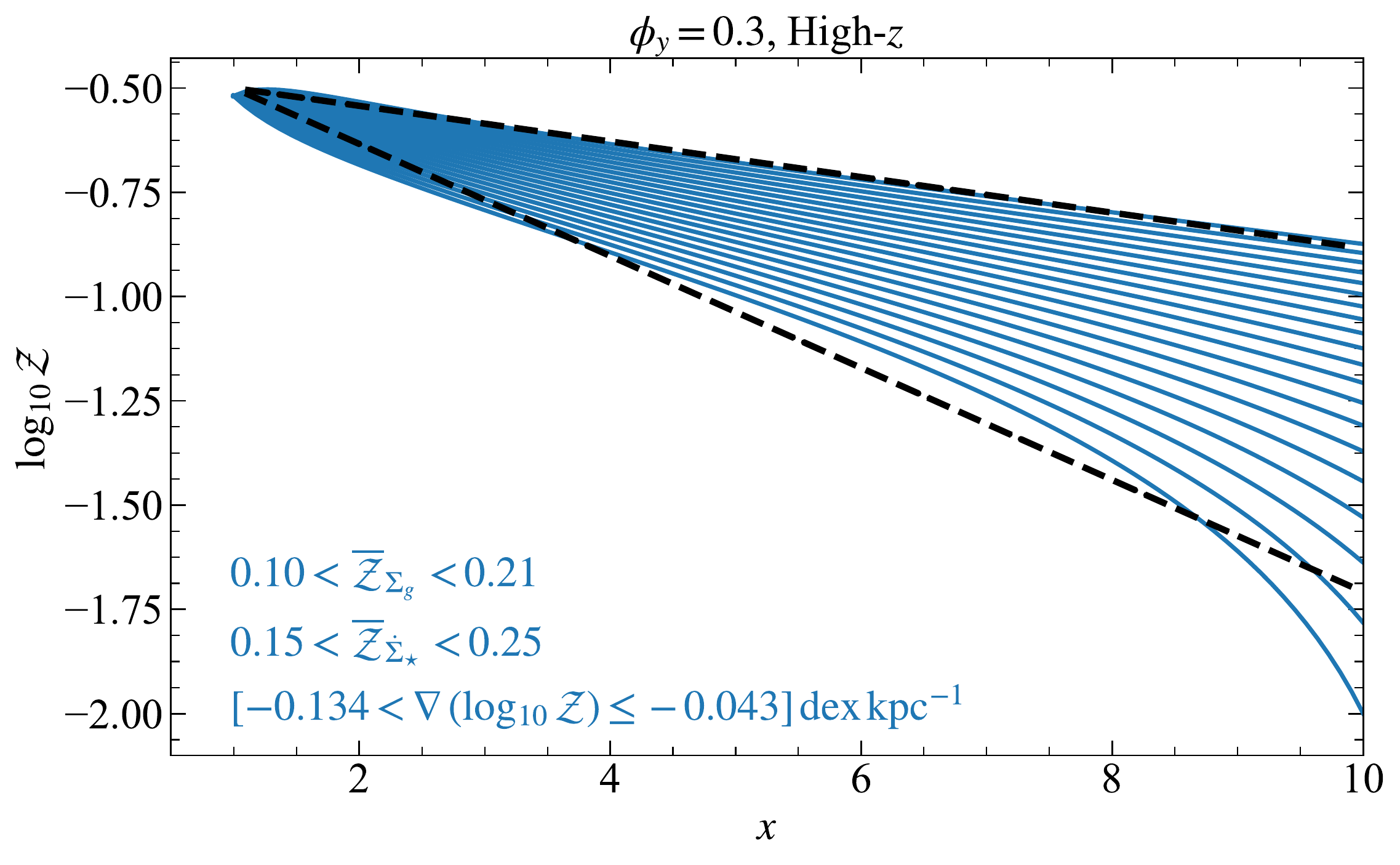}
\includegraphics[width=\columnwidth]{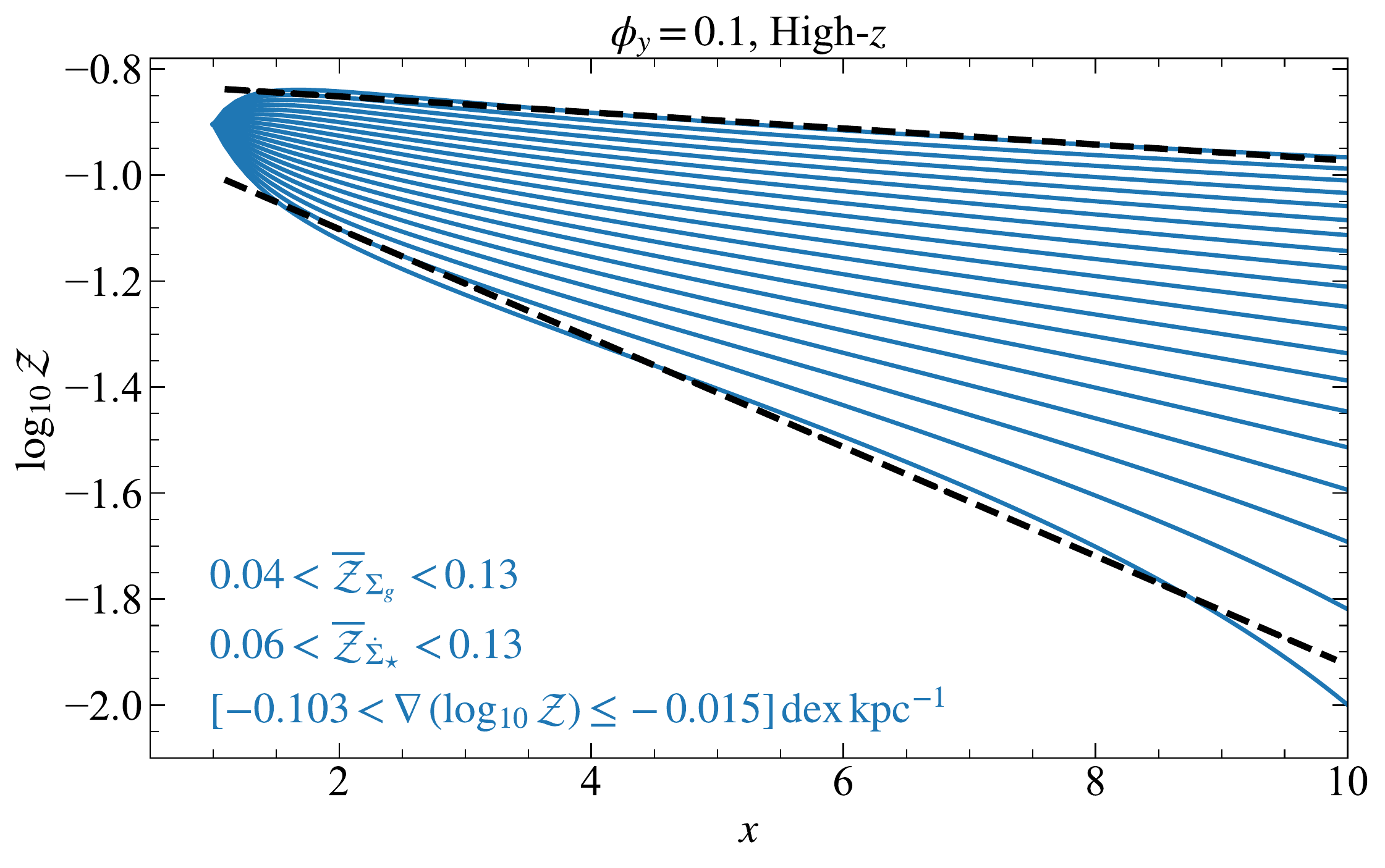}
\caption{Same as \autoref{fig:localspirals}, but for high-$z$ galaxies. Here, $\mathcal{Z}_{r_0}$ is set by the balance between diffusion and advection.}
\label{fig:highz}
\end{figure}

\begin{figure}
\includegraphics[width=\columnwidth]{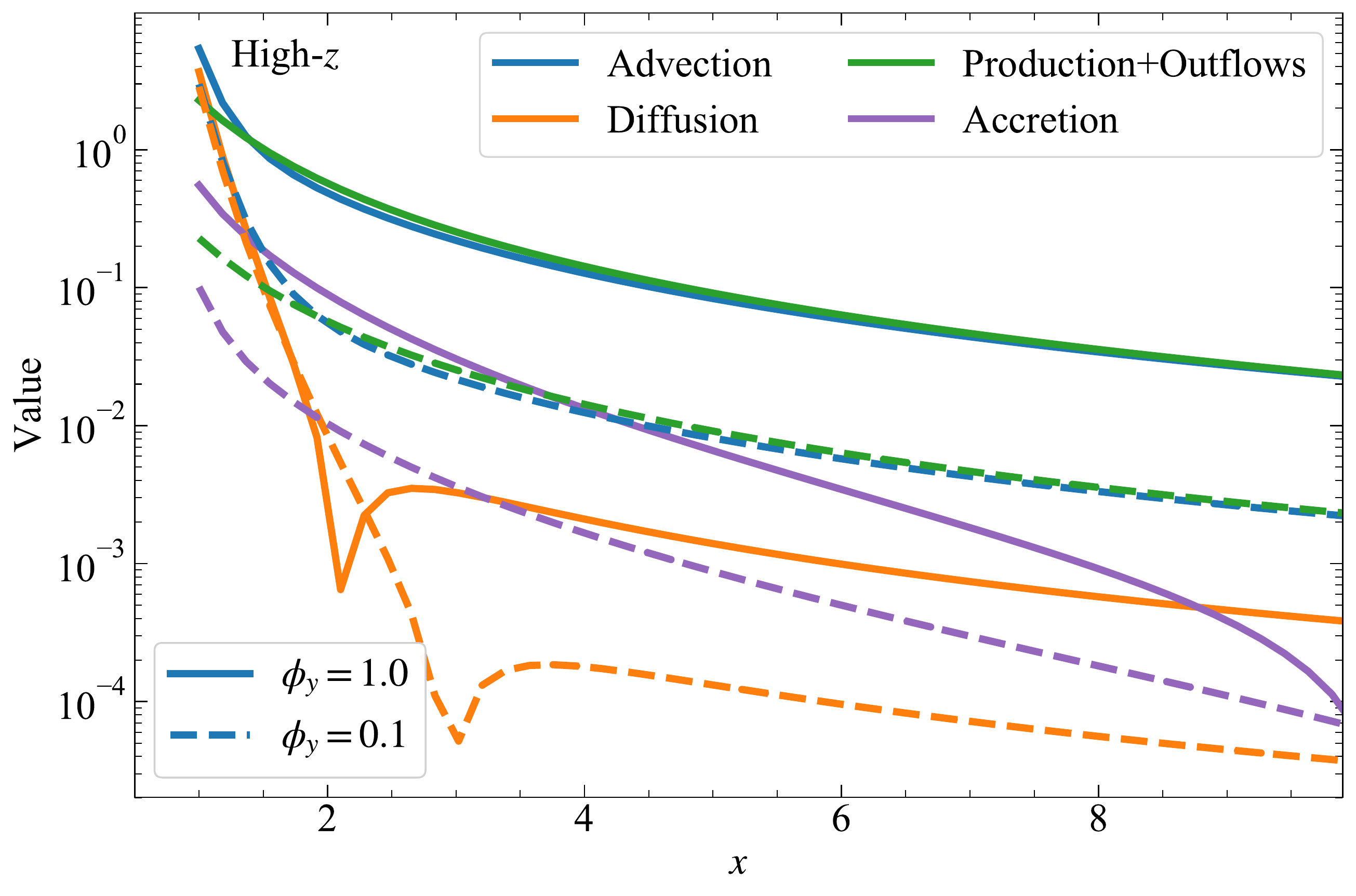}
\caption{Same as \autoref{fig:localspirals_terms}, but for high-$z$ galaxies. Here, the metallicities in the disc are set by the balance between source and advection, due to efficient radial transport of the gas.}
\label{fig:highz_terms}
\end{figure}

\autoref{fig:highz} shows the equilibrium metallicity distributions we obtain for a fiducial high-$z$ galaxy with parameters listed in \autoref{tab:tab1} and \autoref{tab:tab2}, and \autoref{fig:highz_terms} shows our usual diagnostic diagram comparing the importance of different processes. Examining this diagram near $x=1$, it is clear that, as is the case for local dwarfs, the central metallicity $\mathcal{Z}_{r_0}$ is set by the balance between advection and diffusion, which gives
\begin{equation}
    \mathcal{Z}_{r_0} = \frac{\mathcal{S}}{\mathcal{A}} + c_1\left(1 + \frac{\sqrt{\mathcal{P}^2+4\mathcal{A}} - \mathcal{P}^2 - \mathcal{A}}{\sqrt{\mathcal{P}^2+4\mathcal{A}} + \mathcal{P}^2 + \mathcal{A}}\right)\,\,\,\,\left[\mathrm{High}-z\right]\,.
\label{eq:Zr0_highz}
\end{equation}
It varies between $\mathcal{Z}_{r_0} = 0.3$--$0.7$ depending on the value of $\phi_y$, in good agreement with observed metallicities in high-$z$ galaxies in the mass range we consider \citep{2006ApJ...644..813E,2012PASJ...64...60Y}, with $c_1$ same as that in \autoref{eq:c1_localdwarfs}. While the absolute metallicity depends on $\phi_y$, the metallicity gradients for the most part do not -- we find $\nabla (\log_{10}\mathcal{Z}) \approx -0.15$ to $-0.05$ dex kpc$^{-1}$, with order-of-magnitude variations in $\phi_y$ only altering these values by a few hundredths.

The gradients we find for high-$z$ galaxies are steeper than for local spirals, and the distributions are steeper at small radii than at larger radii, the opposite of our finding for local spirals. \autoref{fig:highz_terms} shows why this is the case: gradients over most of the radial extent of high-$z$ galaxies are set by the balance between source and advection, whereas accretion, which dilutes the gradients in local spirals, is sub-dominant. The fundamental reason for this change is due to the vastly higher velocity dispersions of high-$z$ galaxies, which increase the importance of the advection term ($\mathcal{P} \propto (1 - \sigma_{\mathrm{sf}}/\sigma_g)$) while suppressing the accretion term ($\mathcal{A} \propto \sigma^{-3}_g$); this effect is partly diluted by the higher accretion rates found at high-$z$ (\autoref{eq:haloaccr}), but the net change at high redshift is nonetheless toward a smaller role for accretion onto discs and a larger role for transport through them. We discuss this further in detail in \autoref{s:cosmic}.

\section{Cosmic evolution of metallicity gradients}
\label{s:cosmic}
A significant advantage of our model compared to previous analytic efforts, is that it makes meaningful predictions for how galaxy metallicity gradients have evolved across cosmic time. This is the case because we do not have the freedom to adjust parameters such as radial inflow rates and profiles of star formation rate to match any given observed galaxy. Instead, these parameters are either prescribed directly from our galaxy evolution model or depend on parameters that are directly observable (e.g., galaxy velocity dispersions). The basic inputs to our model are the halo mass $M_{\rm{h}}$ and the gas velocity dispersion $\sigma_g$ as a function of $z$. We consider three different ways of selecting galaxies that yield different tracks of $M_h(z)$ (see below for details), while we take the evolution of $\sigma_g(z)$ from the observed correlation obtained by \citet[see their equation~8]{2015ApJ...799..209W}\footnote{As opposed to \cite{2015ApJ...799..209W}, we have explicitly retained the dependence of $\sigma_g(z)$ on $\beta$.} \begin{equation}
    \sigma_g(z) = \frac{v_{\phi}(z)f_{\mathrm{gas}}(z)}{\sqrt{2\left(\beta+1\right)}},
\label{eq:fgas_wisnioski}
\end{equation}
where $f_{\rm{gas}}$ is the molecular gas fraction of the galaxy \citep{2011ApJ...733..101G,2012ApJ...745...11G,2013ApJ...768...74T,2015ApJ...800...20G,2018MNRAS.473.3717F}. This scaling is subject to considerable observational uncertainty, the implications of which we explore in \aref{s:app_obsuncertainties}. We follow \cite{2015ApJ...799..209W} to find $f_{\rm{gas}}$ as a function of $M_{\star}$ and $z$ from \cite{2013ApJ...768...74T} and \cite{2014ApJ...795..104W}, as it is now known that $f_{\rm{gas}}$ decreases with cosmic time and stellar mass \citep{2008AJ....136.2782L,2011MNRAS.415...32S,2011ApJ...730L..19G,2012MNRAS.421...98D,2013ApJ...768...74T,2018ApJ...853..179T,2015MNRAS.454.3792M,2018ApJ...869L..37I}. We note that \citeauthor{2015ApJ...799..209W}'s sample is limited to massive galaxies ($M_{\star}>10^{10}\,\rm{M_{\odot}}$), and there are no observations available for lower-mass galaxies. For this reason we instead follow the results of the IllustrisTNG simulations to obtain $\sigma_g(z)$ \citep[see their Figure~12a]{2019MNRAS.490.3196P} for stellar masses below $10^{10}\,\rm{M_{\odot}}$. Finally, note that all the gradients we produce from the model in this section are in equilibrium across the redshifts we use, since $t_{\rm{eqbm}} < t_{\rm{H(z)}}$. 


\subsection{Trends for a Milky Way-like galaxy across redshift}
\label{s:cosmic_milkyway}
We first study how the gradient in a Milky Way-like galaxy has evolved over time using our model. We only need one parameter to begin with: $v_{\phi}$ at $z=0$. We set this to $220\,\rm{km\,s^{-1}}$ \citep{2016ARA&A..54..529B}. Then, we use \autoref{eq:halomass} to calculate $M_{\mathrm{h}}\,(z=0)$ for a fixed $c=15$. Using $M_{\mathrm{h}}\,(z=0)$ as boundary condition, we integrate \autoref{eq:haloaccr} to find $M_{\mathrm{h}}\,(z)$, keeping in mind that this equation represents an average evolution of $M_{\mathrm{h}}\,(z)$ that may not necessarily apply to the Milky Way. Then, we utilize $M_{\mathrm{h}}\,(z)$ to find $v_{\phi}\,(z)$, by changing the concentration parameter ($c$) as an empirical third-order polynomial fit, following \cite{2009ApJ...707..354Z}. This ensures that as we change $z$, we self-consistently find $M_{\rm{h}}$ and $v_{\phi}$. We adopt a simple linear variation for the outer edge of the star-forming disc, $x_{\rm{max}}$, as a function of $z$ such that it is 15 at $z=0$ and 10 at $z=2$. Similarly, we vary $f_{\rm{sf}}$ between 0.5 and 1 across redshift, keeping in mind that $f_{\rm{sf}}$ cannot be more than 1 at any redshift. For simplicity, we fix the other parameters as follows: $\beta=0$, $f_{g,Q}=f_{g,P}=0.5$, $\sigma_{\mathrm{sf}}=7\,\mathrm{km\,s^{-1}},\,Q_{\mathrm{min}}=1.5$ and $\mathcal{Z}_{\rm{CGM}}=0.1$.

We show the resulting evolution of the gradient in \autoref{fig:depends_redshift_onegalaxy}. The model predicts a steepening of the gradient in Milky Way-like galaxies over time, with the exception of a very recent flattening, between $z\approx 0.15$ and 0. We can understand these trends in terms of the dimensionless parameters $\mathcal{S}$, $\mathcal{P}$, and $\mathcal{A}$ that describe the relative importance of \textit{in situ} metal production, radial advection, and cosmological accretion with diffusion, respectively. The source term $\mathcal{S}$ will always make the gradients steeper because of the steep radial profile of $\dot\Sigma_{\star}$, and it is either $\mathcal{P}$ or $\mathcal{A}$ that balances $\mathcal{S}$ to give rise to flatter gradients. The steepest gradients at $z\approx 0.15$ correspond to when both $\mathcal{P}$ and $\mathcal{A}$ are at their weakest compared to $\mathcal{S}$. We can understand the trends on either side of this maximum in turn.

First, let us focus on the recent epoch, $z\lesssim 0.15$. During this period, cosmological accretion ($\mathcal{A}$) is more important than radial transport ($\mathcal{P}$), and accretion and metal production depend on the galaxy rotational velocity as $\mathcal{A}\propto v_{\phi}^{3.3}$ and $\mathcal{S}\propto v_{\phi}^2$, respectively. Thus, as the galaxy grows in mass, dilution by accretion gets stronger compared to metal production, leading to the recent flattening in the model. However, this can change if the metal production is underestimated, e.g., due to ignoring the contribution from long-term wind recycling \citep{2011ApJ...734...48L}.

During this epoch advection is more important than accretion, $\mathcal{P}>\mathcal{A}$. The ratio of the two effects, $\mathcal{P}/\mathcal{A}$, is large at high redshift, and decreases systematically towards the present day, ultimately reaching $\mathcal{P}/\mathcal{A} \approx 1$ at $z\approx 0.15$. This transition is ultimately driven by the systematic decrease in galaxy velocity dispersions with redshift, as already discussed in the context of our high-$z$ galaxy models (\autoref{s:gradients_highz2}): higher velocity dispersions are strongly correlated with higher rates of radial inflow through a galaxy, so that for a Milky Way progenitor at $z \gtrsim 1$, radial inflow transports metal-poor gas into galaxy centres $\sim 10\times$ faster than cosmological accretion ($\mathcal{P}/\mathcal{A}\approx 10$) -- despite the fact that the absolute accretion rate is higher at $z\gtrsim 1$ than it is today. Similarly, the ratio of radial inflow to metal production, $\mathcal{P}/\mathcal{S}$, scales with velocity dispersion as $\sigma_g^2$ (for $\sigma_g \gg \sigma_{\rm sf}$), so radial inflow also becomes more important relative to metal production as we go to higher redshift and higher velocity dispersion. This explains the flatness of gradients at high redshift\footnote{Note that this is a qualitatively different outcome than our comparison of local spirals and high-$z$ galaxies in \autoref{s:gradients_localspirals1} and \autoref{s:gradients_highz2}, where high-$z$ galaxies were found to have steeper gradients. The difference can be understood by recalling that in \autoref{s:gradients_localspirals1} and \autoref{s:gradients_highz2} we were comparing galaxies with comparable rotation curve speeds $v_\phi$, whereas here we are following a single growing galaxy, so $v_\phi$ is much smaller at high-$z$ than at $z=0$. This reduces $\mathcal{S}$ at high-$z$.}. This transition from radial advection being dominant to being unimportant is mirrored in the transition from gravity-driven to star formation feedback-driven turbulence from high- to low-$z$ \citep{2018MNRAS.477.2716K}, as we noted earlier in \autoref{s:gradients_highz2}. 

Lastly, we find that diffusion is sub-dominant compared to both advection and accretion at all cosmological epochs, because $\mathcal{P}$ and $\mathcal{A}$ are never both less than unity at the same time. Thus, while diffusion can have some effects on the metallicity distributions, particularly towards galaxy centres (cf.~\autoref{fig:localdwarfs_terms}), as well as on metal equilibrium timescales (cf.~\autoref{fig:teqbm_ls}), it is generally unimportant for setting galaxy metallicity gradients.

\subsubsection{Comparison with observations}
There is extensive data on the history of the Galaxy's metallicity gradient, as summarised by \citet[see their Table~1]{2019MNRAS.482.3071M}, and on the history of the gradients in a number of other nearby galaxies. The general outcome of these studies is that gradients measured in \ion{H}{ii} regions (which trace the current-day metal distribution) are steeper than those measured in planetary nebulae or open clusters (which trace older populations) \citep{2010ApJ...714.1096S,2010A&A...521A...3S,2014A&A...567A..88S,2012ApJ...758..133S,2013A&A...552A..12S,2016A&A...588A..91M}. This implies a steepening of the gradient with time in Milky Way-like galaxies, however, this should be treated with caution because measured metallicity gradients in the Galaxy are subject to large errors arising from uncertainties in estimating the ages of the planetary nebulae \citep{2010A&A...512A..19M,2011RMxAA..47...49C}, and due to radial migration that could result in a movement of the planetary nebulae away from their origin \citep{2013A&A...558A...9M}\footnote{Some earlier work reported the opposite trend, whereby the metallicity gradient in the Galaxy was initially steep and has flattened over time  \citep{2003A&A...397..667M,2005MNRAS.358..521M}, while other work found little or no evolution in the gradient over time \citep{2013RMxAA..49..333M}. This is a difficult measurement, and the error bars and uncertainties are large \citep{2010A&A...512A..19M,2011RMxAA..47...49C,2013A&A...558A...9M}.}.

To allow a quantitative comparison of these observations with our model, we show measurements of the metallicity gradient for the Milky Way as a function of lookback time from \cite{2010A&A...521A...3S} as yellow circles in \autoref{fig:depends_redshift_onegalaxy}. The data for the Milky Way (as well as other local spirals, see \citealt{2014A&A...567A..88S}) are in qualitative agreement with the predictions from our model. However, we also note that for our model to agree \textit{quantitatively} with the measurements, we would need $\phi_y$ to be lower at high redshift, and increase towards unity today. Such a change in $\phi_y$ is plausible and is consistent with our expectation that $\phi_y$ should be close to unity in more massive galaxies like the present-day Milky Way, and smaller than unity in less massive galaxies with shallower potential wells, such as the Milky Way's high-$z$ progenitors. However, the exact form of this evolution is not independently predicted by our model.

\subsubsection{Comparison with simulations}
\begin{figure*}
\includegraphics[width=\linewidth]{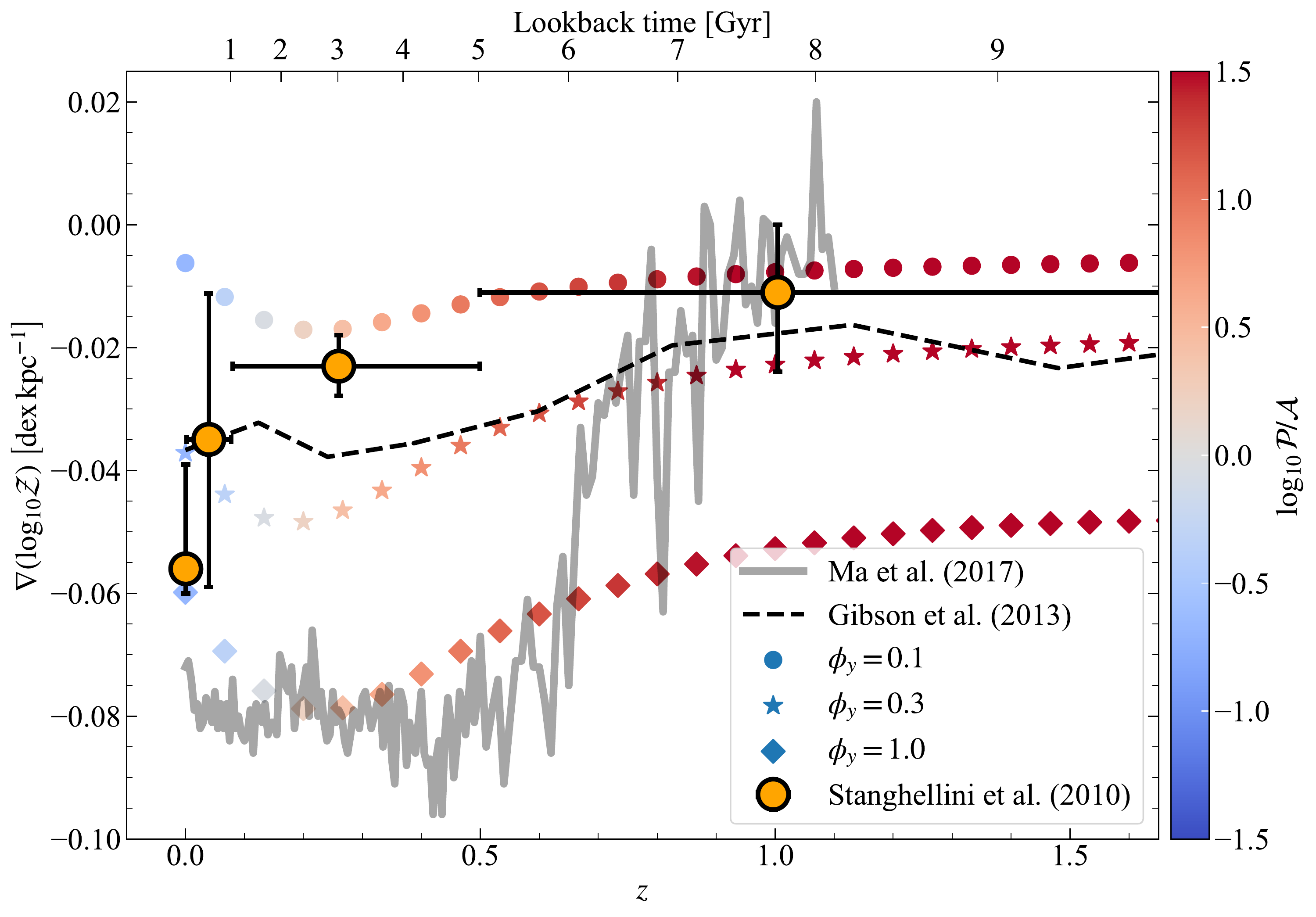}
\caption{Metallicity gradient versus redshift (and lookback time) for a Milky Way-like galaxy. Different symbols show different yield reduction factors, $\phi_y$, while symbol colour shows the ratio of the dimensionless numbers $\mathcal{P}/\mathcal{A}$ that describe the relative importance of radial transport and cosmological accretion, respectively. The grey curve is taken from FIRE simulations of a Milky Way-like galaxy \protect\citep{2017MNRAS.466.4780M} whereas the dashed, black curve is from the MaGICC g1536 simulation by \protect\citet{2013A&A...554A..47G}. The orange points are from observations of \ion{H}{ii} regions, planetary nebulae and open clusters by \protect\cite{2010ApJ...714.1096S}, with horizontal errorbars representing the uncertainties in the ages of planetary nebulae and open clusters. The data, simulations and the model all qualitatively show that gradients in Milky Way-like galaxies have steepened over time, with the model predicting a mild flattening between $z=0.15$ and present-day. In the model, this evolution is driven by a transition from the advection-dominated regime ($\mathcal{P}/\mathcal{A} > 1)$ to the accretion-dominated regime ($\mathcal{P}/\mathcal{A} < 1)$ around $z \approx 0.15$. Such a transition in metallicity gradients is mirrored in the transition in gravity-driven turbulence at high $z$ to star formation feedback-driven turbulence at $z=0$ \protect\citep{2018MNRAS.477.2716K}.}
\label{fig:depends_redshift_onegalaxy}
\end{figure*}

On \autoref{fig:depends_redshift_onegalaxy}, we also overplot results from Feedback In Realistic Environments (FIRE) simulations \citep{2014MNRAS.445..581H,2018MNRAS.480..800H} of a Milky Way-like galaxy (m12i) discussed in \citet{2017MNRAS.466.4780M}. This simulation finds that metallicity gradients are unstable until $z \sim 1$, after which they steepen and stabilise to an equilibrium value. This transition is primarily due to the formation of a robust galactic disc that cannot be disrupted again due to internal or external feedback. While the quantitative trends slightly differ at some redshifts between our model and the simulation, which is not unexpected given that the exact implementation of the feedback and measurements of the gradients are different, there is a very good qualitative match. This match also implies that Milky Way-like galaxies would have had lower $\phi_y$ in the past as compared to the present day, as outflows were more common and stronger in the past due to higher SFR and could have ejected a larger fraction of metals not mixed with the ISM \citep{2015MNRAS.454.2691M,2017MNRAS.466.4780M}; such a scenario has received support from recent high-resolution simulations that spatially resolve multi-phase galactic outflows, and find that the metal enrichment factor in both the cold ($< 2\times10^4\,\rm{K}$) and hot ($> 5\times10^5\,\rm{K}$) outflows increases with the SFR surface density \citep{2020ApJ...900...61K}. We can also compare our results with those of \cite{2013A&A...554A..47G}, where the authors study two identical simulation suites with either weak or enhanced stellar feedback, called MUGS and MaGICC, respectively \citep{2010MNRAS.408..812S}. The authors find that gas phase metallicity gradients are steep at high redshift in MUGS, whereas they are flat in MaGICC, clearly revealing the close correlation between feedback and metallicity gradients in galaxies. One of their simulated galaxies, MaGICC g1536, resembles the Milky Way in terms of its stellar mass, so we also compare our model results to that simulation in \autoref{fig:depends_redshift_onegalaxy}. Again, we find qualitative similarities between the simulations and the model.

\subsection{Trends for matched stellar mass galaxies across redshift}
\label{s:cosmic_identicalmass}

\begin{figure*}
\includegraphics[width=\linewidth]{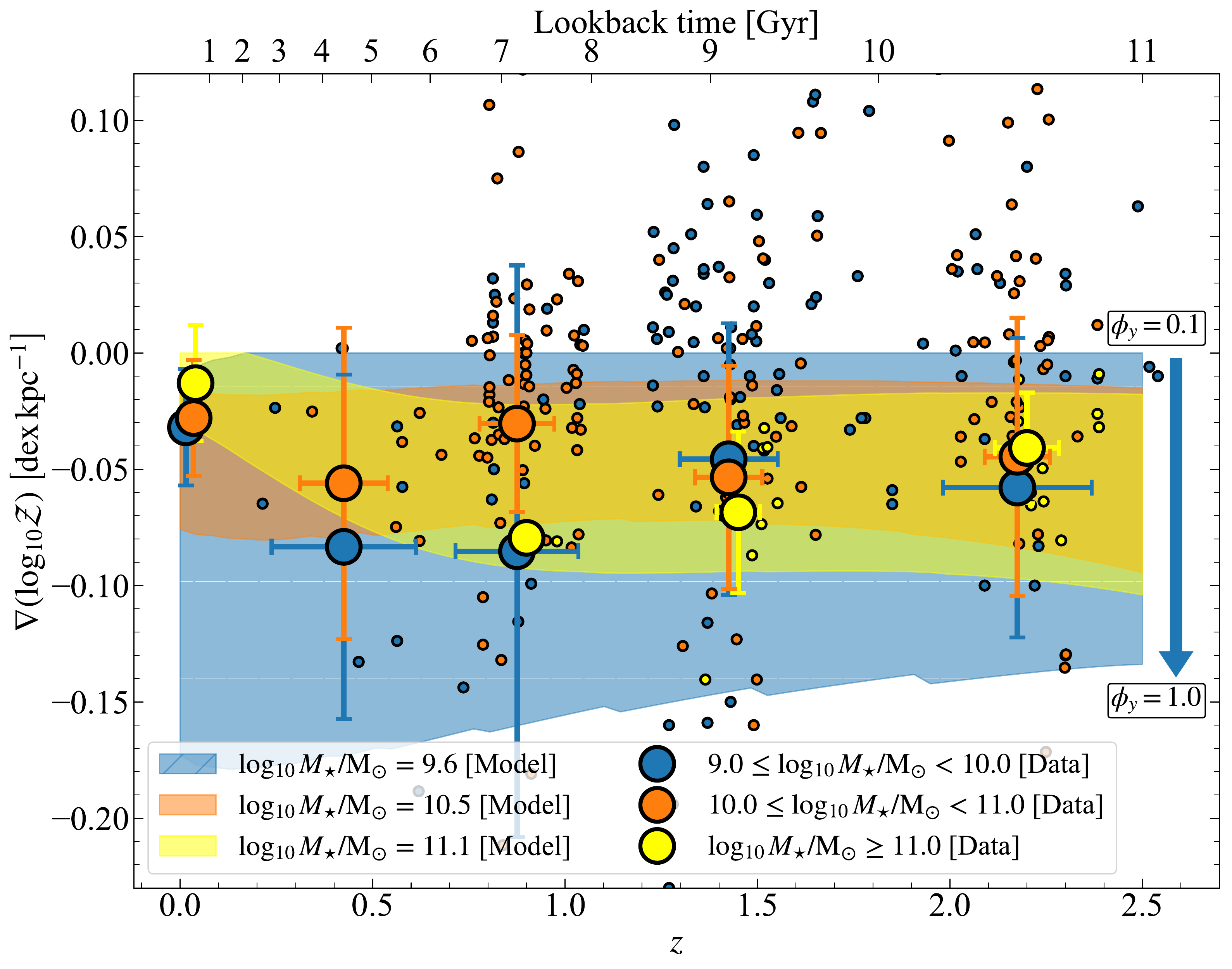}
\caption{Trends in metallicity gradients as a function of redshift and lookback time. Colored markers represent individual galaxies within the three $M_{\star}$ bins as shown in the legend, with bigger markers representing binned averages of non-positive gradients across different redshift bins, and errorbars representing the scatter in the data within each redshift bin. The averages at $z=0$ are taken from local surveys \protect\citep{2014A&A...563A..49S,2016A&A...587A..70S,2017MNRAS.469..151B,2020A&A...636A..42M}. The high-redshift compilation data is taken from \protect\cite{2012A&A...539A..93Q,2012MNRAS.426..935S,2014MNRAS.443.2695S,2016ApJ...820...84L,2016ApJ...827...74W,2017MNRAS.466..892M,2018MNRAS.478.4293C,2018ApJS..238...21F,2020ApJ...900..183W,2020MNRAS.492..821C}, and is inhomogeneous, with systematic issues within the different measurements (see \autoref{s:cosmic_identicalmass}). The colored bands represent models at three $M_{\star}$ values, with the spread resulting from different yield reduction factors $\phi_y$, as marked by the arrow besides the shaded region. This spread in the model is largest for the low mass galaxies. While the general trend of mild evolution of gradients across redshift holds true, the models uncover the underlying variations due to galaxies transitioning from advection- to accretion-dominated regimes between $z=2.5$ and $0$, as is visible in the binned data averages. Some data points lie outside the range of the plot, and we do not include those for the purposes of studying the average trends of the data with the model. }
\label{fig:depends_redshift_samemassgalaxy}
\end{figure*}

In this section, we study the mass-averaged trends of metallicity gradients across cosmic time. For this purpose, we use a compilation of observations of metallicity gradients in (lensed and un-lensed) galaxies spanning $0 \leq z \leq 2.5$ \citep{2012A&A...539A..93Q,2012MNRAS.426..935S,2014MNRAS.443.2695S,2016ApJ...820...84L,2016ApJ...827...74W,2017MNRAS.466..892M,2018MNRAS.478.4293C,2018ApJS..238...21F,2020ApJ...900..183W,2020MNRAS.492..821C}, and we also include results from local surveys \citep{2014A&A...563A..49S,2016A&A...587A..70S,2017MNRAS.469..151B,2020A&A...636A..42M,2020MNRAS.xxx..xxxA}. 

Before proceeding, we warn the reader that there are many uncertainties inherent in comparing metallicity gradients across samples and across cosmic time. For example, most studies in the compiled dataset rely on strong line calibrations that use photoionisation models or electron temperature-based empirical relations to measure metallicity gradients, and the variations between different calibrations can be as high as 0.1 dex per effective half-light radius \citep{2010ApJS..190..233M,2019MNRAS.487...79P,10.1093/mnras/stab205,2020A&A...636A..42M}. Further, since many high-$z$ metallicity gradient measurements rely on nitrogen whereas low-$z$ measurements use a larger set of (optical) emission lines, we also expect some systematic differences in these measurements with redshift \citep{2018MNRAS.478.4293C,2019ARA&A..57..511K}. Using nitrogen can also lead to systematically flatter gradients due to different scalings of N/O with O/H in galaxy centres and outskirts \citep{2020ApJ...890L...3S}. Lastly, it is not yet clear if strong line metallicity calibrations developed for the ISM properties of local galaxies are also applicable at high-$z$, where ISM electron densities, ionisation parameters, N/O ratios, or other conditions may differ from those in local galaxies \citep[e.g.,][]{2014ApJ...787..120S,2016ApJ...816...23S,2016ApJ...822...42O,2017ApJ...835...88K,2017MNRAS.465.3220K,2019ApJ...880...16K,2020arXiv201210445D}. We acknowledge these biases and uncertainties in the measured sample due to different techniques and calibrations or the lack of spatial and/or spectral resolution \citep{2013ApJ...767..106Y,2014A&A...561A.129M,2017MNRAS.468.2140C,2020MNRAS.495.3819A}. We do not attempt to correct for these effects or homogenize the sample because our goal here is simply to get a qualitative interpretation of the data with the help of the model, and not to obtain precise measurements from these data. Future facilities like JWST and ELTs will provide more reliable metallicity measurements, thereby enabling a more robust comparison of the model with the data \citep{2020IAUS..352..342B}.

We bin the data into three bins of $M_{\star}$: $9 \leq \log_{10}\,M_{\star}/\rm{M_{\odot}} < 10$, $10 \leq \log_{10}\,M_{\star}/\rm{M_{\odot}} < 11$ and $\log_{10}\,M_{\star}/\rm{M_{\odot}} \geq 11$. \autoref{fig:depends_redshift_samemassgalaxy} shows the individual data as well as the binned averages of non-positive gradients (represented by bigger markers) with errorbars representing the scatter in the data within different redshift bins. We only select galaxies that show non-positive gradients while estimating the average gradient in different mass bins because our model may not apply to galaxies with positive gradients, as we explore in \autoref{s:noneqbm_inverted_gradients}. We bin the data in redshift such that we can avoid redshifts where there is no data due to atmospheric absorption; such a bin selection in redshift also ensures that the binned averages reflect the true underlying sample for which the averages are calculated. We have verified that our results are not sensitive to the choice of binning the data. For simplicity, we do not overplot measurements for individual galaxies at $z=0$. 

For the model, we select three representative $M_{\star}$ values corresponding to the mean of the three stellar mass data bins as above. Specifically, we use: $\log_{10} M_{\star}/\mathrm{M_{\odot}} = 9.6,\,10.4$ and $11.1\,\rm{M_{\odot}}$ for the model. We start the calculation by selecting rotation curve speeds $v_{\phi}\,(z)$ corresponding to each of these $M_{\star}$ values based on the $M_{\star}-M_{\mathrm{h}}$ relation at all $z$ \citep{2013MNRAS.428.3121M}. Given values of $M_h(z)$ and $v_\phi(z)$ corresponding to each stellar mass $M_\star$ at each redshift $z$, we use our model to predict the equilibrium metallicity gradient exactly as in \autoref{s:cosmic_milkyway}. 

We plot the resulting range of metallicity gradients from the model points in \autoref{fig:depends_redshift_samemassgalaxy}. As in other figures, the spread in the model represents different $\phi_y$ between 0 and 1 (note the arrow besides the shaded regions corresponding to the models). While there is a large scatter within the individual data points, the binned averages are in good agreement with the model. Note that almost one-third of the observed galaxies show inverted gradients, which may not be in metal equilibrium and thus may fall outside the domain of our model, as we explore in detail in \autoref{s:noneqbm_inverted_gradients}. For the most massive galaxies, the model predicts a mild steepening of the gradients from $z=2.5$ to $1$, followed by an upturn (due to the transition from advection- to accretion-dominated regime) and flattening from $z=1$ to $0$. The available data, despite the large scatter and inhomogeneties, also seem to follow the same trend. However, the location where this upturn occurs is unknown because of the lack of data in the most massive galaxy bin around $z = 0.5$. Upcoming large surveys like MAGPI \citep{2020arXiv201113567F} that will observe massive galaxies between $z \approx 0.3-0.5$ will provide crucial data that can be compared against our model in the future to establish whether this upturn is indeed real.

Additionally, we can compare our results with those from the IllustrisTNG50 simulation \citep[Figure~6]{2020arXiv200710993H}. While our results match theirs at low redshifts, there are certain differences at high redshifts where IlustrisTNG50 fails to reproduce the observed flattening, as already noted by the corresponding authors. We explain in a companion paper \citep{2020bMNRAS.xxx..xxxS} that this difference could primarily be due to the gas velocity dispersion $\sigma_g(z)$. At high redshift, IllustrisTNG50 systematically under-predicts galaxy velocity dispersions as compared to, for example, the EAGLE simulations \citep[Figure~12a]{2019MNRAS.490.3196P}, and the empirical relation we use from \cite{2015ApJ...799..209W}.

There is a large diversity of gradients at all redshifts \citep{2020MNRAS.492..821C}, particularly at low stellar mass. This observed scatter can be explained in part due to the range of $\phi_y$ in our model. For example, we notice from \autoref{fig:depends_redshift_samemassgalaxy} that the scatter in the model due to $\phi_y$ for the most massive galaxies is lower at low $z$ than at high $z$. This is consistent with the trend of larger scatter in the gradients of massive galaxies at higher redshift observed in the IllutrisTNG50 simulations \citep[Figure~6]{2020arXiv200710993H}. On the other hand, the scatter in the model is the largest near the upturn, where galaxies transition from advection-dominated to accretion-dominated regime. Between the three models, the scatter due to $\phi_y$ is the highest for the lowest $M_{\star}$, thus reflecting the diverse variety of gradients that can form in low-mass galaxies. This prediction of the model is consistent with observations that find strong evidence for increased scatter in the metallicity gradients in low mass galaxies \citep{2018MNRAS.478.4293C,2020arXiv201103553S}.


\subsection{Trends for abundance-matched galaxies across redshift}
\label{s:cosmic_abundancematch}

\begin{figure*}
\includegraphics[width=\columnwidth]{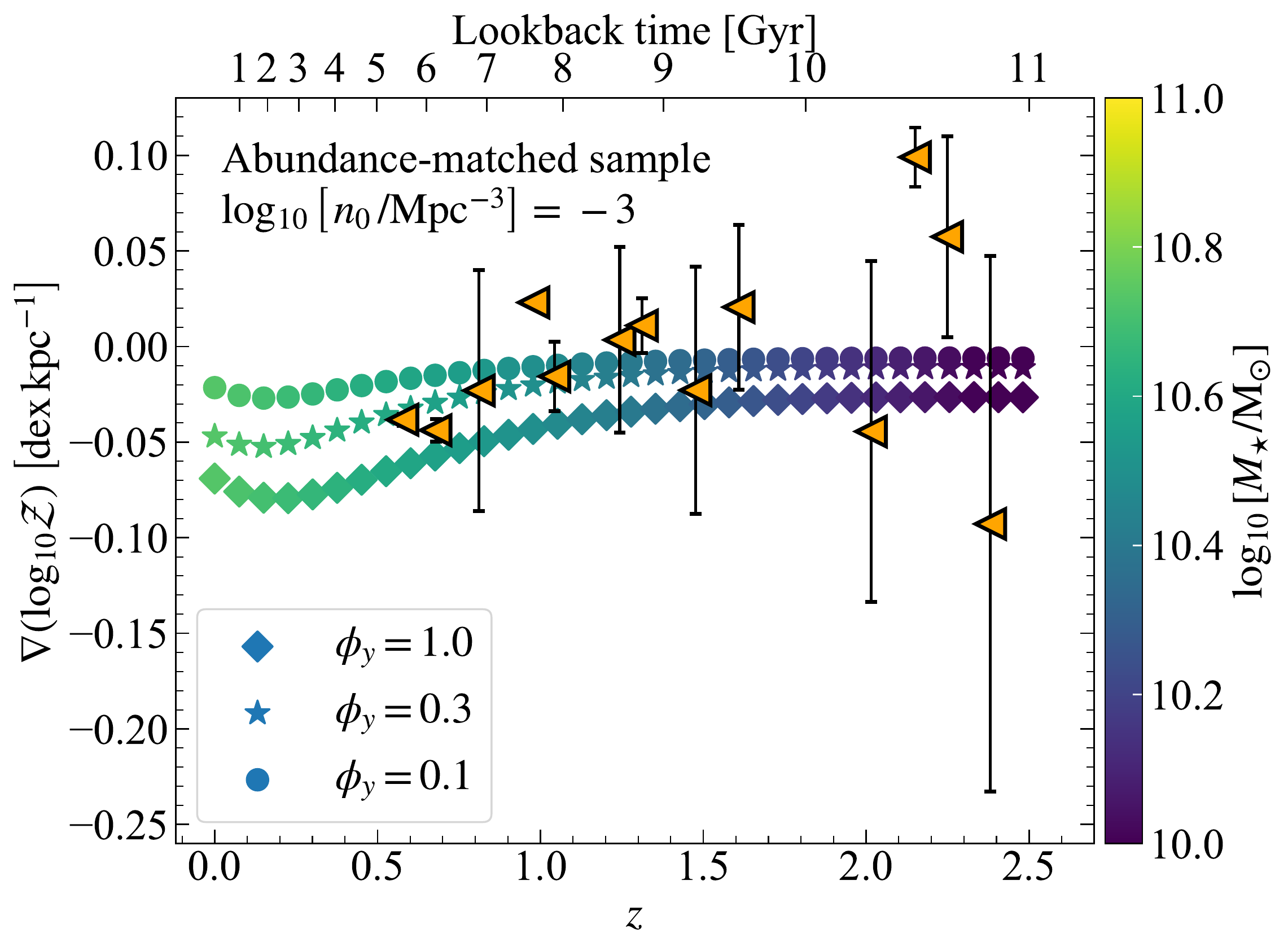}
\includegraphics[width=\columnwidth]{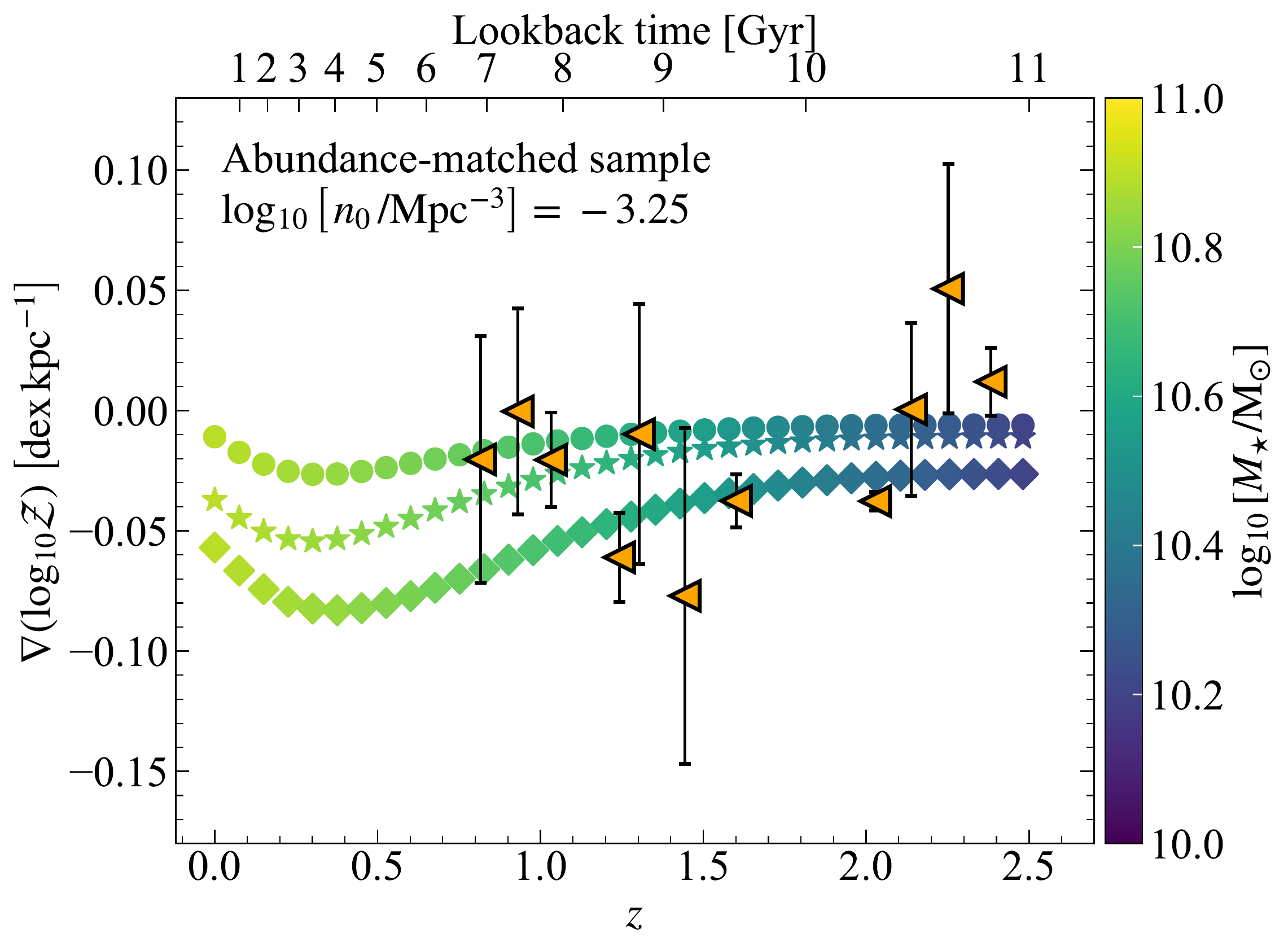}
\includegraphics[width=\columnwidth]{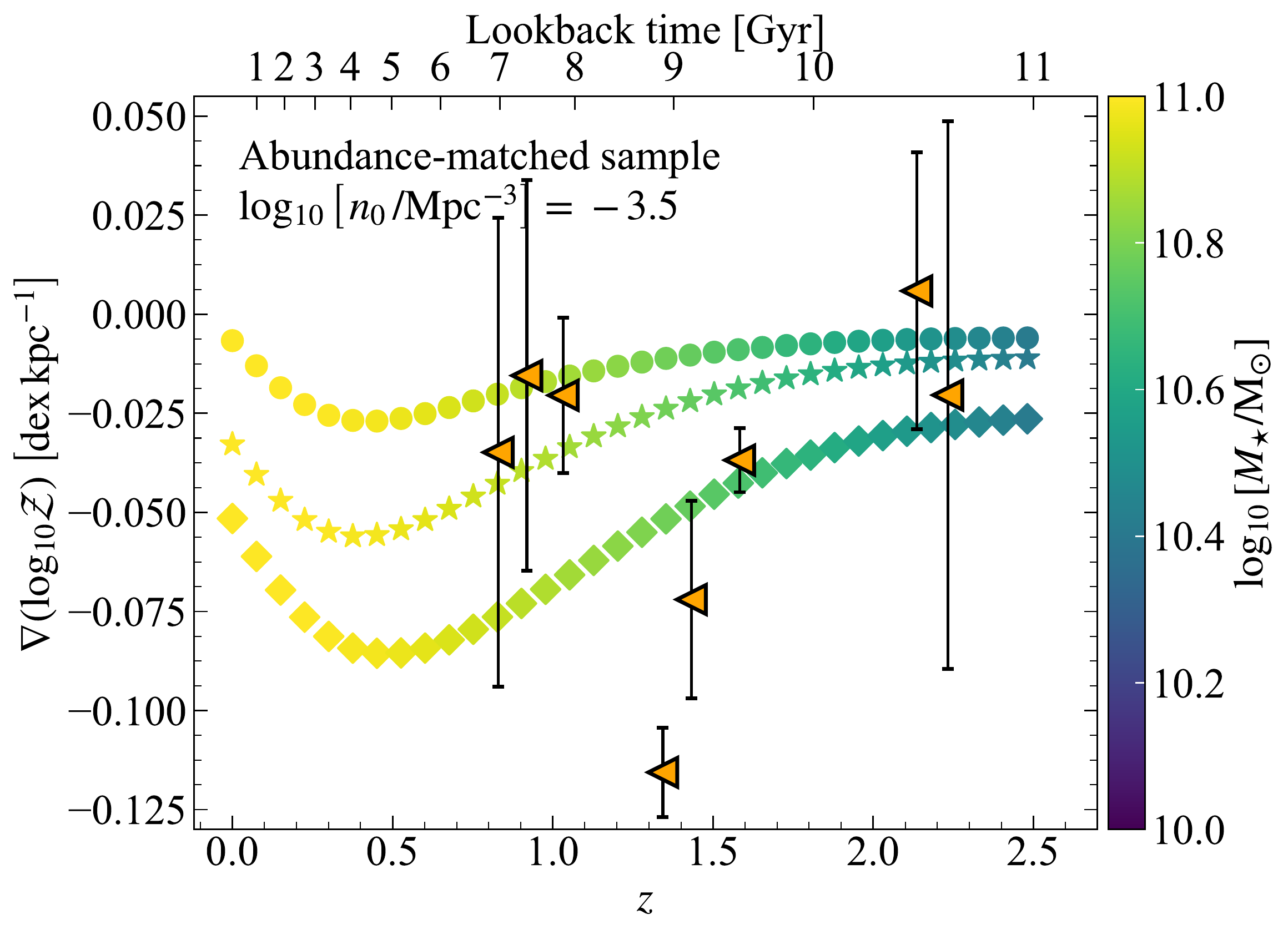}
\includegraphics[width=\columnwidth]{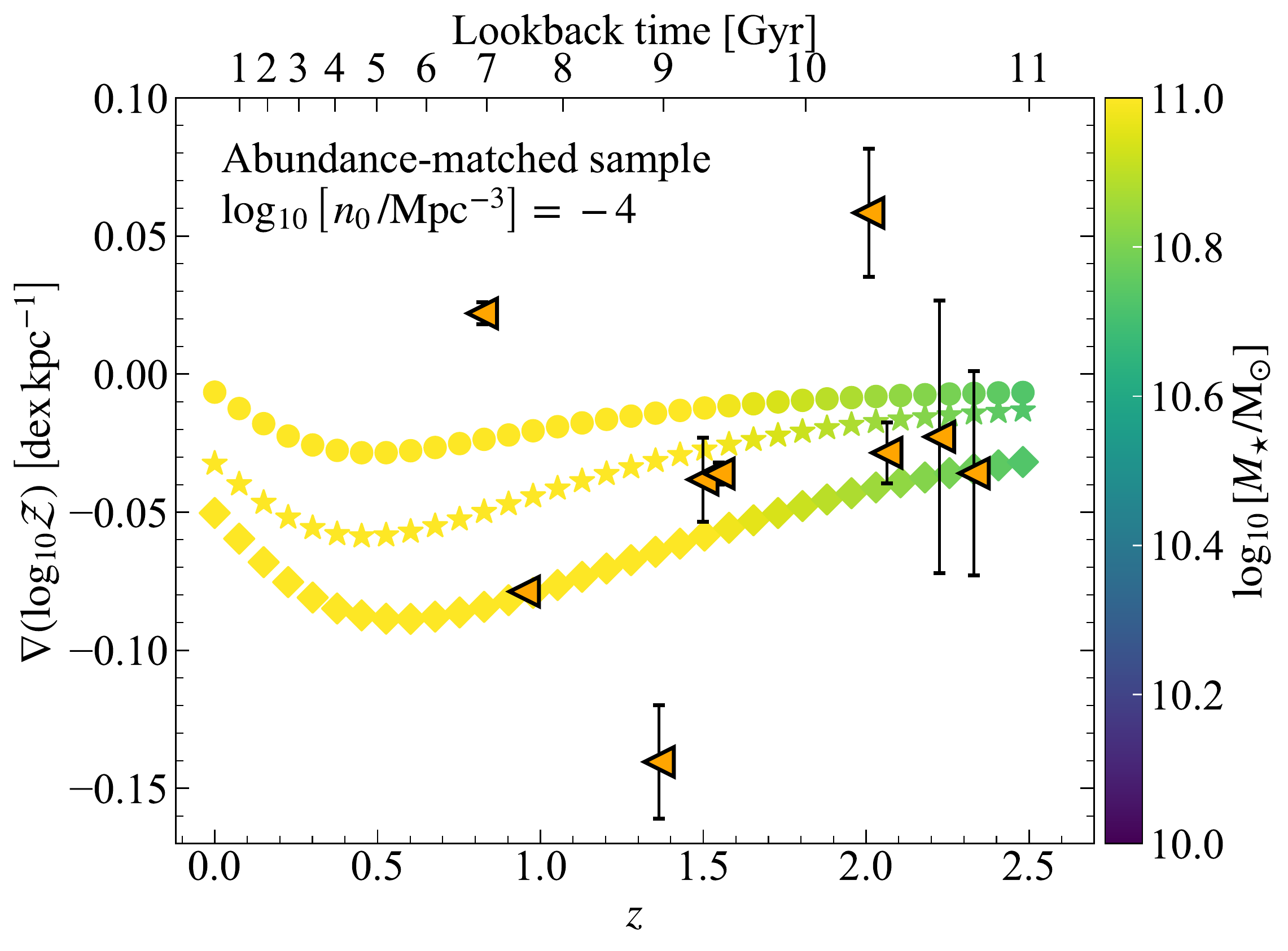}
\caption{Trends in metallicity gradients as a function of $z$ (and lookback time) for four different abundance-matched galaxy samples given a fixed comoving number density of galaxies, $n_0$, color-coded by $M_{\star}$. Abundance matching leads to the selection of more massive galaxies at lower redshifts, and can be used to collectively study gradients in local spirals and their high-$z$ progenitors. The orange data points reflect mean gradients for a constructed abundance-matched sample from available observations, which are the same as that reported in \autoref{fig:depends_redshift_samemassgalaxy}, with errorbars representing the scatter within the data. There is considerable scatter in the data, and the sample is not entirely robust given the \textit{ex post facto} construction. Nonetheless, the model matches the observations reasonably-well.}
\label{fig:cosmic_abundmatch}
\end{figure*}

Finally, we also study the evolution of metallicity gradients across an abundance-matched sample of dark matter haloes spanning a range in $z$\footnote{Abundance in the context of \autoref{s:cosmic_abundancematch} refers to the abundance of galaxies in a given comoving volume in the Universe, and not the metallicity.}. Abundance-matching is based on the premise that the number density of halo progenitors should nearly remain constant across $z$ within a comoving volume in the Universe \citep{1996MNRAS.282.1096M,1996ApJ...467L...9M,2010ApJ...709.1018V}. It has been used to study a range of properties in local galaxies together with their high-$z$ progenitors \citep[e.g.,][]{2009ApJ...701.1765M,2011MNRAS.412.1123P,2011ApJ...742...16T,2012ApJ...753...16K,2013ApJ...766...33L,2019MNRAS.487.5799R}, which is not possible with other selection criteria of galaxies (e.g., selecting galaxies with identical stellar mass, as we do in \autoref{s:cosmic_identicalmass}) as such galaxies evolve in time themselves \citep{2009ApJ...696..620C}. 

Abundance matching involves assigning more massive galaxies to more massive haloes at every $z$; this means selecting galaxies at each $z$ with $M_{\rm{h}}(z)$ that satisfy 
\begin{equation}
\int^{\infty}_{M_{\mathrm{h}}(z)} n\left(M_{\mathrm{h}},z\right) dM_{\mathrm{h}} = n_0
\label{eq:abundancematching}
\end{equation}
where $n_0$ is the target number density\footnote{This approximation of a fixed $n_0$ breaks down if certain galaxies in the abundance-matched sample do not follow the stellar mass rank order, for example, due to an abrupt increment in stellar mass because of mergers, or abrupt decrement due to quenching \citep{2013ApJ...766...33L}.}, and $n\left(M_{\rm{h}},z\right)$ is the number of galaxies per unit mass per unit comoving cubic Mpc given by \citet[equation~14]{2002MNRAS.336..112M} based on the \cite{1999MNRAS.308..119S} modification of the \cite{1974ApJ...187..425P} formalism for the number density of haloes across $z$. Thus, using the functional form for $n$, we can deduce the required $M_{\rm{h}}$ at each $z$ that would correspond to an abundance-matched sample for a given $n_0$. Following \cite{2009ApJ...701.1765M} and \cite{2011MNRAS.412.1123P}, we study four sets of $\log_{10}\,n_0/\rm{Mpc^{-3}} = -3,\,-3.325\,-3.5$ and $-4.0$, respectively. For each of these $n_0$, we find $v_{\phi}(z)$ and $M_{\star}(z)$ using $M_{\rm{h}}(z)$ from \autoref{eq:abundancematching}, and $\sigma_g(z)$ from \autoref{eq:fgas_wisnioski}. We fix $\beta=0$ for all galaxies since our choice of $n_0$ results in massive galaxies with $M_{\star} > 10^{10}\,\rm{M_{\odot}}$ for all $0 \leq z \leq 2.5$. For simplicity, we fix $f_{g,Q}=f_{g,P}=0.5$ and $\sigma_{\rm{sf}}=7\,\rm{km\,s^{-1}}$, the same as that for local spirals. Given that $f_{\rm{sf}}$ varies between 0.5 and 1 as $z$ increases, we use a cubic interpolation to vary it between $z=0$ and $4$. We also fix $\mathcal{Z}_{\rm{CGM}} = 0.1$. 

\autoref{fig:cosmic_abundmatch} shows the cosmic evolution of gradients for an abundance-matched sample of galaxies, each panel representing a different $n_0$. Similar to what we have seen in prior sections, the scatter in the model is the largest at the upturn where gradients start flattening. To the best of our knowledge, there are no existing abundance-matched samples of galaxies across redshift that also contain information on metallicity gradients. However, we can construct an abundance-matched sample from the available data. We caution that constructing an abundance-matched sample from existing observations \textit{ex post facto} is not as accurate as properly constructing the sample to start with. In the absence of the latter, we use our constructed sample to compare against the model to learn about the kinds of metallicity gradients that existed in progenitors of local galaxies. For this purpose, we construct our pseudo-abundance matched sample as follows: for each target value of $n_0$, we first select a redshift, and use \autoref{eq:abundancematching} to estimate the halo mass $M_{\rm{h}}$ corresponding to the target $n_0$ at that redshift. We then estimate the stellar mass of that galaxy $M_{\star}$ using the stellar mass-halo mass relation of \cite{2013MNRAS.428.3121M}. To construct our sample set at that redshift, we then take the data collection described in \autoref{s:cosmic_identicalmass} and select galaxies that have stellar masses within $\pm\,0.05\,\rm{dex}$ of the $M_\star$ from above; this constitutes our pseudo-abundance matched sample for that redshift, from which we then measure the mean and dispersion of metallicity gradient at that redshift bin. We plot these values in \autoref{fig:cosmic_abundmatch}, along with model predictions of the metallicity gradient, which we compute from the halo mass and redshift as in previous sections. The data we obtain in this manner have considerable scatter (shown by the errorbars), but the general trends are reasonably well reproduced by the model. However, given the uncertainties in the procedure we are forced to use to construct the observed sample, it is wiser to regard the model points in \autoref{fig:cosmic_abundmatch} as a prediction for future abundance matching measurements, rather than a rigorous comparison to existing data.


\section{Limitations of the model}
\label{s:modellimitations}
In this section, we describe the limitations of the model, first focusing on physical processes that we have excluded, and then discussing galaxies to which we cannot always apply our assumption of equilibrium.

\subsection{Additional physics}
\label{s:caveats}
Our model omits three possibly-important physical effects: bars, galactic fountains and long term wind recycling. With regard to the first of these, there is some evidence that gas phase metallicity gradients in the presence of bars in local spirals can be systematically shallower than those non-barred galaxies (\citealt{1992MNRAS.259..121V,1994ApJ...420...87Z,2020MNRAS.tmp.2303Z}; see, however, \citealt{2016A&A...587A..70S,2018A&A...609A.119S}). We have not included metal redistribution due to bar-driven flows, and for this reason we limit our study to gradients in the parts of a galaxy where the rotation curve slope ($\beta$) is a constant, which excludes bar-dominated regions \citep{2016ARA&A..54..529B,2020MNRAS.496.1845M}. In fact, even if we wished to include bar-driven mixing, the galaxy formation model that we use as an input in \autoref{s:metalevolve_krumholz2018} is itself not applicable in regions where the bar dominates the dynamics of the galaxy, since it does not include the effects of bar-driven torques on gas and SFR surface density profiles \citep{2018ApJ...860..172S,2020ApJ...901L...8S}. 

With regard to the second issue: we do not explicitly incorporate metal redistribution via galactic fountains \citep{1980ApJ...236..577B}. However, the combination of an enriched $\mathcal{Z}_{\rm{CGM}}$ and low $\phi_y$ essentially constructs a fountain process in the model that we can exploit. Semi-analytic models where the evolution of the CGM is self-consistently followed find that the CGM plays a larger role in the evolution of galaxy metallicity as it gets enriched due to outflows \citep{2020arXiv201104670Y}. We also note that galactic fountains, owing to their short fall-back timescale ($\sim\,100-300\,\rm{Myr}$, \citealt{2017MNRAS.470.4698A}) and short fall-back distance from the starting point ($\sim\,1\,\rm{kpc}$, \citealt{2008A&A...484..743S}) have been shown to play an insignificant role in the metallicity evolution of the local spiral M31 (\citealt{2013A&A...551A.123S}; see, however, simulations by \citealt{2019MNRAS.490.4786G}, where fountains are thought to transport metals to the edge of the star-forming disc). Fountains possibly have a significant effect on the far outskirts of the discs, where there are few or no local sources of metals. 

There is some evidence from simulations that long term wind recycling can provide metals to the disc as it re-accretes the ejected material. These simulations also show that this recycling is independent of the halo mass \citep{2016ApJ...824...57C,2019MNRAS.485.2511T}, and can be the dominant mode of accretion of cold gas at late times. However, this recycling occurs much farther out in the disc than that we consider in our work, thus, its basic features are captured within $\mathcal{Z}_{\rm{CGM}}$ in the model. Additionally, while the above simulations find long term wind recycling timescale to be of the order of a Gyr, results from the EAGLE simulations find it to be comparable to $t_{\rm{H(z)}}$ \citep{2020MNRAS.497.4495M}. Thus, given these findings from simulations and the lack of direct observations, it is currently difficult to determine the importance of wind recycling for metallicity gradients. 

Finally, we caution that our model is intended to apply mainly to metals whose production is dominated by type II SNe, and thus where the injection rate closely follows the star formation rate. We have not attempted to model elements produced by type Ia SNe or AGB stars. This is not a substantial problem for our intended application, however, since type II SNe do dominate production of the $\alpha$ elements that are most easily observable in the gas phase \citep{2018SSRv..214...67N}. The one exception to this statement, where some caution is warranted, is nitrogen, to which AGB stars make a substantial contribution \citep{2002A&A...381L..25M,2005ARA&A..43..435H}. This matters because many of the strong-line diagnostics used at high redshift rely on the \ion{N}{ii} $\lambda6584$ line. While observers who rely on these diagnostics usually attempt to derive the O abundance by calibrating out variations in the N/O ratio (\citealt{2004MNRAS.348L..59P,2016ApJ...828...18M,2017MNRAS.469..151B,2020ApJ...890L...3S}; see also, \citealt{2016MNRAS.458.3466V}), it is nevertheless the case that variations in N abundance may influence the metallicities derived at high-$z$, and that our model does not capture this effect.

\begin{figure}
\includegraphics[width=\columnwidth]{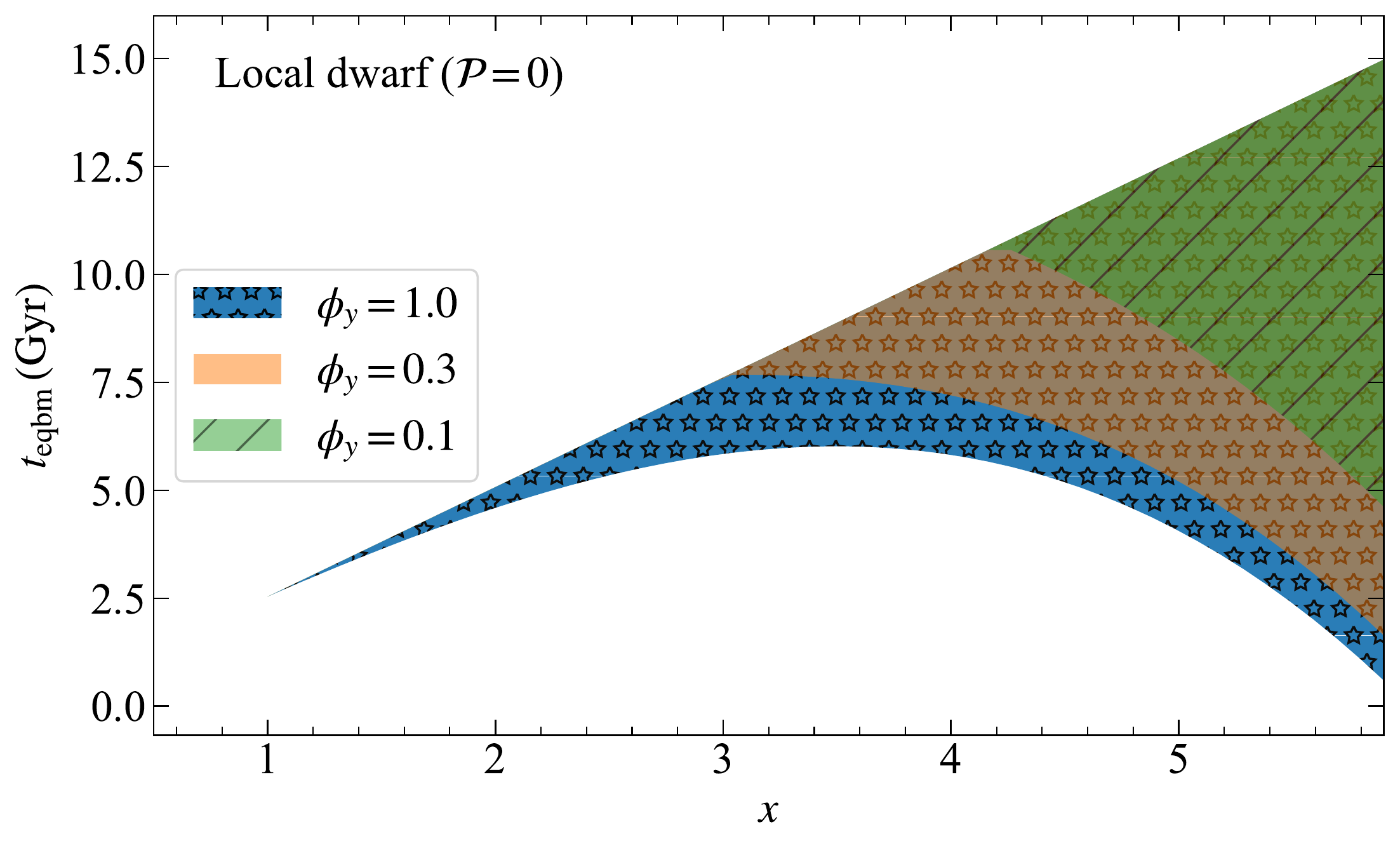}
\caption{Same as \autoref{fig:teqbm_ld}, but without radial inflow such that $\mathcal{P}=0$. Here, $t_{\rm{eqbm}} \gtrsim t_{\rm{H(0)}}$, implying that the metallicity gradients in such cases in local dwarfs may or may not be in equilibrium. Thus, our equilibrium model does not necessarily apply.}
\label{fig:teqbm_ld_noadv}
\end{figure}

\begin{figure}
\includegraphics[width=\columnwidth]{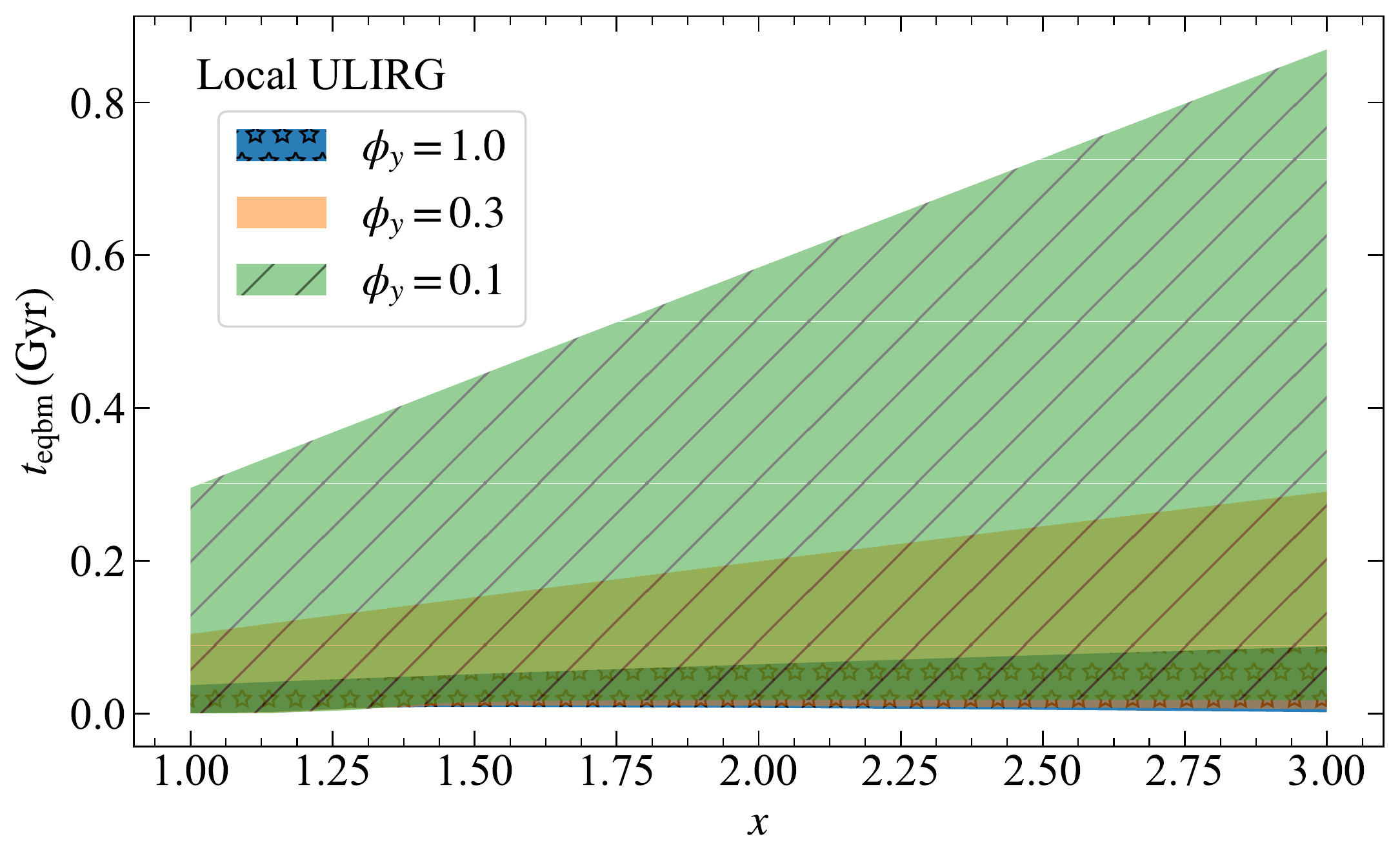}
\caption{Same as \autoref{fig:teqbm_ls}, but for local ultraluminous infrared galaxies (ULIRGs). Here, $t_{\rm{eqbm}} \sim t_{\rm{merge}}$, where the latter is the merger timescale of the order of $\sim\,0.3-1\,\rm{Gyr}$ as seen in models \protect\citep{2008ApJ...675.1095J,2012ApJ...746..108T}. Thus, the metallicity gradients may not be in equilibrium throughout the merger process. In such a case, our equilibrium model for metallicity gradients cannot be applied to local ULIRGs, and the observed gradients, if any, are transient and subject to change as the merger progresses, in line with observations \protect\citep{2010ApJ...723.1255R,2012ApJ...753....5R}.}
\label{fig:teqbm_ul}
\end{figure}

\begin{figure}
\includegraphics[width=\columnwidth]{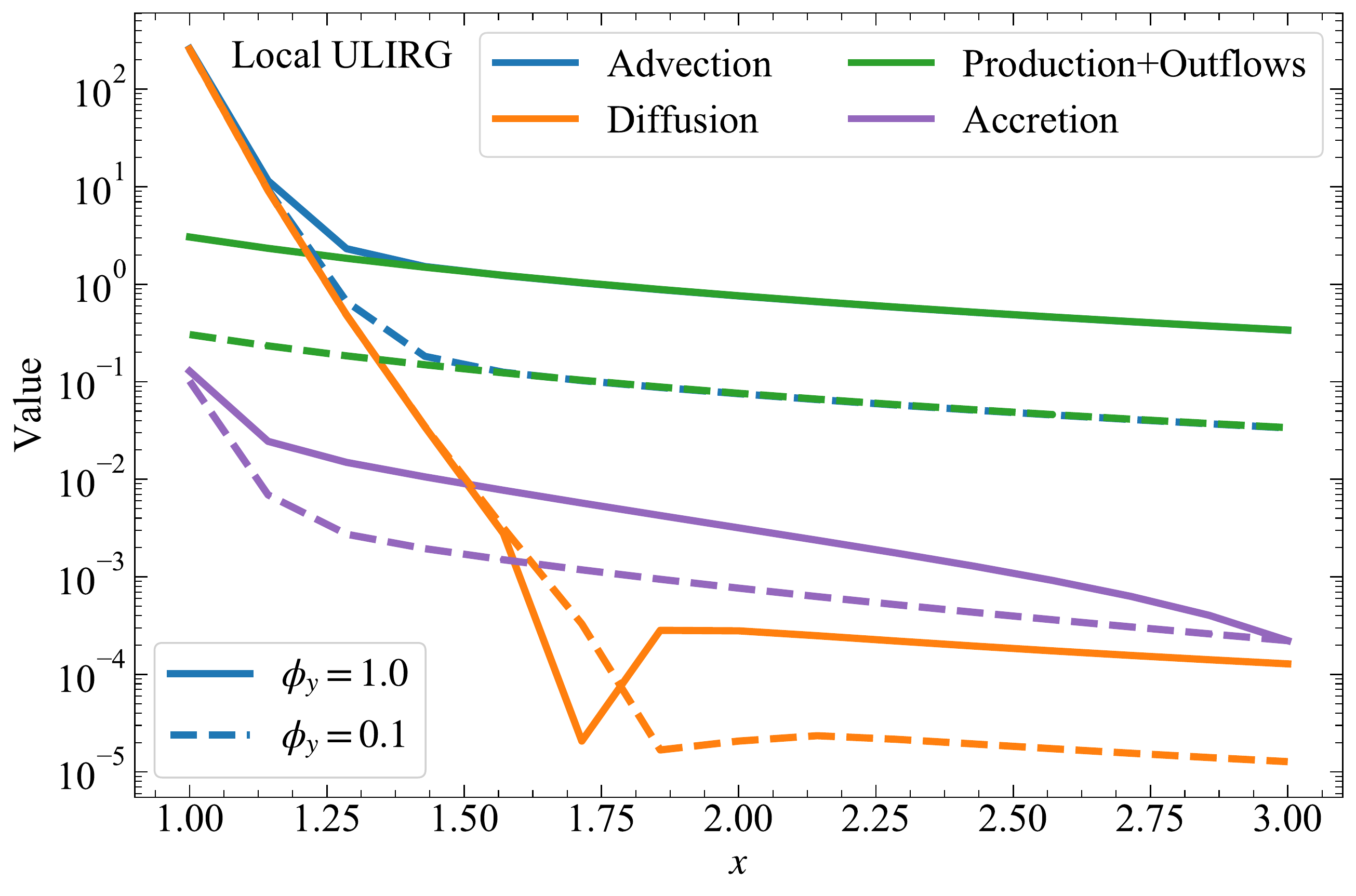}
\caption{Same as \autoref{fig:localspirals_terms}, but for ULIRGs, which are known to be major mergers. The non-equilibrium metallicity distribution is set by advection of gas due to tidal inflows during a merger.}
\label{fig:localULIRG_terms}
\end{figure}

\begin{figure}
\includegraphics[width=\columnwidth]{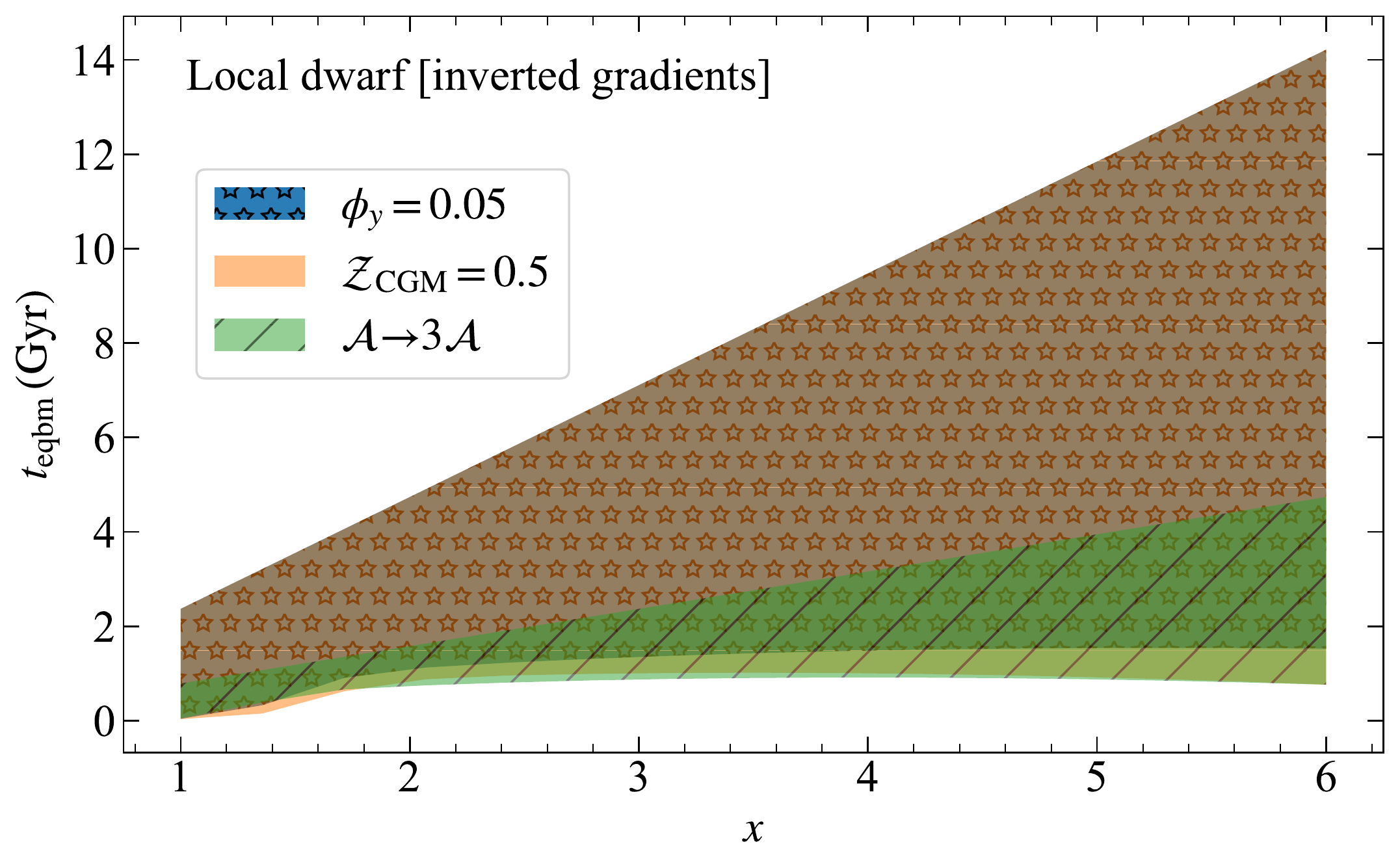}
\includegraphics[width=\columnwidth]{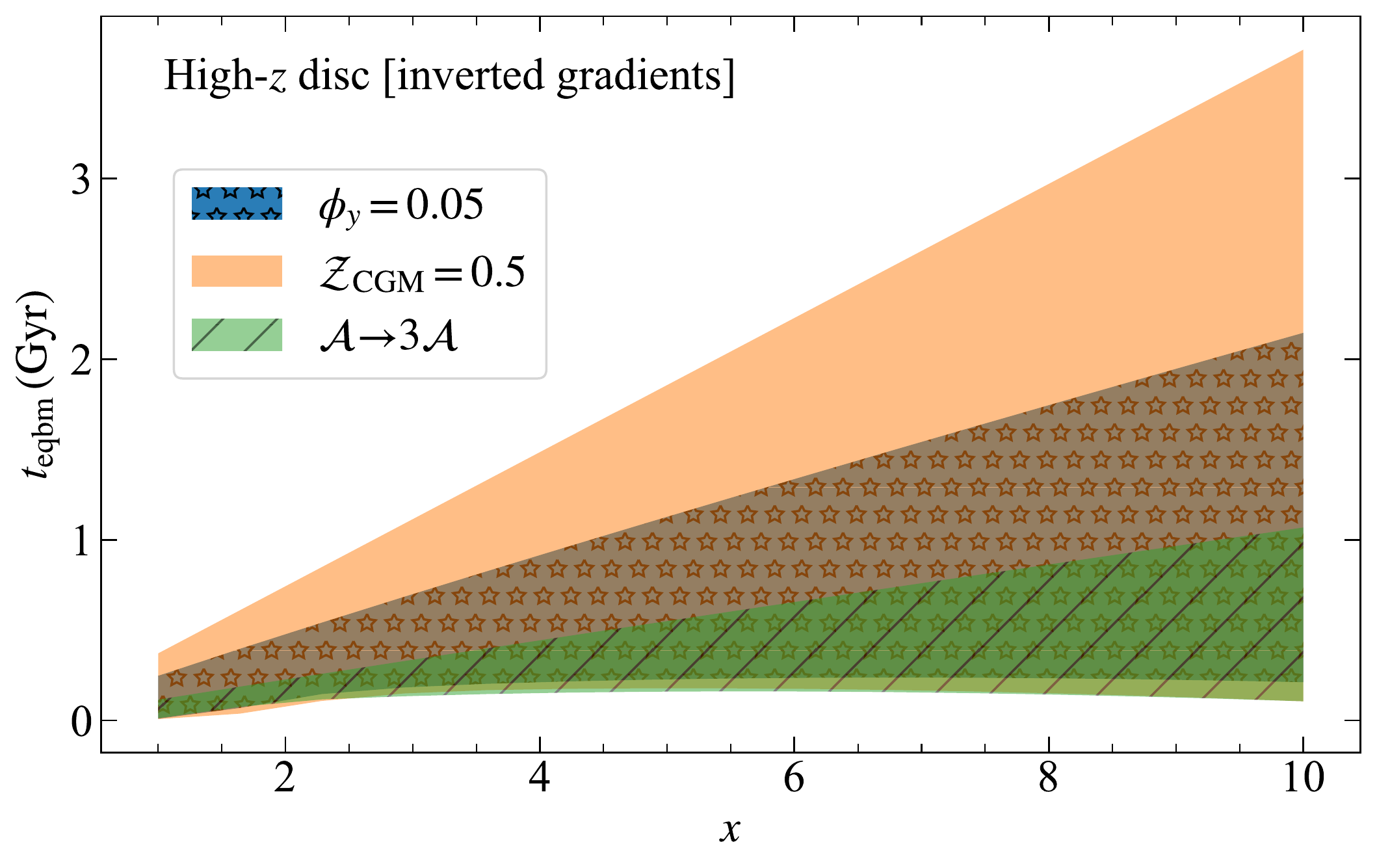}
\includegraphics[width=\columnwidth]{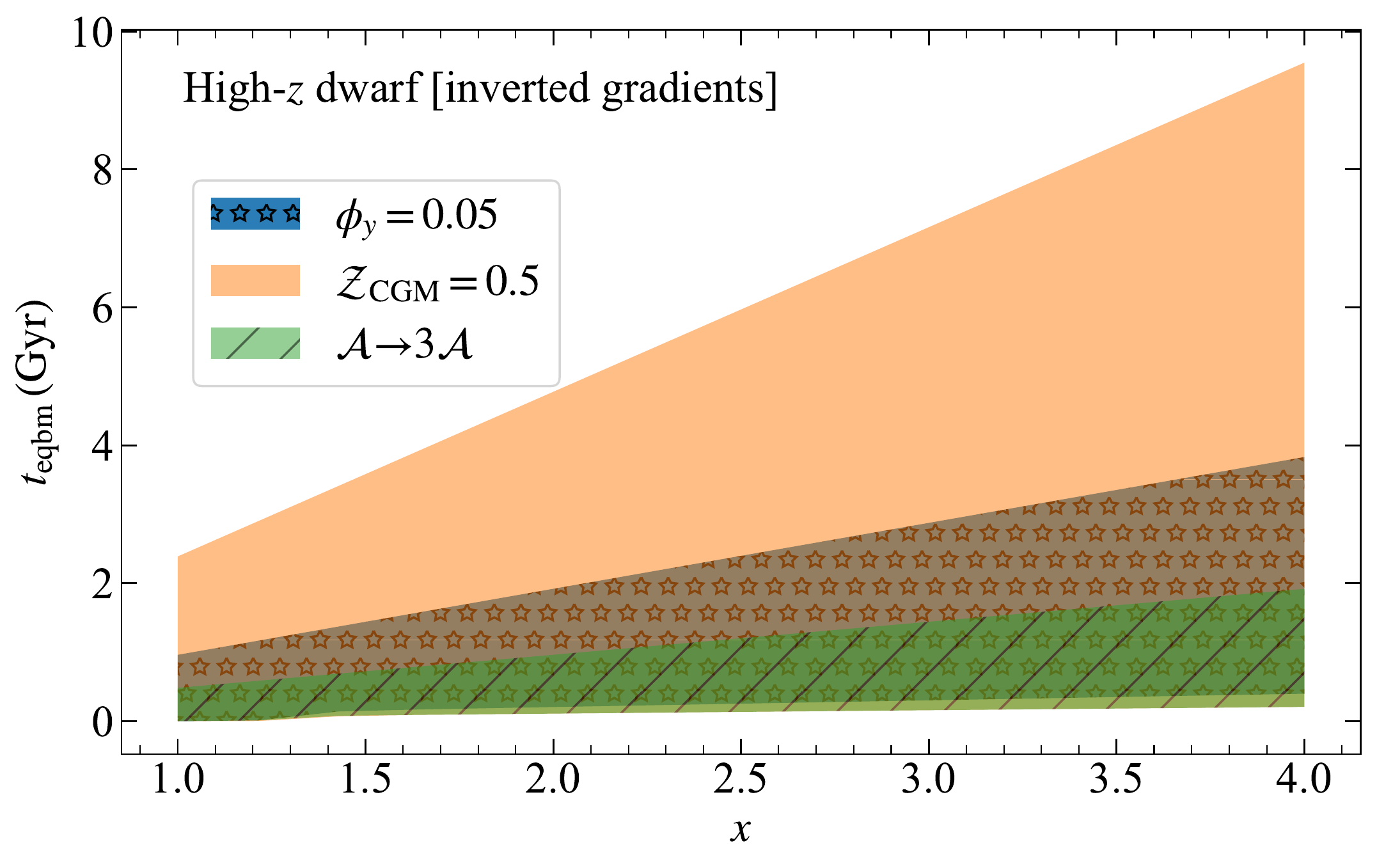}
\caption{Metallicity equilibration timescale $t_{\rm{eqbm}}$ as a function of $x$ in galaxies with inverted gradients. The first panel represents $t_{\rm{eqbm}}$ in local dwarfs. The second panel on high-$z$ discs is identical to the class of high-$z$ galaxies we discuss in \autoref{s:gradients_highz2}. The third panel plots $t_{\rm{eqbm}}$ in the case of high-$z$ dwarfs that we create by combining the fiducial parameters for local dwarfs and high-$z$ galaxies (see \autoref{s:noneqbm_inverted_gradients} for details). The colors correspond to the different ways that can give rise to an inverted gradient in a galaxy: reduction in metal yield due to high preferential metal ejection ($\phi_y=0.05$), enrichment of the CGM due to fountains or metal-rich flows ($\mathcal{Z}_{\rm{CGM}}=0.5$), and excessive cosmic accretion ($\mathcal{A} \to 3\mathcal{A}$). The scatter in the model is due to $c_1$. This plot shows that inverted metallicity gradients may or may not be in equilibrium.}
\label{fig:invertedgradients}
\end{figure}

\subsection{Non-equilibrium metallicity gradients}
\label{s:noneqbm}
There are certain classes of galaxies where we find that the metallicity distribution can be out of equilibrium, \textit{i.e.,} $t_{\rm{eqbm}} \gtrsim t_{\rm{H(z)}}$ or $t_{\rm{eqbm}} \gg t_{\rm{dep,H_2}}$. Hence, the model cannot always be used to predict metallicity gradients in such galaxies. Nonetheless, the limitation of the equilibrium model provides interesting constraints on the evolution of such galaxies. We discuss three such cases below.

\subsubsection{Local dwarfs without radial inflow}
\label{s:noneqbm_ld_noadvec}
The balance between metal production (source) and radial transport of metals through the disc (advection, diffusion) sets the metallicity gradients in local dwarfs (cf. \autoref{s:gradients_localdwarfs}). It has also been suggested that turbulence in these galaxies is mainly driven by star formation feedback and not gravity \citep{2015MNRAS.449.3568M,2018MNRAS.477.2716K}, which gives rise to $\sigma_{\rm{sf}} \sim \sigma_g$, and the low gas velocity dispersions observed in dwarfs \citep{2019MNRAS.486.4463Y,2020MNRAS.495.2265V}. Here, we investigate the case where $\sigma_{\rm{sf}} = \sigma_g$ such that there is no radial inflow of gas through the disc (see \autoref{eq:radialinflow})\footnote{$\sigma_{\rm{sf}} > \sigma_g$ is not possible in equilibrium in the \cite{2018MNRAS.477.2716K} model.}.

\autoref{fig:teqbm_ld_noadv} shows the radial profile of $t_{\rm{eqbm}}$ in this case. It is clear that $t_{\rm{eqbm}} \gtrsim t_{\rm{H(0)}}$ and $t_{\rm{eqbm}} \gtrsim t_{\rm{dep,H_2}}$, especially at low $\phi_y$, however, the exact values are sensitive to the choice of $c_1$. The reason for long metal equilibration timescales in this case is that, in the absence of advection, only diffusion and accretion are available to balance the source term. However, diffusion is weak due to the low gas dispersion ($\kappa_0\Sigma_{g0} \propto \sigma^3_g$), and accretion is weak due to the low halo mass ($\dot\Sigma_{\rm{cos}0} \propto M^{1.1}_{\rm{h}}$). Thus, metallicity gradients may not attain equilibrium in the absence of radial gas inflows in local dwarfs, whereas even a small amount of advection is sufficient to restore metallicity equilibrium (cf.~\autoref{fig:teqbm_ld}). In the case where there is no accretion, one can expect a diverse range of metallicity gradients that are not constrained by the model. Therefore, caution must be exercised while studying metallicity gradients in such dwarfs with an equilibrium model.

At face value, this result might seem consistent with that of \cite{2014MNRAS.443..168F}, where the authors find that dwarf galaxies do not attain statistical equilibrium within a Hubble time (see their Figure~15; see also, \citealt{2015MNRAS.449.3274F,2020A&A...638A.123D}). However, the equilibrium scenarios considered by \citeauthor{2014MNRAS.443..168F} and us are not necessarily the same, and one is not a precondition of the other. \citeauthor{2014MNRAS.443..168F} discuss the equilibrium for the total amount of gas or metals in a galaxy, which is a balance between inflow and outflow. The time required to reach this equilibrium is not necessarily the same as the time to equilibrate the distribution of metals \textit{within} the galactic disc, for a given total metal content. Thus, one equilibrium time can be longer or shorter than the other.

Similarly, comparing $t_{\rm{eqbm}}$ with the metal correlation timescale for local dwarfs from \citet{2018MNRAS.475.2236K}, which is the time required for diffusion alone to smooth out the metallicity distribution in the azimuthal direction, reveals that azimuthal metal distribution in these galaxies reaches equilibrium substantially more quickly than the radial distribution that we study here. This is consistent with the findings of \citet{Petit15a}, who also find that metal distributions equilibrate much more quickly in the azimuthal than the radial direction.


\subsubsection{Local ULIRGs}
\label{s:noneqbm_ulirgs}
Local ULIRGs are very dynamically active, and are well-known to be undergoing major mergers or have companions \citep{1989MNRAS.240..329L,1990A&A...231L..19M,1994MNRAS.267..253L,1996MNRAS.279..477C,1999ApJ...522..113V}. These galaxies are often characterized by strong starburst and/or AGN-driven outflows \citep{1995ApJS...98..171V,2013ApJ...776...27V,2012ApJ...757...86S,2014A&A...568A..14A}. They also have extremely short orbital timescales (of the order of $\sim\,5\,\mathrm{Myr}$, \citealt{2018MNRAS.477.2716K}). Local ULIRGs are very compact, with discs extending out only to $2$--$3\,\rm{kpc}$ \citep{1998ApJ...507..615D,2011ApJ...726...93R}. It is quite challenging to extract gas metallicities in these galaxies because the ionised gas emission lines are often dominated by shocks \citep{2006ApJ...637..138M,2010A&A...517A..28M} and AGN activity \citep{2013MNRAS.435.3627E}, which interfere with traditional photoionisation-based metallicity diagnostics. In addition, high levels of dust obscuration make it difficult to model the emission line spectra \citep{2009A&A...505.1017G,2011A&A...526A.149N,2013A&A...553A..85P,2014ApJ...790..124S}. For these reasons, there are only a handful of studies that have been able to extract gas metallicities in local ULIRGs (e.g., \citealt{2006AJ....131.2004K,2007A&A...472..421M,2008A&A...479..687A,2008ApJ...674..172R,2012MNRAS.424..416W,2014ApJ...797...54K,2017MNRAS.470.1218P}), and to the best our knowledge the only published studies of the metallicity gradient in ULIRGs are those of 
\citet[see their Figure~2]{2012ApJ...753....5R} and \citet{2019MNRAS.482L..55T}.

The short orbital timescales of ULIRGs ensure that they return to dynamical equilibrium quickly compared to their merger timescales, which based on simulations are estimated to be $t_{\rm{merge}} \sim 0.3-1\,\rm{Gyr}$ \citep{2008ApJ...675.1095J,2012ApJ...746..108T}. Thus our dynamical equilibrium model from \cite{2018MNRAS.477.2716K} is applicable to them. We investigate whether the metallicity distribution is also in equilibrium in \autoref{fig:teqbm_ul}, which shows $t_{\rm{eqbm}}$ for local ULIRGs. It is clear that $t_{\rm{eqbm}} \sim t_{\rm{merge}}$, thus, metallicity may or may not be in equilibrium during the entire process of a merger. Our results corroborate those of \cite{2012MNRAS.421...98D}, who argue that merging galaxies should not be in equilibrium because tidal flows will fuel star formation \citep{2000ApJ...530..660B,2006AJ....131.2004K,2009ApJ...691.1005R,2011MNRAS.417..580P,2013MNRAS.435.3627E,2020MNRAS.tmp.2894M}, making cosmic accretion irrelevant. We show this quantitatively in \autoref{fig:localULIRG_terms}, where advection (radial transport of gas due to tidal inflows) is the dominant term that sets the non-equilibrium metallicity distribution, and cosmic accretion is insignificant in comparison. Our results are also in line with those from simulations and observations where metallicity gradients in local ULIRGs are observed to continuously evolve and flatten as the merger progresses \citep[Figure~4]{2012ApJ...753....5R}, implying that the metallicity distribution is not in a steady-state. This also implies that non-equilibrium gradients in local ULIRGs are transient; assuming the galaxy settles back to being a quiescent disc after the merger, the metallicity gradient will return to the equilibrium value for a spiral galaxy on the $\sim$ few Gyr equilibrium timescale for local spirals (cf.~\autoref{fig:teqbm_ls}).

Given a merger rate, we can estimate the fraction of galaxies as a function of redshift that are expected to be out of metal equilibrium as $1-e^{-\theta}$, where $\theta$ is the product of the merger rate and the metallicity equilibration timescale. Following \citet[Figure 9]{2015MNRAS.449...49R}, we see that the observed average merger rate for massive galaxies ($M_{\star} \geq 10^{10}\,\rm{M_{\odot}}$) at $z=0$ is less than $0.06\,\rm{Gyr^{-1}}$ \citep{2011ApJ...742..103L}, so we expect less than 20 per cent of massive galaxies to be out of metal equilibrium at redshift zero. Similarly, based on available observational results that find a merger rate of $0.5\,\rm{Gyr^{-1}}$ at $z \approx 2$ \citep{2009MNRAS.394L..51B,2012ApJ...747...34B,2012ApJ...744...85M}, we expect less than 40 per cent of the most massive galaxies ($M_{\star} \geq 10^{11}\,\rm{M_{\odot}}$) to be out of metal equilibrium at redshift two. The larger fraction of galaxies that are expected to be out of metal equilibrium at high redshift could explain the inverted gradients seen in high-$z$ observations, a topic we explore in \autoref{s:noneqbm_inverted_gradients}.

\subsubsection{Galaxies with inverted gradients}
\label{s:noneqbm_inverted_gradients}
Recent observations have discovered the presence of inverted (positive) gas phase metallicity gradients in galaxies \citep{2014A&A...563A..49S,2017MNRAS.469..151B,2020A&A...636A..42M}, especially at high redshift \citep{2010Natur.467..811C,2012A&A...539A..93Q,2014MNRAS.443.2695S,2018MNRAS.478.4293C,2019ApJ...882...94W,2020MNRAS.492..821C,2020ApJ...900..183W,2020arXiv201103553S}. Inverted gradients reflect the possibility of galaxies deviating from the classical, inside-out formation picture, at least temporarily. The three leading mechanisms that are believed to give rise to an inverted gradient are: (1.) substantial metal mass loading or merger-induced tidal flows of metal-poor gas that deprives the galaxy centre of metals, especially in dwarfs (\citealt{2005MNRAS.363....2K,2006AJ....131.2004K,2018ApJ...869...94E,2019MNRAS.482.1304E,2018MNRAS.481.1690C,2019MNRAS.482.2208T}; see, however, \citealt{2019ApJ...874...18W}), (2.) re-accretion of ejected metals at the outer edge of the disc from the CGM through cold, metal-rich flows or galactic fountains \citep{2003MNRAS.345..349B,2006MNRAS.368....2D,2009Natur.457..451D,2010Natur.467..811C,2013ApJ...776L..18C,2015MNRAS.448..895S}, and (3.) cosmic accretion of metal-poor gas at the centre that dilutes the central metallicity \citep{2010Natur.467..811C}.

Corresponding to these three scenarios, we can produce inverted gradients in our model by coupling a moderate or high value of $\mathcal{Z}_{\rm GCM}$ (\textit{i.e.,} addition of metal-rich gas to galaxy outskirts) with small values of $\phi_y$ or large values of $\mathcal{A}$ (corresponding to depressed central metallicity due to heavy metal loss or rapid dilution by metal-poor gas, respectively). However, any inverted gradients that we get from the model are sensitive to our choice of $\mathcal{Z}_{\rm{CGM}}$, in the sense that we never get an inverted gradient for a sufficiently low value of $\mathcal{Z}_{\rm{CGM}}$. Nevertheless, regardless of the value of $\mathcal{Z}_{\rm{CGM}}$ that we adopt, the resulting inverted gradients may or may not be in equilibrium. We illustrate this in \autoref{fig:invertedgradients}, where we plot $t_{\mathrm{eqbm}}$ for local dwarfs, high-$z$ discs (identical to high-$z$ galaxies we discuss in \autoref{s:gradients_highz2}), and high-$z$ dwarfs. We introduce the latter category by combining fiducial parameters for local dwarfs and high-$z$ galaxies from \autoref{tab:tab2} in the following manner: $\beta=0.5,\,\sigma_{\mathrm{sf}}=7\,\mathrm{km\,s^{-1}},\,\sigma_g = 40\,\mathrm{km\,s^{-1}},\,v_{\phi}=80\,\mathrm{km\,s^{-1}},f_{g,Q}=f_{g,P}=0.9, f_{\mathrm{sf}}=0.4$, and $x_{\rm{max}}=4$ at $z=2$. The three colors in all the panels in \autoref{fig:invertedgradients} correspond to low $\phi_y = 0.05$ (with $\mathcal{Z}_{\rm{CGM}}=0.1$), high $\mathcal{Z}_{\rm{CGM}}=0.5$ (with $\phi_y=0.1$), and high accretion where we multiply our fiducial values of $\mathcal{A}$ by 3 (with $\phi_y=0.1,\,\mathcal{Z}_{\rm{CGM}}=0.1$), respectively. The shaded regions correspond to the allowed values of $c_1$ based on the constraints we introduced in \autoref{s:modelevolve_solution}.

We see that whether galaxies with inverted gradients are likely to be in equilibrium or not depends largely on what produces the inversion. Galaxies where the gradient inverts due to rapid accretion (high $\mathcal{A}$) have relatively short values of $t_{\rm eqbm}$, and may be in equilibrium as long as the accretion lasts, while those that invert due to an influx of metal-rich gas at their outskirts (high $\mathcal{Z}_{\rm GCM}$) are almost certainly out of equilibrium; galaxies with extremely efficient metal loss (low $\phi_y$) are intermediate, and may or may not be in equilibrium. Regardless of these details, the fact that many inverted gradients are not in equilibrium also hints at the possibility of them being transient \citep[see also,][]{2017MNRAS.467.1154S}. This is because subsequent star formation in the galaxy centre (due to cold gas accretion or re-accretion of enriched gas from the CGM) will replenish the metal supply on timescales comparable to the star formation timescale, thus leading to the formation of a negative gradient again. Hence, we expect inverted gradients to be erased within a star formation timescale ($\lesssim 2\,\rm{Gyr}$ for massive galaxies, \citealt{2008AJ....136.2782L}) unless they are re-established on a similar timescale. Since the processes that can cause inverted gradients (strong fountains, mergers, sudden accretion events, etc.) tend to wane with redshift, we expect that most massive galaxies will establish negative gradients by $z=0$, though some dwarfs, which have longer equilibration (and star formation) timescales, might retain their inverted gradients to $z=0$ or close to it.

\section{Conclusions}
\label{s:conclusions}
In this work, we present a new theoretical model to explain the occurrence and diversity of gas phase metallicity gradients in galaxies. Starting from the conservation of metal mass, we incorporate major physical processes that can impact the distribution of metals in galaxies, namely, metal production, consumption, loss, advection, accretion and diffusion. Our first-principles based model shows that the radial metallicity gradients observed in galaxies are a natural consequence of inside-out galaxy formation. The equilibrium metallicity evolution model we present is a standalone model, but it requires inputs from a galaxy evolution model to set the galaxy properties that control metallicity. This intricate link between gas and metallicity lets us directly predict the evolution of metallicity gradients without ad hoc assumptions about galaxy properties.

The evolution of metallicities in our model depends on four dimensionless ratios: $\mathcal{T}$, $\mathcal{P}$, $\mathcal{S}$, and $\mathcal{A}$. These describe the ratio of the orbital timescale to the diffusion timescale, advection to diffusion, production (and metal ejection) to diffusion, and cosmic accretion to diffusion, respectively. Based on the input galaxy evolution model \citep{2018MNRAS.477.2716K}, we show how these ratios depend on various properties of the gas (cf. \autoref{eq:physicalChi} $-$ \autoref{eq:physicalA}). The resulting second order differential equation of the radial distribution of metallicity has a simple analytic solution given by \autoref{eq:main_nondimx_solution} that we use to predict a possible range of metallicity gradients as a function of galaxy properties. We use this capability to predict the metallicity gradients of local spirals, local dwarfs, and high-redshift disc galaxies, and to predict the evolution of metallicity gradients in galaxies with redshift. Below, we list our main results:

\begin{enumerate}
    \item The time required for the metal distribution within a galaxy to reach equilibrium is smaller than the Hubble time and comparable to the molecular gas depletion time in local spirals, (most) local dwarfs, and rotation-dominated high-$z$ galaxies. Thus, for most galaxies over most of cosmic time, the gas phase metallicity gradient is in equilibrium. Exceptions to this general trend \textit{can} include merging galaxies, galaxies with inverted metallicity gradients, and some very low-mass local dwarf galaxies.
    \item Galaxies tend to approach a particular value of central metallicity, dictated by the balance between the two dominant processes that depend on the properties of the galaxy (see below). The central metallicities we predict agree well with observations.
    \item In local spirals, the two dominant processes shaping the metallicity gradient are metal production ($\mathcal{S}$), which tries to steepen the gradient, and accretion of metal-poor gas ($\mathcal{A}$), which tries to flatten it. On the other hand, metallicity gradients in local dwarfs and high-$z$ galaxies are set by the balance between $\mathcal{S}$ and advection of metal-poor gas from the outer to the inner parts of galaxies ($\mathcal{P}$).
    \item One crucial free parameter that emerges from our model is the ``yield reduction factor'' $\phi_y$, defined as the fraction of supernova-produced metals that mix with the ISM rather than being lost immediately in metal-enhanced galactic winds. While metallicity gradients in local spirals are not tremendously sensitive to $\phi_y$, it has a significant effect on the metallicity gradients in local dwarfs and high-$z$ galaxies. $\phi_y$ also impacts the absolute metallicities in all galaxies. Comparison of the model with observations reveals that massive galaxies prefer a high value of $\phi_y$, whereas low-mass galaxies prefer a lower value of $\phi_y$. Thus, the model predicts that low-mass galaxies undergo more preferential metal ejection, and should have more metal-enriched winds than massive galaxies. Future work should thus focus on constraining $\phi_y$ from observations.
\end{enumerate}

As a first application of our model, we study the evolution of metallicity gradients with redshift, both within a single galaxy and over samples of galaxies at different redshifts selected to have matching stellar masses or comoving densities. Our model shows that gradients in Milky Way-like galaxies have steepened over time, in qualitative agreement with recent observations; quantitative agreement between the model and the data requires a scaling of $\phi_y$ such that $\phi_y$ was low for the Galaxy in the past as compared to today, consistent with that seen in simulations. We also predict the existence of specific signatures for the evolution of metallicity gradient with redshift as a function of stellar mass that can be tested with future surveys. We show that both the Milky Way in particular and disc galaxies in general transition from the advection-dominated ($\mathcal{P} > \mathcal{A}$) to the accretion-dominated ($\mathcal{P} < \mathcal{A}$) regime from high to low redshifts. This transition mirrors the transition from gravity-driven to star formation-driven turbulence from high to low redshifts \citep{2018MNRAS.477.2716K}. In companion papers, we show that this transition (along with $\phi_y$) is also responsible for driving the shape of the mass-metallicity relation and the mass-metallicity gradient relation \citep{2020aMNRAS.xxx..xxxS} in the local Universe, and we also apply our model to explain the relationship between metallicity gradients and gas kinematics in high redshift galaxies \citep{2020bMNRAS.xxx..xxxS}.

\section*{Acknowledgements}
We dedicate this work to the medical services personnel who have been working tirelessly to combat COVID-19 across the world; this work would not have been possible without their sincere efforts in keeping the community safe. We thank the anonymous reviewer for a thorough referee report that helped improve the manuscript. We also thank Meridith Joyce, Stephanie Monty, J. Trevor Mendel, Lisa Kewley, Kenneth Freeman, and Raymond Simons for several useful discussions. We are grateful to Xiangcheng Ma for sharing results of the FIRE simulations, and to Mirko Curti for sharing their data compilation on metallicity gradients in high-$z$ galaxies. PS is supported by the Australian Government Research Training Program (RTP) Scholarship. MRK and CF acknowledge funding provided by the Australian Research Council (ARC) through Discovery Projects DP190101258 (MRK) and DP170100603 (CF) and Future Fellowships FT180100375 (MRK) and FT180100495 (CF). MRK is also the recipient of an Alexander von Humboldt award. PS, EW and AA acknowledge the support of the ARC Centre of Excellence for All Sky Astrophysics in 3 Dimensions (ASTRO 3D), through project number CE170100013. JCF is supported by the Flatiron Institute through the Simons Foundation. Analysis was performed using \texttt{numpy} \citep{oliphant2006guide,2020arXiv200610256H} and \texttt{scipy} \citep{2020NatMe..17..261V}; plots were created using \texttt{Matplotlib} \citep{Hunter:2007} and \texttt{astropy} \citep{2013A&A...558A..33A,2018AJ....156..123A}. This research has made extensive use of NASA's Astrophysics Data System Bibliographic Services. This research has also made extensive use of Wolfram|Alpha\footnote{\url{wolframalpha.com}} and \texttt{Mathematica} for numerical analyses, the \cite{2006PASP..118.1711W} cosmology calculator, and the image-to-data tool \texttt{WebPlotDigitizer}\footnote{\url{https://automeris.io/WebPlotDigitizer}}.

\section*{Data availability statement}
No data were generated for this work. 

\bibliographystyle{mnras}
\bibliography{references} 

\appendix

\section{Functional form of cosmic accretion rate surface density}
\label{s:app_cosmicaccr}
Here, we describe how the solutions change if we pick a different functional form for the radial profile of cosmological accretion, $\dot c_{\star}(x)$. Note that we must numerically solve for these functional forms, because analytic solutions either do not exist or are so complex in functional form that a numerical integration is preferable. Specifically, we experiment with $\dot c_{\star}(x) = 1/x$ and $\dot c_{\star}(x) = 1$.

\begin{figure}
\includegraphics[width=\columnwidth]{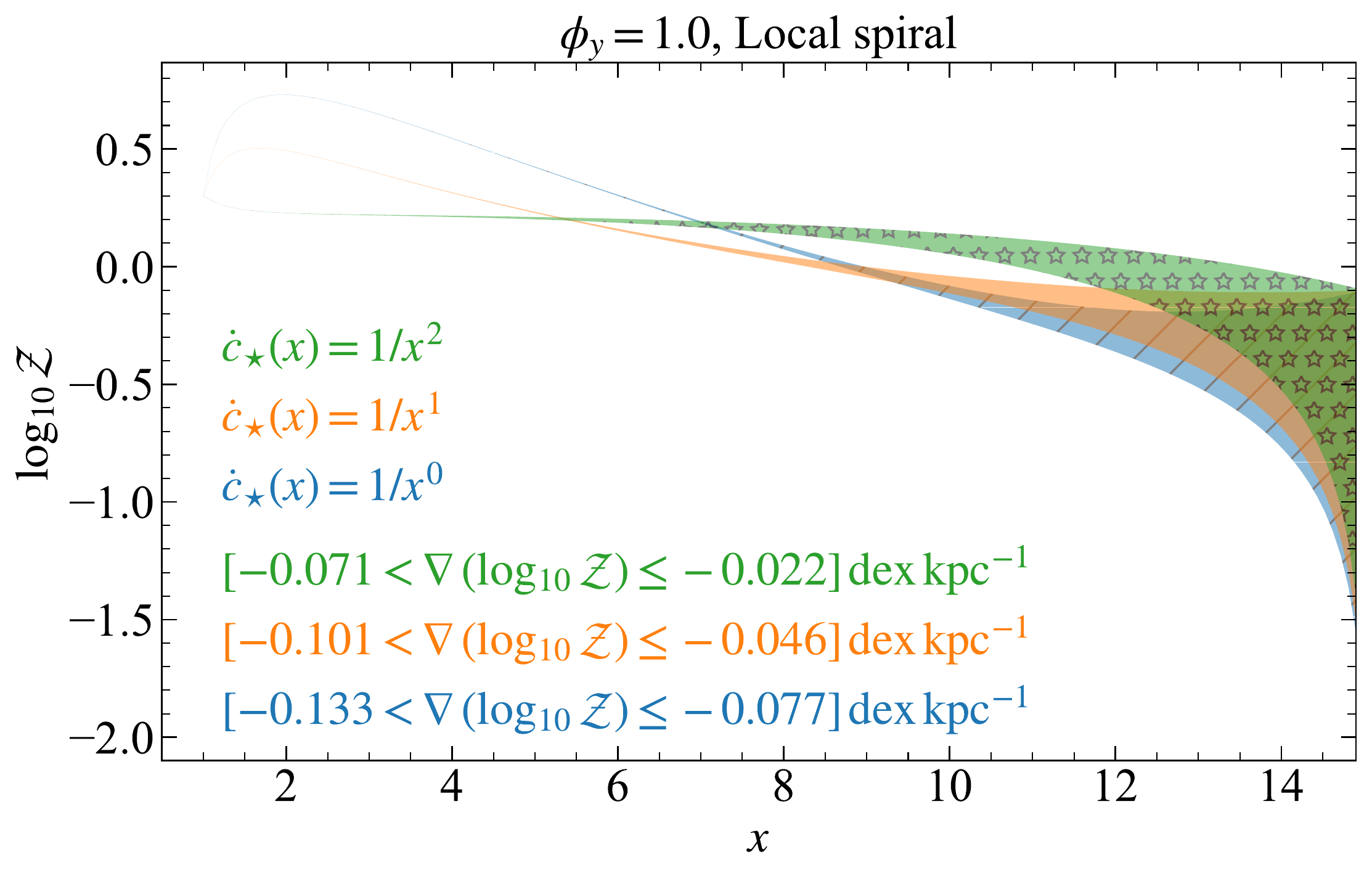}
\caption{Same as \autoref{fig:localspirals} but with different functional forms for cosmic accretion, namely, $\dot c_{\star} = 1/x^2$ (the one we use in the main text), $1/x$ and $1$, respectively. Flatter cosmic accretion profiles make the gradients steeper (within a factor of 2).}
\label{fig:app_cosmicaccr_localspirals}
\end{figure}

\begin{figure}
\includegraphics[width=\columnwidth]{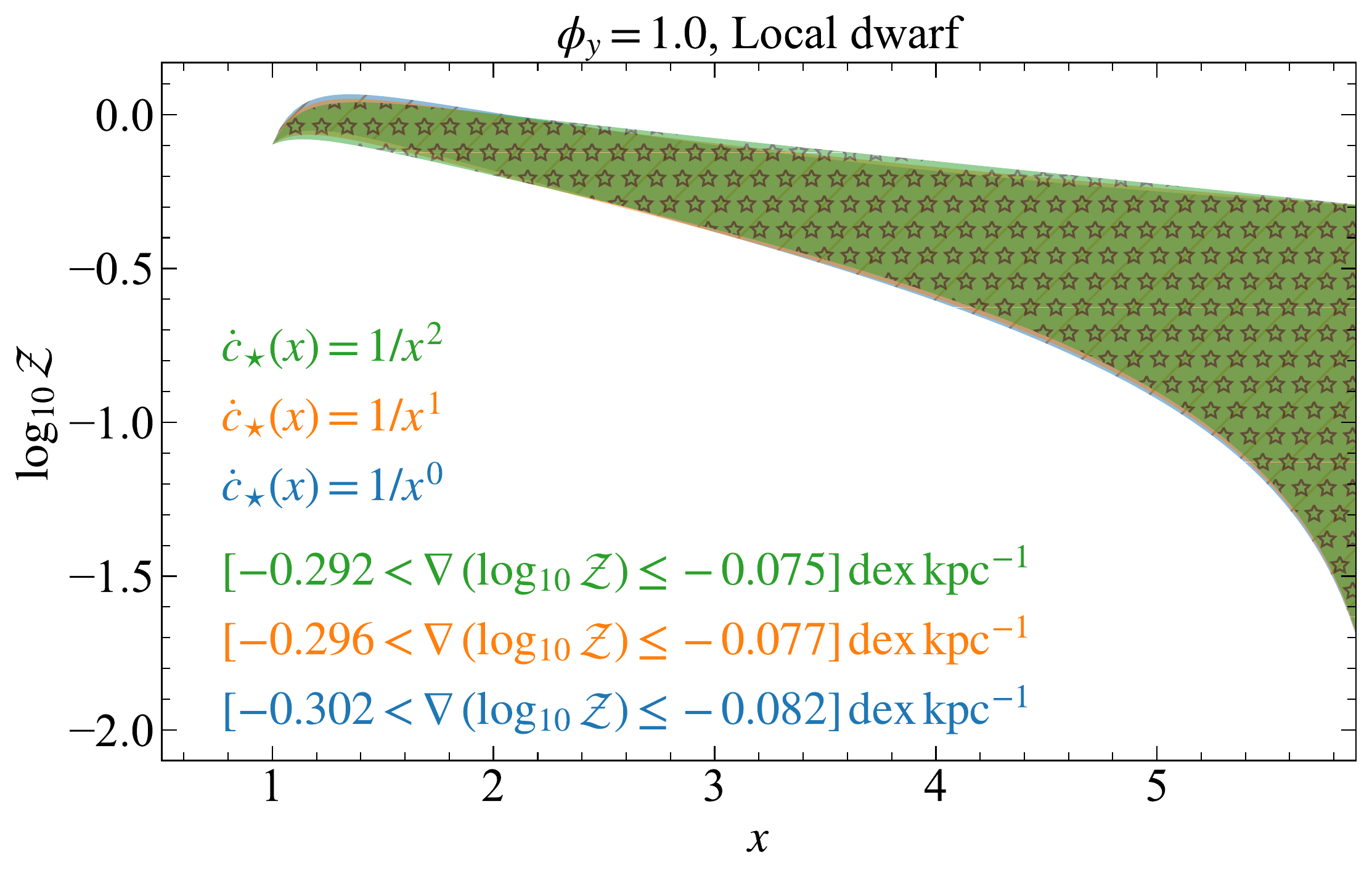}
\caption{Same as \autoref{fig:app_cosmicaccr_localspirals} but for local dwarfs. Changing the functional form of $\dot c_{\star}(x)$ has no impact on the metallicity distributions in local dwarfs.}
\label{fig:app_cosmicaccr_localdwarfs}
\end{figure}

\begin{figure}
\includegraphics[width=\columnwidth]{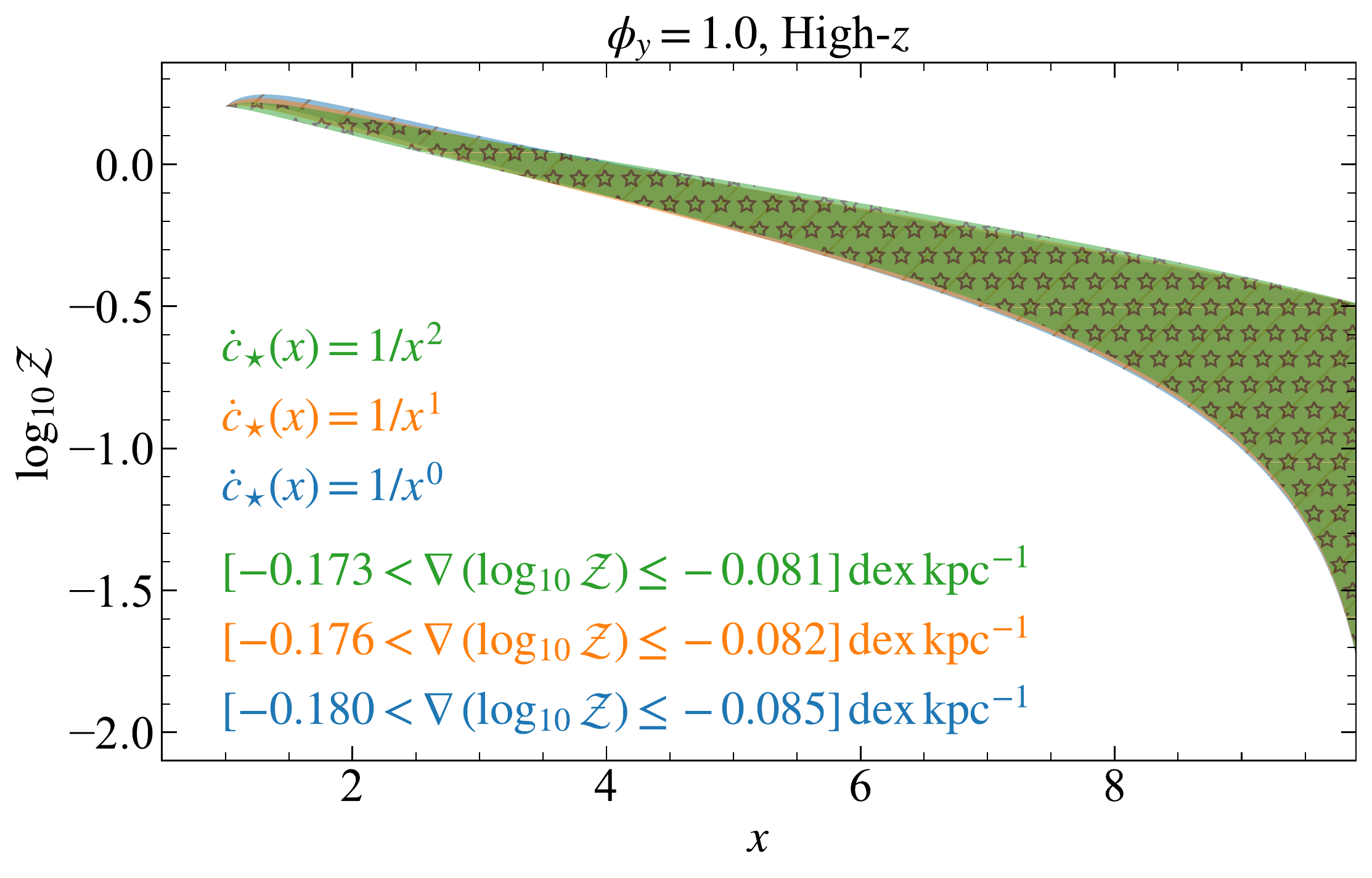}
\caption{Same as \autoref{fig:app_cosmicaccr_localspirals} but for high-$z$ galaxies. Similar to local dwarfs, changing the functional form of $\dot c_{\star}(x)$ has no impact on the metallicity distributions in high-$z$ galaxies.}
\label{fig:app_cosmicaccr_highz}
\end{figure}

\autoref{fig:app_cosmicaccr_localspirals}, \autoref{fig:app_cosmicaccr_localdwarfs}, and \autoref{fig:app_cosmicaccr_highz} show metallicity profiles with different $\dot c_{\star}(x)$ for local spirals, local dwarfs, and high-$z$ galaxies with $\phi_y = 1$, respectively; we use $\phi_y = 1$ because this maximises the dependence on $\dot{c}_\star(x)$ -- smaller $\phi_y$ values suppress variations. Note that the dimensionless parameters $\mathcal{P}$ and $\mathcal{S}$ are identical to that used in the \autoref{s:gradients} for the corresponding galaxies, but that $\mathcal{A}$ differs due to its dependence on $\dot{c}_\star$ (\autoref{eq:physicalA}). We see that changing the profile of $\dot{c}_\star(x)$ has no noticeable effect for local dwarfs or high$-z$ galaxies. This is because cosmic accretion is not a dominant term in the metallicity model for these galaxies, and the metallicities are instead mainly set by source, advection and diffusion.

The profile of $\dot{c}_\star(x)$ does matter for local spirals; as $\dot c_{\star}(x)$ flattens, metallicities in the inner regions of the disc reach higher values whereas the outer regions of the disc become more metal poor, thus leading to somewhat steeper gradients. For the most extreme case of constant $\dot{c}_\star(x)$, gradients are $\approx -0.05$ dex kpc$^{-1}$ steeper than our fiducial model. Thus, if the cosmic accretion profiles in local spirals are flatter than that we use in the main text, we expect slightly steeper gradients from the metallicity model, which is largely due to the SFR that we input from the galaxy evolution model. This is because under the input galaxy evolution model of \cite{2018MNRAS.477.2716K} where the SFR varies as $1/x^2$, flatter accretion profiles will dilute the metallicity in the metal-deficient outer regions by the same amount as that in the metal-rich inner regions, thus giving a larger difference between metallicities in the inner and outer regions in the disc. 

\section{Uncertainties in cosmological evolution of accretion and velocity dispersion}
\label{s:app_obsuncertainties}
Our predictions of metallicity gradient evolution depend on a scaling of $\sigma_g$ with $z$ derived from high-$z$ galaxy observations \citep{2015ApJ...799..209W}, and a scaling of $\dot M_{\rm{h}}$ with $z$ derived from cosmological simulations \citep{2008MNRAS.383..615N,2011MNRAS.417.2982F}. Since these two methods of deriving the scaling are very different, it is important to comment on any possible discrepancies between the two, and if they affect our results. Indeed, there is what appears at first glance to be an inconsistency: as we noted in \autoref{s:metalevolve_krumholz2018}, cosmological equilibrium demands that the inflow rate $\dot M$ through the disc (and, the star formation rate $\dot M_{\star}$) be similar to or less than the accretion rate onto the galaxy $\dot M_{\rm{ext}}$, in order to conserve the total mass. In terms of our model, the above condition translates into $\mathcal{P}/\mathcal{A} \lesssim \ln x_{\rm{max}}$ for $\dot c_{\star} = 1/x^2$. However, in many cases, our adopted scalings of $\sigma_g$ and $\dot{M}_h$ with $z$ give considerably larger values of $\mathcal{P}/\mathcal{A}$ at high-$z$. This discrepancy is simply a manifestation of the known problem that galaxies at $z\sim 2$ have star formation rates $\dot{M}_{\star} > \dot M_{\rm{ext}}$ \citep{2008ApJ...674..151E,2013ApJ...762L..31B,2017ApJ...837..150S}; high-$z$ galaxies obey the same observed scaling between star formation rate and velocity dispersion as local galaxies \citep{2018MNRAS.477.2716K, 2020MNRAS.495.2265V}, and since the inflow rate is directly set by $\sigma_g$, $\dot{M}_* > \dot M_{\rm{ext}}$ directly implies $\dot{M} > \dot{M}_{\rm ext}$.

The discrepancy between star formation rates (and velocity dispersions) and expected cosmological accretion rates has several possible explanations, but from the standpoint of our model for metallicity gradients, we can divide these into two main categories. One is that galaxies near the epoch of peak star formation do in fact form stars and move mass inward faster than their mean cosmological accretion rates, either because mass that was ejected at an earlier epoch falls back onto the galactic outskirts \citep[e.g.,][]{2016ApJ...824...57C,2019MNRAS.485.2511T}, or because of large angular momentum mismatch between the infalling material and the disc that triggers a sufficiently large radial inflow \citep{1981A&A....98....1M,2012MNRAS.426.2266B,2016MNRAS.455.2308P}, or because galaxies accumulate large gas reservoirs at $z\gtrsim 2$, which then flow into the star-forming portion of the disc due to compaction events \citep{2009ApJ...703..785D,2014MNRAS.438.1870D}, interactions \citep{2010ApJ...710L.156R} or mergers \citep{2018MNRAS.475.1160H} at $z\sim 2$. In these cases, the model we present is sufficient and we do not need to make any changes, since in such cases galaxies can maintain a large $\mathcal{P}/\mathcal{A}$ for a considerable time.

\begin{figure}
\includegraphics[width=\columnwidth]{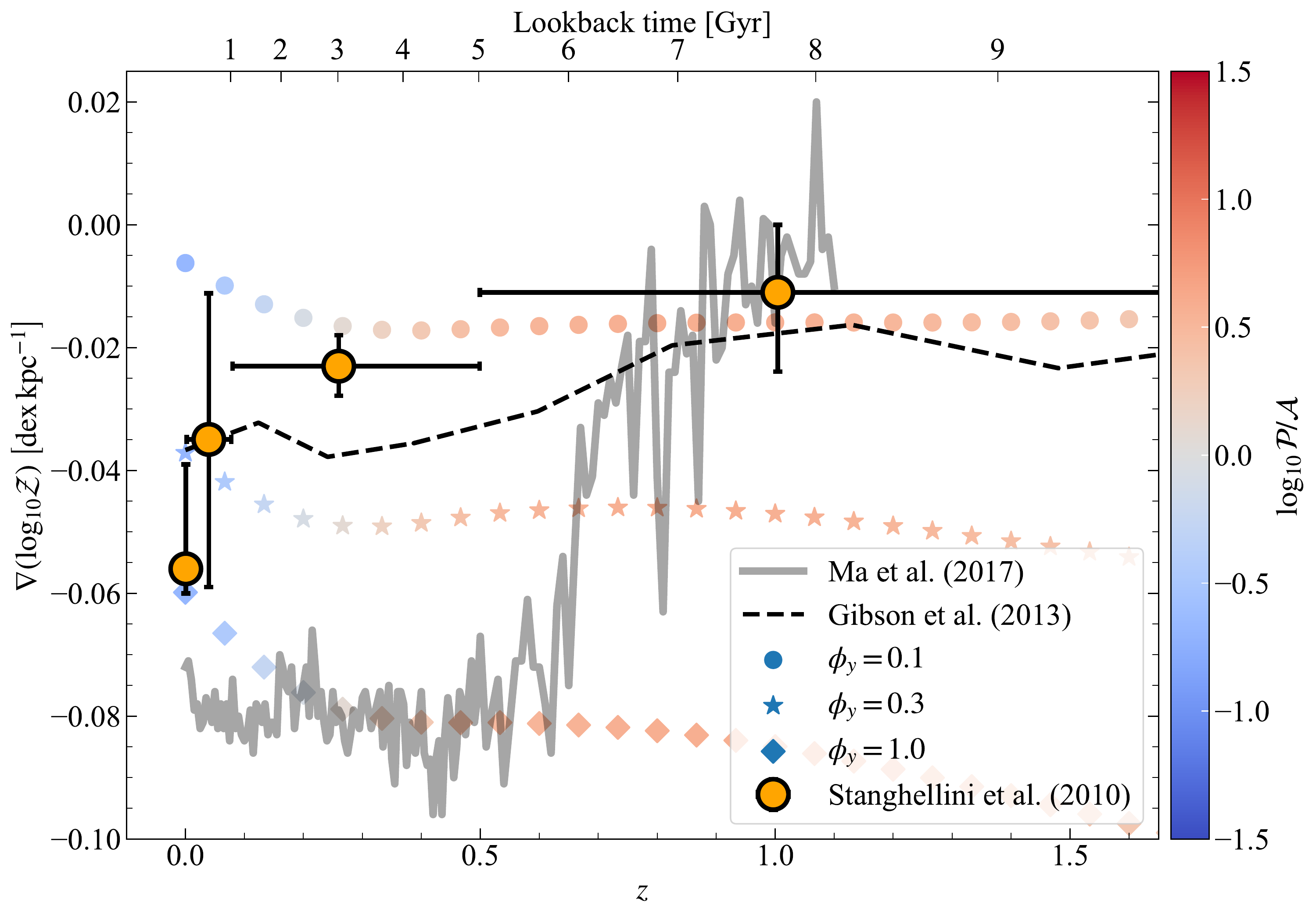}
\caption{Same as \autoref{fig:depends_redshift_onegalaxy}, but with $\sigma_g(z)$ reduced by a factor of $1+z$ as compared to \autoref{eq:fgas_wisnioski} (see \aref{s:app_obsuncertainties} for a discussion).}
\label{fig:app_milkyway}
\end{figure}

The other possibility is that the $\dot M_{\star}$ measured at high-$z$ are overestimated (e.g., \citealt{2019ApJ...877..140L,2020ApJ...893..111L}), thus altering the $\sigma_g-\dot{M}_*$ relationship; this is functionally equivalent to overestimating $\sigma_g$ at fixed $\dot M_{\star}$. Such an overestimate in high-redshift galaxies could plausibly be due to beam smearing, inclination uncertainty, and similar resolution-dependent (and thus redshift-dependent) factors \citep{2011ApJ...741...69D,2015MNRAS.451.3021D}. This would have a significant impact on $\mathcal{P}/\mathcal{A}$ because $\mathcal{P}/\mathcal{A} \propto \sigma^3_g$. To study the effects this can have on our results, we reproduce the expected cosmic evolution of metallicity gradient for the Milky Way in the model (cf. \autoref{fig:depends_redshift_onegalaxy}), but using $\sigma_g$ obtained from \autoref{eq:fgas_wisnioski} reduced by a factor of ($1+z$), in line with the redshift scaling of the limits of spectral resolution of different instruments \citep[Figure~2.3]{2020arXiv200909242M}. \autoref{fig:app_milkyway} presents the resulting metallicity gradients from the model, where the maximum $\log\mathcal{P}/\mathcal{A}$ is only 0.6. The qualitative shape of the model changes slightly for $z > 0.5$. However, $\phi_y$ remains the primary factor that drives the model gradients closer to the observed gradients across cosmic time. The main difference from our fiducial model is that steep gradients become possible at high-$z$ if $\phi_y$ is close to unity, because a smaller $\mathcal{P}/\mathcal{A}$ implies weaker homogenisation of the ISM by inward advection of metal-poor gas.

\begin{figure}
\includegraphics[width=\columnwidth]{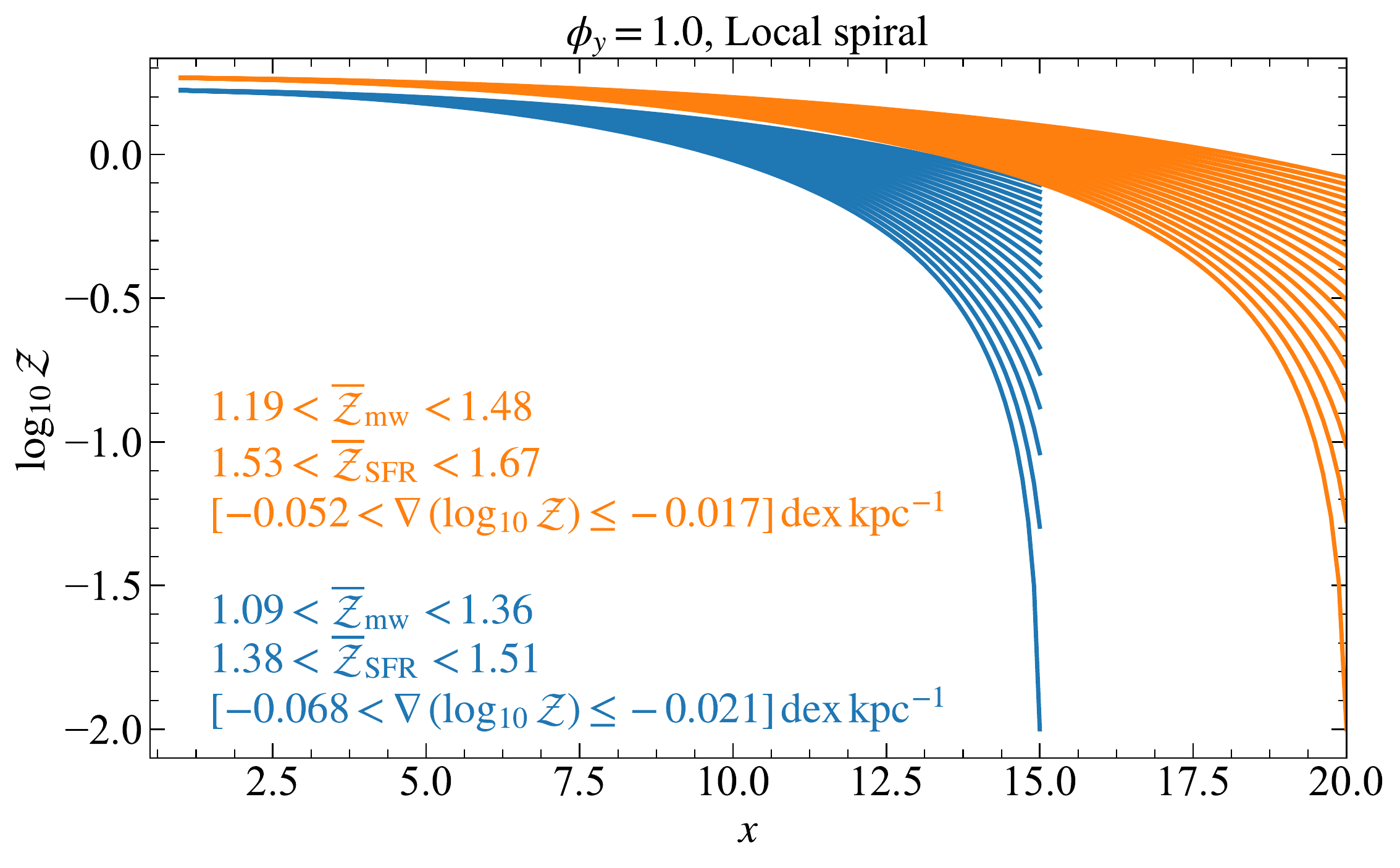}
\caption{Metallicity gradients as shown in the main text in \autoref{s:gradients}, but with an alternate value of the edge of the star-forming disc, $x_{\rm{max}}$. The profile of the distribution is preserved in each case, with slight variations in the absolute metallicities and metallicity gradients, with diminishing differences for decreasing $\phi_y$. Larger galaxies (in each galaxy class) show higher mean metallicity and flatter gradients.}
\label{fig:app_gradients_xmax_localspirals}
\end{figure}

\section{Dependence on the location of disc edges}
\label{s:app_xmax}
In the main text, we non-dimensionalize the solution in terms of $x$, where $x = r/r_0$ and we adopt a fiducial value of $r_0 = 1\,\rm{kpc}$, finding solutions for $x$ in the range $(x_{\rm min},\,x_{\rm max})$, with $x_{\rm min} = 1$ and $x_{\rm max}$ chosen based on observations of the sizes of the star-forming region for different galaxies. To explore the sensitivity of our results to the choice of our range in $x$, we show in \autoref{fig:app_gradients_xmax_localspirals} how the equilibrium gradients change in local spirals if we increase $x_{\rm{max}}$ from the fiducial 15 used in the main text to 20, noting that the qualitative trends remain the same across all types of galaxies. Increasing $x_{\rm{max}}$ leads to higher metallicities at each location in the disc, with slightly higher mean metallicities and shallower metallicity gradients. This is because increasing $x_{\rm{max}}$ decreases $\mathcal{A}$ since it means that there is a larger disc for the same total cosmic accretion rate (see \autoref{eq:haloaccr} and \autoref{eq:physicalA}). Thus, the ratio $\mathcal{S}/\mathcal{A}$ that appears in $\mathcal{Z}$ (see \autoref{eq:main_nondimx_solution}) increases, giving higher $\mathcal{Z}(x)$. Further, this increment in $\mathcal{S}/\mathcal{A}$ is reduced if $\phi_y < 1$ because $\mathcal{S} \propto \phi_y$. Thus, for lower $\phi_y$, changing $x_{\rm{max}}$ does not lead to any appreciable change in $\mathcal{Z}(x)$. Similarly, if we shift the inner edge of the galactic disc (where the rotation curve flattens out) by decreasing it to $x_{\rm{min}} = 0.5$ (or increasing it to $2$), the solution allows for slightly higher (lower) mean metallicities, and steeper (shallower) gradients.

Thus, this analysis implies that in galaxies where the transition from the star-forming disc to the bulge occurs at smaller galactic radius, the galaxy could potentially build slightly steeper metallicity gradients. Conversely, in galaxies where the star-forming disc is larger, we expect slightly shallower metallicity gradients. However, these variations remain insignificant compared to the scatter introduced due to other parameters, particularly $\phi_y$. Thus we do not regard variations in $x_{\rm min}$ or $x_{\rm max}$ as a substantial uncertainty in the model.

\bsp	
\label{lastpage}
\end{document}